%% file: main.tex
\documentclass[12pt]{article}

\input{preamble}

\title{Contextual Visual Distinctiveness in Online Product Search}
\author{Felicia Nguyen}
\date{\today}

\begin{document}

\maketitle

\begin{abstract}
\noindent \onehalfspacing
In online product search, returned alternatives often look alike. We investigate when a product's visual separation from its closest look-alike in a returned set increases its choice probability. We introduce two occasion-level constructs: \textit{contextual visual distinctiveness} (image distance from the nearest returned neighbor) and \textit{relative fit} (compatibility with the search query). We hypothesize that distinctiveness favors selection, yields a premium that rises with relative fit, and matters most when text descriptions fail to differentiate options. Analyzing over 800,000 e-commerce search events using dense representations, we apply search-event and product fixed effects to evaluate the exact same product alongside varying visual neighbors. Results show a product is significantly more likely to be clicked when it lacks a close look-alike. This distinctiveness premium increases with relative fit and roughly doubles when competing descriptions are highly similar. Consistent with a model where distinctiveness aids in standing out pre-evaluation rather than increasing inherent utility, the extra clicks distinctiveness recruits convert 5--7\% less often downstream. Ultimately, the
evidence characterizes visual differentiation as a local attention allocation mechanism whose
value depends on query fit and information from competing cues.\\

\medskip

\end{abstract}

\clearpage

\input{sections/01_introduction}
\input{sections/02_theory}
\input{sections/03_data_measurement}
\input{sections/04_empirical_strategy}
\input{sections/05_results}
\input{sections/06_discussion}

\clearpage
\singlespacing
\bibliographystyle{aer}
\bibliography{references}

\ifinlineexhibits\else
  \clearpage
  \input{exhibits/main_exhibits}

  \FloatBarrier
\fi

\clearpage
\appendix
\setcounter{table}{0}
\setcounter{figure}{0}
\renewcommand{\thetable}{A\arabic{table}}
\renewcommand{\thefigure}{A\arabic{figure}}
\renewcommand{\theequation}{A\arabic{equation}}
\setcounter{equation}{0}
\section*{APPENDICES}
\input{appendices/web_appendix}

\end{document}

%% file: preamble.tex
\usepackage[T1]{fontenc}
\usepackage[utf8]{inputenc}

\usepackage[letterpaper,margin=1in]{geometry}
\usepackage{microtype}
\usepackage{setspace}
\usepackage[compact]{titlesec}

\usepackage{amsmath}
\usepackage{amssymb}
\usepackage{mathtools}
\usepackage{amsthm}
\usepackage{bm}
\usepackage{newpxtext, newpxmath}
\newtheoremstyle{hypothesisstyle}
  {9pt}{9pt}
  {\itshape}{}
  {\bfseries}{.}
  {0.6em}{}
\theoremstyle{hypothesisstyle}
\newtheorem{hypothesis}{Hypothesis}
\newtheorem{proposition}{Proposition}
\newtheorem{corollary}{Corollary}
\usepackage{tikz}
\usetikzlibrary{matrix,positioning,arrows.meta}
\usepackage{graphicx}
\usepackage{booktabs}
\usepackage{array}
\usepackage{tabularx}
\usepackage{longtable}
\usepackage{threeparttable}
\usepackage{adjustbox}
\usepackage{float}
\usepackage{placeins}
\usepackage{caption}
\usepackage{xcolor}
\newif\ifinlineexhibits
\inlineexhibitstrue

\newcolumntype{Y}{>{\raggedright\arraybackslash}X}
\newcommand{\TBD}{\textcolor{gray}{\textit{TBD}}}

\newcommand{\ExhibitPlaceholder}[2]{%
  \fbox{\begin{minipage}[c][#1][c]{0.90\textwidth}
  \centering\small\textcolor{gray}{#2}
  \end{minipage}}}

\usepackage[round,authoryear]{natbib}
\definecolor{linkcol}{HTML}{A63A2B}
\usepackage[
  colorlinks=true,
  linkcolor=linkcol,
  citecolor=linkcol,
  urlcolor=linkcol,
  breaklinks=true
]{hyperref}
\usepackage[nameinlink,noabbrev]{cleveref}
\hypersetup{
  pdftitle={Contextual Visual Distinctiveness in Online Product Search},
  pdfauthor={Felicia Nguyen},
  pdfsubject={Quantitative marketing manuscript},
  pdfkeywords={product search, visual differentiation, semantic fit, ranking}
}

\usepackage{fancyhdr}
\allowdisplaybreaks

%% file: sections/01_introduction.tex

\section{Introduction}
\label{sec:introduction}

A shopper searches an online store for a black leather office chair. The engine returns two dozen products, and most of them are photographed the same way: a dark chair, three-quarter view, white background. For any one of these products, the practical question is not whether it looks unusual in the abstract but whether the page contains a near twin. A product displayed beside an almost identical alternative is easy to skip and easy to confuse; the same product, returned an hour later in a set that contains nothing like it, stands alone as the recognizable version of what the shopper asked for. And how much that separation is worth should depend on the query: standing apart helps a product that answers the question, while a poor match that stands apart is merely conspicuous.

This paper studies when visual separation from the other returned products helps a product get chosen. Two constructs describe a product's position on a search occasion. \emph{Relative fit} is the product's standing, within the returned set, on compatibility with the goal the query expresses: whether it is among the best available answers or the worst. \emph{Contextual visual distinctiveness} is the visual distance between the product's image and its nearest visual substitute, the most similar product returned in the same search. Both are properties of the occasion rather than of the product. The same product is the best answer to one query and mid-pack for another, and it has a near twin in one returned set and none in the next. We test three predictions. First, distinctiveness from the nearest substitute favors selection: the same product is chosen more often on occasions when the returned set leaves it without a close look-alike. Second, this advantage carries what we call a \emph{goal-consistency premium}: in probability units it grows with the product's relative fit, because visual separation matters only for products whose merits give the consumer something to act on. Third, the advantage is largest where other cues fail to separate the alternatives, because a consumer leans on whichever channel discriminates. A simple model in which a product must be individuated, perceived as a distinct option, before its merits can bear on choice turns these into propositions and adds a sharp prediction about estimation scale: the premium should appear in a linear probability model and vanish in a conditional logit, because distance operates on individuation rather than on utility. All three predictions concern how clicks are allocated among returned products. A descriptive companion analysis at the level of the search event, using the zero-click searches the choice analysis must exclude, asks what a returned set's visual composition signals about the occasion that produced it.

Existing work supplies the pieces of this argument without assembling them. Research on goals shows that an active objective determines which products count as plausible and which attributes matter \citep{ratneshwar1996goal,haubltrifts2000consumer,kim2011mapping,bronnenberg2016zooming}. Research on context-dependent choice shows that surrounding alternatives change which comparisons are made and how much weight distinguishing features receive \citep{tverskysimonson1993context,dharsherman1996common,rooderkerk2011context}. Research on product design documents effects of aesthetics, prototypicality, novelty, and visual uniqueness measured against a category or a marketplace \citep{bloch1995ideal,landwehr2011gut,feng2025visual,peng2026metaanalysis}, and visual-search research links a target's similarity to its most similar distractors to attention and identification \citep{duncan1989visual,huang2021now,vanderlans2021online}. What these literatures do not establish is whether separation from the products actually returned for a query improves selection, and whether that advantage depends on the product's standing on the stated goal. Search and recommendation systems take a position on the second question by default: standard diversification objectives add a relevance term and a nonredundancy term, so visual novelty earns the same premium whether or not the product fits the query \citep{carbonellgoldstein1998mmr,ziegler2005diversification}. Our framework gives nonredundancy a value of its own and predicts where it is worth most.

We test the predictions with the Coveo SIGIR e-commerce dataset, which records more than 800,000 search events from a mid-size online retailer \citep{tagliabue2021sigir}. Each search event lists the products returned to the client and the products subsequently clicked, and the release supplies dense representations of the query, of each product's description, and of each product's image, together with hashed catalog categories. Semantic compatibility between a query and a product is measured as an aligned cosine similarity, using a retrieval-estimated map that carries query vectors into description space (Section~\ref{subsec:semantic-fit}); relative fit is the product's rank on that similarity within its returned set. Contextual distinctiveness is one minus the product's highest image similarity with any returned peer. Because the raw text and images are withheld, we cannot inspect what the vectors encode. Because the realized returned sets are preserved, we can measure each product's position relative to its actual competitors on each occasion.

The empirical design treats each single-click search as a choice among the returned products. Search-event fixed effects absorb the query, the session state, the time, and every other condition common to the returned set. Product fixed effects absorb stable appeal, image quality, price position, popularity, and global visual distinctiveness. Flexible indicators for recorded position absorb the placement gradient documented in online search \citep{ghose2014ranking,ursu2018rankings}. What remains is variation in the same product's fit and visual position across the searches in which it appears. The preferred specification separates distinctiveness into a product mean and an occasion-specific deviation and asks the deviation to carry the effect: is a product more likely to be chosen when the current set leaves it farther from its nearest substitute than usual? The premium specification interacts fit with that deviation and, separately, with the product-mean component, so that a fit-dependent pattern among enduringly unusual products cannot masquerade as a contextual one. Category-homogeneous sets, same-category neighborhoods, textual differentiation controls, price controls, position restrictions, a conditional logit, and a representation placebo probe the interpretation, and a pre-stated block of measurement contrasts re-estimates the same models under alternative constructions of both measures, asking how far the operative visual field extends beyond the nearest substitute and which fit quantity carries the association.

The evidence answers the first question clearly, and it answers it in a way that a catalog-level analysis would have missed. Pooled across products, looking unlike one's neighbors predicts nothing about being chosen. Hold the product fixed, so that the same image is compared with itself across the different sets retrieval places it in, and a positive association appears: a product is more likely to be clicked on the occasions when the returned page happens to contain no close look-alike, by roughly five percent of a random-choice click, and no draw of a placebo that reassigns images across products reproduces it. The two facts belong together rather than in tension. What helps a product is not being unusual but being locally unsubstitutable, and because retrieval reshuffles a product's neighbors from one query to the next, the same image is both of those things on different occasions.

The advantage is concentrated where the framework says it should be, and the manner of the concentration says what kind of advantage it is. It grows with the product's standing on the query, so separation is worth most to products that are already plausible answers and close to nothing to products that are not. But it grows only in probability units and vanishes in log-odds, which is the signature that a model of individuation predicts and that a taste for goal-consistent distinctiveness does not. The remaining results follow the same logic. The visual channel does about twice as much work in sets whose descriptions sit close together as in sets whose descriptions already separate the alternatives, which is what one expects of a cue that carries information and not of one that merely decorates. And when two returned products are near-identical, the higher-placed one wins about two thirds of their contests while the better-fitting one wins barely more often than a coin toss, which is what happens when options that have been conflated are resolved by salience rather than by merit. Visual difference, on this reading, buys a product the chance to be weighed on its merits; it does not make the product more attractive.

What that chance is worth is bounded, and the purchase funnel is what bounds it. The clicks that distinctiveness recruits convert into carts and purchases proportionally less often at every stage, abandonment included, and the total downstream activity within a returned set is indistinguishable from unchanged, so inside the page distinctiveness moves attention between products rather than creating demand. A chronological holdout points the same way: adding distinctiveness to a relevance score does not reorder held-out clicks. The picture is of a mechanism that is real, local, and narrow. It decides which of several plausible answers absorbs a click; it decides this most where the query is answered well and the words fail to discriminate; and it does not decide whether the search succeeds. That is a smaller claim than diversification objectives implicitly make, and a more actionable one, because it says where visual difference is worth engineering and where it is not.

The paper makes three contributions. First, it joins goal compatibility to relational differentiation and identifies when each cue decides. Design research asks whether a product looks attractive, typical, or unusual; context-dependent choice research asks how surrounding options alter comparison. The goal-consistency premium ties the two together: standing apart from one's closest competitor is worth more when the product that stands apart still answers the question the consumer asked. The channel-substitution result completes the pair by showing the reverse dependence, that appearance matters most where the returned descriptions stop separating the alternatives, which is the signature of a cue that carries information rather than decoration. Second, it changes the reference set against which visual differentiation is measured. Global uniqueness compares an image with a category, a marketplace, or a learned population \citep{feng2025visual}; our measure compares it with the single most substitutable product returned in the current search, and the within-product design separates that local position from stable visual appeal. The paper uses existing representation spaces for a measurement problem in demand analysis, in the spirit of recent work that extracts economically meaningful features from images \citep{liu2020visual,zhang2022image,dew2022logos,burnap2023aesthetic}; it does not propose a new image model. Third, it turns the additive relevance-diversity objective into a testable restriction, evaluates the alternative out of sample, and asks at the level of the returned set whether redundancy costs the platform engagement or merely moves it between sellers. The results speak to how a search system should allocate visual diversity: toward reducing near-duplication among strong answers to the query, and with most to gain where the candidates' descriptions leave them hardest to tell apart, not uniformly across the candidate list.

The scope of the evidence is deliberately narrow. Returned products form a platform-generated candidate set rather than an observed consideration set \citep{abaluckadamsprassl2021consider}, the platform chooses retrieval and position with signals absent from the public release, and a click records selection from the returned products rather than purchase. The estimates are conditional associations within realized search environments under the platform's historical retrieval and display policy. Within that boundary, the paper isolates a recurring feature of digital choice: consumers meet each product both as an answer to a query and as a member of a competitive visual field, and a product's difference is useful when it distinguishes a plausible answer from other plausible answers.

The rest of the paper proceeds as follow: Section~\ref{sec:theory} develops the constructs and predictions. Section~\ref{sec:data} describes the data and measures, Section~\ref{sec:empirical} the estimands and design, Section~\ref{sec:results} the confirmatory results, Section~\ref{sec:robustness} the mechanisms, robustness, and out-of-sample validity, and Section~\ref{sec:discussion} the implications and limitations. The Web Appendix contains additional tables and figures, further robustness and exploratory analysis, descriptions of the data and transformations, and the proofs of the analytical model.

%% file: sections/02_theory.tex

\section{Theoretical Framework}
\label{sec:theory}

\subsection{Choice within returned sets and relative fit}
\label{subsec:goal-compatibility}

Choice requires a standard for deciding which attributes matter, and an active consumption goal often supplies it. Goals organize products by their usefulness for the task at hand, pulling together options from different conventional categories and separating products that otherwise look alike \citep{ratneshwar1996goal}. A search query makes part of this goal observable. It expresses a product type, a use, an attribute, or a combination of these with varying precision, and the retrieval system uses it to select a candidate set from the catalog. Retrieval lowers search costs but does not end comparison: consumers continue through a narrow region of attribute space, and their paths depend on which alternatives appear and in what order \citep{haubltrifts2000consumer,kim2011mapping,bronnenberg2016zooming,chen2017sequential}.

The choice that follows is comparative. A consumer who clicks selects from the products the platform returned, not from the catalog, so the operative question about any candidate is not how well it matches the query in the abstract but how well it matches relative to the alternatives displayed with it. We define \emph{relative fit} as a product's standing, within its returned set, on compatibility with the goal the query expresses. Relative fit is specific to the search occasion twice over: the same product answers different queries differently, and the same answer occupies a different standing depending on the strength of the products returned beside it. A moderately compatible product can be the best available answer to a vague query, and a closely compatible product can sit mid-pack when retrieval returns a strong set. Framing fit relationally also respects a measurement reality of search: queries differ in breadth, vocabulary, and precision, so levels of query-product similarity are not calibrated across occasions, while a product's standing among the candidates scored against the same query is. Relative fit differs from category prototypicality, which measures resemblance to a category representation \citep{veryzerhutchinson1998unity,landwehr2011gut}: an unusual category member can be the closest available answer to a query, and a prototypical product can rank poorly when it lacks a requested attribute. Higher relative fit should raise the probability of selection, but that relationship is not the paper's claim; it validates the measure. Our questions concern the product's visual position among the returned alternatives and how fit changes its value.

Two clarifications about the choice environment carry through the paper. The platform, not the consumer, generates the candidate set, and returned products need not all enter the consumer's consideration set \citep{abaluckadamsprassl2021consider}. We use \emph{returned set} for the products the platform returned and \emph{selection} for the returned product that received the click, conditional on a click occurring.

\subsection{Contextual distinctiveness}
\label{subsec:contextual-differentiation}

Visual differentiation has meaning only relative to a reference set. Similarity judgments depend on which objects are compared and which features that comparison makes salient \citep{tversky1977features}, so a product can be distinctive in one assortment and ordinary in another. Within a grid of product images, moreover, the cost of looking similar is not spread evenly over the assortment; it is concentrated in the closest look-alike. Visual-search research shows that identifying a target slows as the target comes to resemble its distractors, with the most similar distractors doing the damage \citep{duncan1989visual}, and salience, proximity, and position govern which products attract inspection in shopping displays \citep{milosavljevic2012saliency,atalay2012center,huang2021now}. Choice research reaches a parallel conclusion: features shared with close alternatives provide no basis for discrimination, which shifts weight onto the features that separate them \citep{dharsherman1996common}, and a product flanked by a near twin competes for the same moment of attention and the same slot in the consideration set. Assortment research treats visual structure as relational in the same way: perceived variety depends on distances among products and their arrangement, not only on their number \citep{hoch1999variety,kahnwansink2004assortment,deng2016wide}, image-based presentation raises perceived variety while making large assortments harder to process \citep{townsendkahn2014visual}, and visual frames that separate products encourage by-alternative comparison and reduce deferral \citep{jia2025framing}.

We therefore define \emph{contextual visual distinctiveness} as the visual distance between a focal product and its nearest visual substitute, the most similar product returned in the same search. The definition is local, because the returned products are the reference group; product-specific, because products in the same set have different nearest neighbors; and occasion-varying, because the same image acquires a new nearest substitute whenever retrieval assembles a new set. It differs from global visual uniqueness, which compares an image with a broad marketplace, category, or learned population \citep{feng2025visual}: a globally ordinary chair is highly distinctive in a set that contains nothing like it, and a globally unusual chair is redundant beside its twin. The nearest-substitute form is the sharpest theoretical object, because the peer that competes for identification and substitution is the most similar one, and because the pairwise analyses below require an identified nearest substitute. How far the operative visual field extends beyond that single peer is an empirical question rather than something the theory settles: interference in visual search is dominated by the most similar distractors, but crowding from the surrounding field also impairs identification, and an average distance over all returned peers weights exactly that field. Section~\ref{subsec:robustness} estimates the same models under the all-peer and trimmed aggregations, and Section~\ref{subsec:robustness-results} reports what the comparison shows.

Why should distinctiveness from the nearest substitute favor selection? Retrieval has already screened the returned set for rough goal relevance, so within such a set visual separation mostly distinguishes plausible answers rather than marking intruders. A product with no near twin is easier to individuate during a fast visual scan, is less likely to be confused with or substituted by a neighbor, and offers the consumer a discriminating reason for choosing it over alternatives that otherwise blend together \citep{chernev2005feature}; differentiation can acquire value even when the differentiating attribute is substantively minor \citep{carpenter1994meaningful}. The prediction is not unconditional as a matter of aesthetics: fluency, unity, and prototypicality improve evaluation \citep{veryzerhutchinson1998unity,landwehr2011gut}, and novel designs impose processing costs and uncertain inferences \citep{hekkert2003maya,rubera2015design,feng2025visual}. Those forces, however, attach mainly to the image itself and to the product's enduring look, which product fixed effects hold constant. The occasion-level claim is about the same image meeting different neighbors: a product should be selected more often on the occasions when the returned set leaves it without a close visual substitute than on the occasions when it sits next to one.

\subsection{The goal-consistency premium}
\label{subsec:goal-consistent}

Relative fit and distinctiveness describe two relations, one between the product and the query and one between the product and its closest competitor. Fit gives visual separation its interpretation. When a product ranks among the best answers to the query, standing apart from its nearest look-alike separates one plausible answer from other plausible answers: the product remains anchored to the goal, so its visual separation aids individuation, clarifies comparison, and supplies a reason for choosing among otherwise acceptable options. When the product ranks poorly, the same separation carries less assurance; the product may attract inspection while appearing inconsistent with what was asked for, and the inspection it attracts is spent confirming that it is not the answer. Distinctiveness should therefore command a premium among the products that fit: the association between distinctiveness and selection should become more favorable as relative fit rises. We call this the \emph{goal-consistency premium}. It is a statement about joint evaluation, not about a sequence of mental stages; consumers need not judge relevance first and appearance second for fit to change what visual separation is diagnostic of.

The premium is also what separates the account from a pure salience story. If visual separation worked only by capturing attention, it would help wherever it occurred, and the distinctiveness slope would not vary with the product's standing on the goal. If instead separation works by distinguishing a credible answer from its close competitors, the slope should rise with relative fit. The two accounts agree about the average effect and disagree about its gradient, which makes the gradient the informative test.

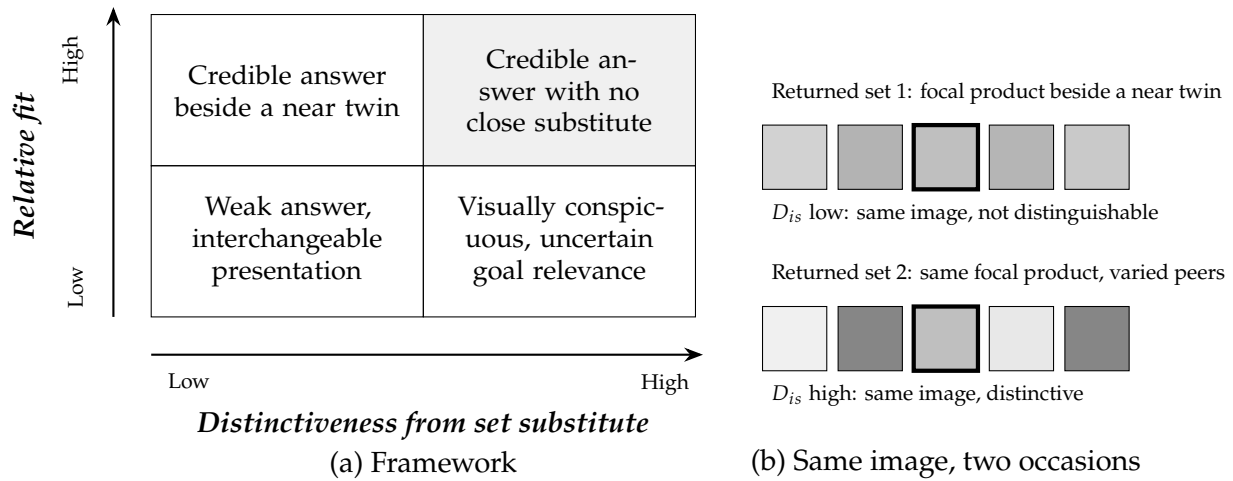
\begin{figure}[!ht]
\centering
\begin{minipage}[b]{0.56\textwidth}
\centering
\begin{tikzpicture}[
  cell/.style={draw, rectangle, minimum width=3.6cm, minimum height=2.0cm,
               text width=3.2cm, align=center, inner sep=3pt, font=\footnotesize},
  ax/.style={-{Stealth[length=2.5mm]}, thick},
  axtitle/.style={font=\small\bfseries\itshape},
  tick/.style={font=\scriptsize}
]
\node[cell] at (1.8,1.0) {Weak answer, interchangeable presentation};
\node[cell] at (5.4,1.0) {Visually conspicuous, uncertain goal relevance};
\node[cell] at (1.8,3.0) {Credible answer beside a near twin};
\node[cell, fill=gray!12] at (5.4,3.0) {Credible answer with no close substitute};
\draw[ax] (-0.5,0) -- (-0.5,4.1);
\draw[ax] (0,-0.5) -- (7.3,-0.5);
\node[tick, rotate=90, anchor=south] at (-0.75,0.4) {Low};
\node[tick, rotate=90, anchor=south] at (-0.75,3.4) {High};
\node[tick, anchor=north] at (0.5,-0.6) {Low};
\node[tick, anchor=north] at (6.8,-0.6) {High};
\node[axtitle, rotate=90, anchor=south] at (-1.3,2.0) {Relative fit};
\node[axtitle, anchor=north] at (3.6,-1.1) {Distinctiveness from set substitute};
\node[font=\small] at (3.6,-1.95) {(a) Framework};
\end{tikzpicture}
\end{minipage}\hfill
\begin{minipage}[b]{0.42\textwidth}
\centering
\begin{tikzpicture}[
  prod/.style={draw, rectangle, minimum width=0.85cm, minimum height=0.85cm, inner sep=0pt},
  lab/.style={font=\scriptsize, align=left}
]
\node[lab, anchor=west] at (0,4.55) {Returned set 1: focal product beside a near twin};
\foreach \x/\shade in {0/35,1/60,2/55,3/58,4/42} {
  \node[prod, fill=gray!\shade] at (0.45+\x*1.0,3.7) {};
}
\node[prod, fill=gray!50, ultra thick, draw=black] at (0.45+2*1.0,3.7) {};
\node[lab, anchor=west] at (0,2.95) {$D_{is}$ low: same image, not distinguishable};
\node[lab, anchor=west] at (0,2.15) {Returned set 2: same focal product, varied peers};
\foreach \x/\shade in {0/12,1/95,2/50,3/18,4/95} {
  \node[prod, fill=gray!\shade] at (0.45+\x*1.0,1.3) {};
}
\node[prod, fill=gray!50, ultra thick, draw=black] at (0.45+2*1.0,1.3) {};
\node[lab, anchor=west] at (0,0.55) {$D_{is}$ high: same image, distinctive};
\node[font=\small] at (2.45,-0.35) {(b) Same image, two occasions};
\end{tikzpicture}
\end{minipage}
\caption{Goal-consistent contextual visual differentiation. Panel (a) crosses relative fit with distinctiveness from the nearest visual substitute in consideration set; the shaded cell is the configuration in which distinctiveness should be worth most. Panel (b) shows the same focal product (bold border) returned in two sets; shading stands for image content, and the product's distinctiveness $D_{is}$ differs across the two occasions although the image is unchanged.}
\label{fig:conceptual-framework}
\end{figure}

Figure~\ref{fig:conceptual-framework} summarizes the argument. Panel~(a) crosses relative fit with contextual distinctiveness. Low-fit, low-distinctiveness products offer neither a strong answer nor a basis for comparison; low-fit, high-distinctiveness products are conspicuous with uncertain goal relevance; high-fit, low-distinctiveness products are plausible but interchangeable with a close look-alike; and high-fit, high-distinctiveness products combine credibility with comparative separation, the configuration in which distinctiveness should be worth most. Panel~(b) illustrates why the measure is contextual: the same product image is redundant in one returned set and distinctive in another, so its distinctiveness varies across occasions while the image itself is fixed. This occasion-level variation is what the empirical design uses.

The framework departs from an additive relevance-diversity objective. Diversification methods in information retrieval and recommendation score candidates with separate relevance and nonredundancy components \citep{carbonellgoldstein1998mmr,ziegler2005diversification,agrawal2009diversifying}, which gives visual separation the same marginal value at every level of fit. Our account gives nonredundancy a value of its own and then says where it should be worth most: among the products that remain credible answers to the query.

\subsection{A simple model of individuation and selection}
\label{subsec:model}

A small model makes the argument precise and yields predictions the verbal account does not, including one about estimation scale. Conditional on a click occurring, let \(m_{is}\) denote product \(i\)'s \emph{merit}: the probability that a consumer who perceives every returned product as a distinct option would select \(i\). Merits sum to one within the set and follow any standard random-utility rule \citep{mcfadden1974conditional}; relative fit, position, and tastes enter here, and the model is silent about their weights. The friction is upstream of evaluation. Scanning a grid of listings, a product must first be \emph{individuated}, perceived as a distinct option, before its merits can bear on the choice, and individuation is what similarity to the nearest distractor impairs \citep{duncan1989visual}. Formally, product \(i\) fails individuation with probability
\begin{equation}
q_{is}=q_v\!\left(D_{is}\right)\,q_t\!\left(T^{NN}_{is}\right),
\label{eq:model-failure}
\end{equation}
where \(D_{is}\) is visual distance from the nearest substitute, \(T^{NN}_{is}\) the corresponding textual distance, and \(q_v,q_t\in(0,1)\) are decreasing: either channel can separate the product from its neighbor, and failure requires both to fail. A product that fails individuation is conflated with its nearest substitute. The conflated pair competes as a single option whose merit is that of its better member, because a duplicated answer does not double its appeal, and a click landing on the pair falls on its more salient member, with salience share \(\sigma_{is}\) governed by display position. The choice stage then operates on the individuated options and conflated clusters. This is a stochastic consideration structure in the sense of \citet{manzinimariotti2014stochastic}, with the consideration friction tied to a measurable perceptual quantity rather than left free. Web Appendix~\ref{app:model} derives exact choice probabilities; in a large set they reduce, to first order, to
\begin{equation}
P_{is}=\bigl(1-q_{is}\bigr)\,m_{is}+r_{is},
\label{eq:model-reduced}
\end{equation}
where the spillover \(r_{is}\) collects mass received from other products' failures, is of order \(1/J_s\), and does not depend on \(i\)'s own distances except in the mutually nearest case treated below. Everything the framework predicts comes from reading Equation~\ref{eq:model-reduced} in four directions, and Web Appendix~\ref{app:model} proves each claim.

Take first the direct effect of moving a product away from its closest look-alike.

\begin{proposition}[Distinctiveness]
\label{prop:distinctiveness}
Selection probability increases in distance from the nearest visual substitute:
\(\partial P_{is}/\partial D_{is}=-q_v'\,q_t\,m_{is}>0\).
\end{proposition}

The derivative runs through \(q_{is}\), which is a property of the occasion rather than of the product, so the prediction concerns within-product variation: the same image should do better on the occasions when the returned set happens to leave it isolated. Nothing in the statement implies that habitually unusual products are chosen more often, and this is why the empirical design decomposes distance into a product mean and a deviation from it rather than using the level.

That derivative also carries a factor of merit, and following the factor rather than the sign produces the sharpest thing the model contributes, a prediction about the estimation scale itself.

\begin{proposition}[Two scales]
\label{prop:scales}
The distinctiveness slope is proportional to merit, so on the probability scale the cross-partial with any merit component, including relative fit, is positive: \(\partial^2 P_{is}/\partial D_{is}\,\partial m_{is}=-q_v'\,q_t>0\). In latent-utility units, however, \(\log P_{is}=\log m_{is}+\log(1-q_{is})\) up to the spillover term, so the distance effect is additive and carries no first-order interaction with merit.
\end{proposition}

The two halves of that statement are why the paper estimates the premium twice on two scales. A positive fit-by-distinctiveness interaction in a linear probability model together with a null interaction in a conditional logit is not a discrepancy to be explained away but the exact signature of a mechanism operating on individuation: the linear interaction is an implication of the level effect rather than a second force acting alongside it. A genuine preference for goal-consistent distinctiveness would appear on both scales, so the pair of estimates discriminates between the two accounts in a way that neither estimate alone can.

Because the two channels enter Equation~\ref{eq:model-failure} multiplicatively, each becomes more valuable as the other becomes less informative.

\begin{proposition}[Channel substitution]
\label{prop:channels}
The magnitude of the visual-distance effect is increasing in \(q_t\), and symmetrically the magnitude of the textual effect is increasing in \(q_v\).
\end{proposition}

Visual separation should therefore matter most where words fail to separate the alternatives. This is the prediction that distinguishes a cue carrying information from one that merely decorates, since a decorative attribute has no reason to become more consequential precisely where the substantive channel stops discriminating.

The remaining two results concern the limits of the mechanism rather than its presence, and the first describes what happens when individuation fails almost surely.

\begin{proposition}[Near twins]
\label{prop:twins}
For a mutually nearest pair at small distance, conditional on the pair receiving the click, the probability that the focal member wins tends to its salience share \(\sigma\) as \(q\to1\); merit differences enter only through the vanishing \((1-q)\) branch.
\end{proposition}

Between near-identical listings, in other words, display position decides and merit breaks only a small part of the remaining tie. The implication is testable and unflattering to the mechanism itself: the model predicts its own irrelevance in exactly the cases where visual similarity is most extreme, which makes the near-twin pairs a useful place to look for a failure of the account.

The second limit concerns how the effect scales with the size of the returned set.

\begin{corollary}
\label{cor:setsize}
Because merits are of order \(1/J_s\), both the distinctiveness slope and the probability-scale interaction shrink with returned-set size.
\end{corollary}

This bears on interpretation as much as on prediction. A mechanical artifact of set size would produce a superficially similar pattern, since distance to the nearest of many neighbors falls as sets grow while every within-set slope scales inversely with the number of alternatives, and separating the two is what Section~\ref{subsec:premium-diagnostics} is for.

The model also places boundaries on what claims we can make. Since our framework is conditional on a click, it says nothing about whether a returned set produces one, and the set-level analysis below remains descriptive. We treat \(q_v\) and \(q_t\) as decreasing but otherwise free, so no functional forms are imposed. And it locates the entire effect of visual separation in individuation rather than in preference: consumers in this model do not necessarily value distinctiveness more \emph{per se}, they are simply able to act on the merits of what they can tell apart, i.e., cognitive friction-reducing instead of utility-enhancing.

\subsection{Two further margins}
\label{subsec:further-margins}

The premium says which product inside a returned set benefits from visual separation. Two further questions follow, one about which sets make visuals decisive and one about what redundancy costs the set as a whole.

Consider first what the consumer has to work with. A product can be told apart from its neighbors through more than one channel: its image can separate it, and so can its description. A consumer leans on whichever cue discriminates \citep{feldmanlynch1988self,chernev2005feature}, and in the model of Section~\ref{subsec:model} the channels enter multiplicatively, so the visual channel carries the most weight exactly where the textual channel fails to separate the alternatives. The prediction is therefore about the composition of the returned set: the distinctiveness effect should be larger in sets whose descriptions sit close together, where words cannot do the separating, and smaller where the descriptions already discriminate. This channel claim is distinct from a claim about the dispersion of query compatibility. In the model, how far apart the set's products stand on the goal operates through a different slot, the merit at stake, and Section~\ref{subsec:contest-results} reports both moderators and finds that they behave as the two slots imply rather than as a single contested-set story would.

Consider next what a returned set's visual composition says about the search occasion itself. Conditioning on a click, redundancy can only redistribute: if one look-alike loses, another gains, and the click still lands somewhere. Whether the total is also affected is a question the choice model cannot reach, because it conditions on a click having occurred, and it can only be asked at the level of the search event, where zero-click searches are observed. Two forces push in opposite directions there. Interference is one: a consumer who cannot separate the alternatives may inspect nothing, reformulate, or leave \citep{iyengarlepper2000choice,townsendkahn2014visual,jia2025framing}. Retrieval quality is the other, and it is the stronger candidate. A visually tight set is often what good retrieval looks like, because the query picked out a coherent product family, whereas a visually dispersed set is frequently what a broad or ambiguous query produces. The composition of a returned set is therefore a signal about the query as much as a treatment applied to the consumer, and we approach the set-level analysis as a description of that signal rather than as a test of a causal claim.

\subsection{Predictions}
\label{subsec:hypothesis}

Let \(F_{is}\) denote relative fit and \(D_{is}\) distinctiveness from the nearest visual substitute for product \(i\) in search \(s\), let \(C_s\) be the returned set, and let \(Y_{is}\) indicate the clicked product. The first prediction is the distinctiveness effect itself.

\begin{hypothesis}
\label{hyp:distinctiveness}
Within a platform-returned set, a product's distinctiveness from its nearest visual substitute is positively associated with its conditional probability of selection, holding relative fit and display position fixed.
\end{hypothesis}

In continuous form, Hypothesis~\ref{hyp:distinctiveness} states
\begin{equation}
\frac{\partial
\Pr\!\left(Y_{is}=1 \mid i\in C_s,\ \textstyle\sum_{j\in C_s}Y_{js}=1\right)}
{\partial D_{is}}
>0.
\label{eq:conceptual-main}
\end{equation}
This is Proposition~\ref{prop:distinctiveness}. Because the claim is contextual, its test must be within-product: the prediction is that the same product is selected more often on the occasions when the returned set leaves it without a close visual substitute, not that unusual-looking products do better in general. The empirical design decomposes \(D_{is}\) into a product mean and an occasion-specific deviation and asks the deviation to carry the effect.

\begin{hypothesis}
\label{hyp:premium}
Within a platform-returned set, the positive association between distinctiveness and selection strengthens as the product's relative fit with the query increases.
\end{hypothesis}

Hypothesis~\ref{hyp:premium} states
\begin{equation}
\frac{\partial^2
\Pr\!\left(Y_{is}=1 \mid i\in C_s,\ \textstyle\sum_{j\in C_s}Y_{js}=1\right)}
{\partial D_{is}\,\partial F_{is}}
>0.
\label{eq:conceptual-interaction}
\end{equation}
The prediction concerns the cross-partial, and Proposition~\ref{prop:scales} sharpens it in a way a verbal account cannot: because distance operates on individuation rather than on utility, the cross-partial is positive on the probability scale as an implication of the level effect, while in latent-utility units the distance effect is additive and the interaction is zero to first order. The model therefore predicts a specific two-scale pattern, a positive interaction in the linear probability model together with a null interaction in the conditional logit, and both scales are estimated. Under a genuine preference for goal-consistent distinctiveness, by contrast, the interaction would appear on both scales. Together the two hypotheses describe a distinctiveness gradient that is positive on average and steepest, in probability units, among the products that best answer the query.

The third prediction is Proposition~\ref{prop:channels} taken to the set. Let \(\bar T_s\) denote the textual separation of the returned set, the mean contextual textual distance among its products, which proxies how well the description channel separates each product from its neighbors.

\begin{hypothesis}
\label{hyp:contest}
The association between distinctiveness and selection weakens as the textual separation of the returned set increases:
\begin{equation*}
\frac{\partial^2
\Pr\!\left(Y_{is}=1 \mid i\in C_s,\ \textstyle\sum_{j\in C_s}Y_{js}=1\right)}
{\partial D_{is}\,\partial \bar T_{s}}
<0 .
\end{equation*}
\end{hypothesis}

Equivalently, visual separation matters most where words fail: in sets whose descriptions sit close together, the image is the channel that can still tell products apart. This is what distinguishes an account in which appearance carries information from one in which appearance is merely decorative, since a decorative attribute would not become more consequential precisely where the substantive channel stops discriminating. A neighboring moderator, the dispersion of query-product compatibility within the set, is deliberately not part of the hypothesis: in the model it operates through the merit at stake rather than through a discrimination channel, so its sign is not restricted, and Section~\ref{subsec:contest-results} reports it as a variant alongside the confirmatory channel moderator.

The three predictions concern the allocation of clicks among returned products and are silent about their total. Section~\ref{subsec:deferral-model} takes up the total in an explicitly descriptive set-level analysis, asking how the visual composition of a returned set relates to whether the set produces any click. We state no directional prediction there. The interference logic points one way and retrieval quality points the other, the two are not separately identified in observational retrieval logs, and the analysis is reported for what it reveals about the search occasions themselves rather than as a test of the framework.

A natural further question is whether these relations depend on the categorical composition of the returned set, and we examine it in an explicitly exploratory analysis rather than as a hypothesis, because the direction is theoretically ambiguous. Let \emph{category coherence} be the share of a product's returned peers that share its fine-grained catalog category. Coherence could amplify the premium: when peers are categorically comparable, visual distance is unlikely to reflect a difference in product type, so a well-fitting product's separation is more clearly a separation among substitutes. But coherence could equally mute it: in a set of same-type products, visual separation already distinguishes substitutes at every level of fit, whereas in a mixed set fit is precisely what tells a distinctive credible answer apart from a distinctive intruder. Because both mechanisms are plausible, we estimate the coherence moderation with a full interaction hierarchy, report it two-sided and unadjusted, and treat it as a boundary-describing estimate rather than a test of the framework.

The predictions separate the account from its neighbors, and each rival account implies an attenuation or a displacement that the design tests. A global-uniqueness account predicts that products with consistently unusual images perform differently everywhere, which product fixed effects and the within-product decomposition absorb. A category-mismatch account predicts that visual distance mainly marks a different product type, which category-homogeneous sets and same-category neighborhoods test directly. A textual-differentiation account predicts that the visual pattern is a proxy for being an outlier in description space, which a textual distinctiveness control tests. A salience-only account predicts a distinctiveness slope that neither varies with fit nor rises when the goal stops discriminating, which Hypotheses~\ref{hyp:premium} and~\ref{hyp:contest} test. Finally, the constructs themselves pose aggregation questions that the design answers empirically: whether the operative visual field is the single nearest substitute or the broader local field, and whether the operative fit quantity is the within-set standing or the global similarity level. Section~\ref{subsec:robustness} estimates the same models under each construction and reports the comparison alongside the confirmatory tests. All three hypotheses are predictions about conditional associations within realized returned sets; Section~\ref{sec:empirical} states the estimands and their limits.

%% file: sections/03_data_measurement.tex

\section{Data and measurement}
\label{sec:data}

\subsection{Setting and data}
\label{subsec:data-setting}

We use the public Coveo SIGIR e-commerce dataset, released by Coveo, an enterprise search and recommendation provider, for the SIGIR eCom 2021 Data Challenge \citep{tagliabue2021sigir}. The data come from a single mid-size online retailer and contain about 36 million browsing events from roughly 4.9 million anonymized shopping sessions, more than 800,000 search events, and 57,483 catalog products. A session groups events that occur no more than 30 minutes apart. Timestamps are shifted by an undisclosed number of weeks, which preserves event order and within-week patterns but removes calendar dates.

Three tables describe the search environment, the catalog, and browsing activity. The search table records one row per query event with a session identifier, a shifted timestamp, a dense query representation, the array of product identifiers returned to the client, and the array of product identifiers clicked after the query. The catalog table maps each product to a hashed category hierarchy, a price decile, a dense description representation, and a dense image representation. The dataset documentation states that the query and description representations are compatible, so their similarity is directly interpretable; the image representation is a separate space. The browsing table records product-detail, add-to-cart, and purchase events, which we use to reconstruct session order and to form a supplementary downstream outcome. The release withholds raw queries, descriptions, images, retailer identity, and calendar dates. That protection leaves intact the two features the design requires, the realized returned sets and separate textual and visual representations, but it limits construct validation: we cannot inspect what the vectors encode or determine which returned products appeared in the viewport.

\subsection{Returned sets and analytic sample}
\label{subsec:analytic-sample}

The unit of analysis is product \(i\) returned in search event \(s\). Expanding the returned array of a search with \(J_s\) distinct products yields \(J_s\) product-search observations. The outcome is
\begin{equation}
Y_{is}=\mathbb{1}\{\text{product } i \text{ is clicked following search } s\}.
\label{eq:click-outcome}
\end{equation}
The primary sample contains searches with exactly one distinct clicked product,
\begin{equation}
\sum_{i\in C_s}Y_{is}=1,
\label{eq:one-click}
\end{equation}
where \(C_s\) is the returned set. Each such search contributes one selected product and \(J_s-1\) nonselected products, and no negative sampling is used. Zero-click searches concern deferral rather than item-level selection and are excluded from the choice analysis. Multiclick searches enter a robustness analysis that assigns outcome mass \(1/K_s\) to each of the \(K_s\) clicked products.

An eligible search has a valid query representation, at least five distinct returned products, and a clicked product that appears in the returned array. Every returned product must map to one catalog row with valid description and image representations and a usable top-level category. When any candidate fails these checks we drop the entire search rather than the candidate, because deleting one candidate would change the competitive neighborhood used to measure distinctiveness and coherence for every other product in the set. Searches whose returned array repeats a product identifier are excluded for the same reason. Appendix~\ref{app:data-audit} gives the parsing rules, representation checks, and the sequential attrition in Table~\ref{tab:app-attrition}.

Recorded position is the product's one-indexed location in the returned array. The dataset documentation describes the array as the results returned to the client but does not state that it is ordered; Appendix~\ref{app:position-verification} records the documentary basis on which we treat the array as ordered and the specifications that bound the consequence if it is not, and Figure~\ref{fig:app-position-gradient} shows that conditional click incidence declines steeply with array position, as it does in ranked search interfaces \citep{ghose2014ranking,ursu2018rankings}. Section~\ref{subsec:robustness} reports specifications that drop position controls and that restrict the sample to top positions.

Two samples follow from these rules. The \emph{choice panel} adds the single-click condition of Equation~\ref{eq:one-click} and is the estimation sample for Hypotheses~\ref{hyp:distinctiveness} through~\ref{hyp:contest}. The \emph{search-event sample} drops that condition and retains every search meeting the structural rules together with the composition of its returned set, whether or not a click followed; it is the estimation sample for the descriptive set-level analysis in Section~\ref{subsec:deferral-model}. The two are nested by construction, so the set-level analysis describes the environments from which the choice panel is drawn rather than a different population.

The choice panel contains 56,931 searches from 53,334 sessions, 1,049,523 product-search observations, and 24,383 unique products; the search-event sample contains 333,694 searches, of which 75.0\% received no click. Table~\ref{tab:sample-descriptives} reports descriptive statistics, measurement validity, and identifying support; Appendix Table~\ref{tab:app-attrition} reports the attrition path and compares retained with excluded searches.

\subsection{Semantic compatibility and relative fit}
\label{subsec:semantic-fit}

Let \(q_s\) be the query representation and \(t_i\) the description representation, both normalized to unit length after removing malformed and zero-norm vectors. The release supplies both families of vectors but does not guarantee that they share a coordinate system, and empirically they do not: the raw cosine \(\cos(q_s,t_i)\) is no higher for the products the engine returned than for random catalog products (\(-0.039\) versus \(-0.004\)), and the category nearest the query in raw-cosine terms predicts the returned set's dominant category below the majority-class baseline (Appendix~\ref{app:fit-validation} reports the full diagnostic). Measurement of compatibility therefore proceeds in two steps. First, an alignment map \(\widehat{W}\) carries queries into description space; it is estimated by ridge least squares of returned-set description centroids on query vectors,
\begin{equation}
\widehat{W}=\arg\min_{W}\sum_{s\in\mathcal{S}_0}\Bigl\lVert \bar t_{s}-W^{\top}\tilde q_s\Bigr\rVert_2^2+\lambda\lVert W\rVert_F^2,
\qquad \bar t_{s}=\frac{1}{J_s}\sum_{j\in C_s}t_j,
\label{eq:query-alignment}
\end{equation}
where \(\tilde q_s\) appends an intercept and \(\mathcal{S}_0\) contains only searches in the first three quarters of the observation window, strictly before the chronological holdout of Section~\ref{sec:empirical}. The supervision is retrieval alone: the map never sees clicks or image vectors, so it cannot manufacture the hypothesized relations. Second, semantic compatibility is the cosine in the aligned space,
\begin{equation}
\phi_{is}=\cos\bigl(\widehat{W}^{\top}\tilde q_s,\;t_i\bigr).
\label{eq:semantic-fit}
\end{equation}
Higher values indicate closer query-product compatibility. The unaligned cosine is retained as a variable, and its failed validation is reported alongside the diagnostic and the map's out-of-window accuracy in Appendix~\ref{app:fit-validation}.

Relative fit converts compatibility into the product's standing within its returned set,
\begin{equation}
F_{is}=\operatorname{rank}_{j\in C_s}\!\left(\phi_{js}\right)\big/\,J_s ,
\label{eq:relative-fit}
\end{equation}
the within-set fractional rank of \(\phi_{is}\), with tied values assigned their average rank. The rank form implements the construct of Section~\ref{subsec:goal-compatibility}: choice within a returned set is comparative, so the operative quantity is whether the product is among the best available answers, and queries differ in breadth and vocabulary, so levels of \(\phi_{is}\) are not calibrated across occasions while standings within a set scored against the same query are. Because the same product ranks differently under different queries and against different competitors, \(F_{is}\) measures current standing on the goal rather than stable quality or prototypicality. Two alternative forms appear in the robustness suite: compatibility centered on the set mean, which preserves spacing information while remaining relational, and the globally standardized level of \(\phi_{is}\), a measurement contrast that removes the cross-query calibration and thereby asks whether the within-set standing or the raw similarity level is the operative quantity.

Relative fit describes a product's standing; the channel moderation of Hypothesis~\ref{hyp:contest} and its variants need set-level measures of how far apart the returned products stand. The confirmatory moderator, the set's textual separation \(\bar T_s\), lives in description space and is defined in Section~\ref{subsec:visual-differentiation} together with its product-level building block. On the compatibility side we define the set's \emph{compatibility dispersion} as
\begin{equation}
S_s=\operatorname{sd}_{j\in C_s}\!\left(\phi_{js}\right),
\label{eq:fit-spread}
\end{equation}
the standard deviation of aligned compatibility across the returned products, with the gap between the set's best and second-best value, raw and as a share of the set's compatibility range, as variants that look at the top of the distribution rather than its spread; the range-normalized form is invariant to any per-query affine rescaling of the score. A low dispersion marks a contested set in which the query does not separate the alternatives; a high value marks a set with a clear winner. In the model these compatibility-based measures move the merit at stake as well as the discrimination channel, which is why Section~\ref{subsec:hypothesis} leaves their sign unrestricted and reserves the directional prediction for \(\bar T_s\). Each is a property of the set, so it is constant across the products within a search, and the rank transform in Equation~\ref{eq:relative-fit} is by construction uninformative about any of them, which is what allows fit and a set moderator to enter the same model as distinct terms.

Behavioral convergence validates the measure. Within each search we compare the clicked product's compatibility with the mean of its nonclicked peers, compute the clicked product's within-set fit percentile, and plot click incidence by within-search fit decile (Figure~\ref{fig:model-free-evidence}, panel~a). The clicked-minus-nonclicked gap is 0.126 standard deviations (95\% CI 0.121 to 0.131), and the clicked product's median standing is the 61st percentile of its set. A steadily rising click gradient shows that the supplied vectors carry economically relevant goal information; it does not validate every dimension of the proprietary representation. Fit is standardized in the estimation sample with equal-event weights. The holdout exercise standardizes with development-period constants only.

\subsection{Contextual visual distinctiveness}
\label{subsec:visual-differentiation}

Let \(v_i\) be the unit-normalized image representation of product \(i\). Contextual visual distinctiveness is the product's distance from its nearest visual substitute, one minus its highest image similarity with any returned peer,
\begin{equation}
D_{is}=1-\max_{\substack{j\in C_s\\ j\neq i}}v_i^{\top}v_j .
\label{eq:nearest-neighbor}
\end{equation}
The maximum implements the construct of Section~\ref{subsec:contextual-differentiation}: the peer that competes for identification and substitution is the most similar one, so the measure weights that peer entirely and ignores separation from products that are obviously different. Larger values mean the current set contains no close look-alike. The measure changes when the returned set changes even though \(v_i\) is fixed, which is what separates it from global uniqueness \citep{feng2025visual}.

Three alternative aggregations serve as contrasts and probes. The all-peer average,
\begin{equation}
D^{avg}_{is}=1-\frac{1}{J_s-1}\sum_{\substack{j\in C_s\\ j\neq i}}v_i^{\top}v_j ,
\label{eq:visual-differentiation}
\end{equation}
spreads the comparison over the whole set, so its contrast with Equation~\ref{eq:nearest-neighbor} asks how far the operative visual field extends: a strictly local account concentrates the association in the nearest-substitute form, while a crowded field, in which several close neighbors each contribute, lets the broader aggregate carry it as well. Section~\ref{subsec:robustness} estimates the contrast together with a trimmed average that drops the single farthest peer (Appendix~\ref{app:efficient-differentiation} gives the computational identities and their audit). Position-local distinctiveness restricts the peer set to products within \(k\) recorded positions of the focal product, which approximates the products displayed nearest to it on the page and probes whether the whole returned set or the local display neighborhood is the operative reference group.

Three by-products of Equation~\ref{eq:nearest-neighbor} support the mechanism analyses. The identity of the nearest substitute names the peer the measure is defined against; its recorded position gives the display separation between a product and its closest look-alike, which Appendix~\ref{app:adjacency} uses to ask whether interference depends on proximity on the page; and a pair is \emph{mutually nearest} when each product is the other's nearest substitute, which defines the near-identical pairs used in Appendix~\ref{app:twin-duel}. We call a product a \emph{near twin} when its nearest-substitute distance falls in the lowest decile of the estimation sample. The decile rule keeps the definition on the empirical scale of these representations rather than on an absolute cosine threshold whose meaning we cannot inspect.

Products differ in their enduring tendency to look unusual. We separate that tendency from the current neighborhood with
\begin{equation}
\bar D_i=\frac{1}{N_i}\sum_{s:\, i\in C_s}D_{is},
\qquad
D^W_{is}=D_{is}-\bar D_i,
\label{eq:within-differentiation}
\end{equation}
where \(N_i\) is the number of eligible searches in which product \(i\) appears. The within-product component \(D^W_{is}\) is positive when the current set leaves product \(i\) farther from its nearest substitute than the product's other appearances do; it carries both confirmatory hypotheses. For the holdout exercise, product means use development observations only, and products first seen in the test period receive the development grand mean and are reported separately.

We also compute a textual analogue, contextual textual differentiation \(T_{is}\), by replacing image vectors with description vectors in the average form (Equation~\ref{eq:visual-differentiation}). It serves two purposes. As a control: if the visual estimates survive conditioning on \(T_{is}\), its within-product component, and their interactions with fit, the pattern is not simply that the focal product is an outlier in every representation space. And, aggregated to the set, as the confirmatory moderator of Hypothesis~\ref{hyp:contest}: the \emph{textual separation} of a returned set, \(\bar T_s=\frac{1}{J_s}\sum_{j\in C_s}T_{js}\), measures how far apart the set's descriptions stand and hence how much separating work the textual channel can do; it lives entirely in description space, so it requires no query-specific calibration. Image distance remains a computational measure. Image similarities can track consumer judgments when validated independently \citep{satomura2014copy,feng2025visual}, but the images themselves are unavailable here, so we test alternative neighborhoods rather than treat cosine distance as a complete measure of perceived uniqueness.

\subsection{Category coherence and category-restricted distance}
\label{subsec:category-coherence-measure}

Let \(g_i\) denote a catalog category code for product \(i\). Category coherence is the share of returned peers in the focal product's category,
\begin{equation}
G_{is}=\frac{1}{J_s-1}\sum_{\substack{j\in C_s\\ j\neq i}}\mathbb{1}\{g_j=g_i\}.
\label{eq:category-coherence}
\end{equation}
It ranges from zero to one and varies across products in the same set: a product from a minority category has low coherence in a set whose majority shares another category. The exploratory moderation analysis uses the second level of the hashed hierarchy, because the top level is nearly degenerate in this catalog: 74.4\% of searches are fully homogeneous at the top level, so top-level coherence equals one for most observations and separates almost nothing. Second-level coherence is defined for the 99.8\% of observations whose returned sets carry usable second-level codes throughout, and the moderation analysis re-standardizes on that subsample. The catalog hierarchy contains 8 top-level and 38 second-level categories. Three variants probe the definition: the top-level version of \(G_{is}\), the search-level dominant-category share, whose interactions are identified although its level is absorbed by search effects, and a search-level indicator for fully homogeneous sets.

Coherence proxies for the categorical comparability of the competitive field, not perceived substitutability. Hashed categories may be coarse, and consumers compare across category boundaries. The measure's value is in separating visual distance among similar product types from distance created by obvious type differences, which is exactly the distinction on which the two exploratory mechanisms of Section~\ref{subsec:hypothesis} disagree.

Same-category distance holds the comparison group's top-level category fixed. Let \(M_{is}=\{j\in C_s\setminus\{i\}: g_j=g_i\}\). For observations with at least two same-category peers,
\begin{equation}
D^{SC}_{is}=1-\frac{1}{|M_{is}|}\sum_{j\in M_{is}}v_i^{\top}v_j .
\label{eq:same-category-differentiation}
\end{equation}
Together with the homogeneous-set restriction, it tests whether the confirmatory relations survive a competitive neighborhood in which visual distance cannot mark a different product type.

\subsection{Descriptive evidence and identifying variation}
\label{subsec:measurement-diagnostics}

Table~\ref{tab:sample-descriptives} summarizes returned-set size, position, relative fit, nearest-substitute distinctiveness, its within-product component, textual differentiation, category coherence, and relative price position, together with product recurrence, representation coverage, category-homogeneous searches, and the support available for same-category distance. Four facts determine whether the data can carry the tests, and Table~\ref{tab:sample-descriptives} and Figures~\ref{fig:model-free-evidence} and~\ref{fig:identifying-variation} report each of them: recorded position must show a plausible click gradient; fit must predict selection within returned sets; distinctiveness must vary for the same product across occasions, because the within-product component carries both hypotheses; and second-level coherence must have joint support with fit and distinctiveness so that the exploratory moderation does not rest on sparse cells.

We quantify the identifying variation in distinctiveness with the two-way decomposition
\begin{equation}
D_{is}=\alpha_s+\mu_i+u_{is},
\label{eq:differentiation-variance}
\end{equation}
where \(\alpha_s\) and \(\mu_i\) are search and product effects and \(u_{is}\) is the part of a product's distinctiveness explained by neither its stable position nor the common structure of the set. Figure~\ref{fig:identifying-variation} reports the Shapley-attributed variance shares (Appendix~\ref{app:variance-support}), the distribution of within-product standard deviations of \(D_{is}\), and the trajectories of \(D_{is}\) across appearances for a sample of recurring products. Figure~\ref{fig:model-free-evidence}, panel~(b), plots click incidence relative to the random-choice benchmark \(1/J_s\) across within-search fit and distinctiveness bins, with cell counts, and shows the raw pattern that the fixed-effects models then adjust. A support table in Appendix~\ref{app:variance-support} cross-tabulates fit, within-product distinctiveness, and coherence and marks the cells above the minimum event count; every reported contrast is evaluated inside that region.

%% file: sections/04_empirical_strategy.tex

\section{Empirical strategy}
\label{sec:empirical}

\subsection{Estimands and identifying variation}
\label{subsec:estimand}

The analysis concerns selection within the returned set, conditional on one observed click. For product \(i\) in search \(s\), the Hypothesis~\ref{hyp:distinctiveness} estimand is the within-product distinctiveness slope,
\begin{equation}
\theta_1=\frac{\partial\operatorname{E}\!\left[Y_{is}\mid i\in C_s,\ \textstyle\sum_{j\in C_s}Y_{js}=1\right]}{\partial D^W_{is}},
\label{eq:estimand}
\end{equation}
where \(D^W_{is}\) is the occasion-specific deviation of distinctiveness from the product's own mean, defined in Section~\ref{subsec:visual-differentiation}; the prediction is \(\theta_1>0\). The quantity describes whether the same product is selected more often when the current set leaves it without a close visual substitute; it averages over neither zero-click searches nor products the platform did not return. The Hypothesis~\ref{hyp:premium} estimand is
\begin{equation}
\theta_2=\frac{\partial\theta_1}{\partial F_{is}},
\label{eq:premium-estimand}
\end{equation}
with prediction \(\theta_2>0\): the distinctiveness slope should be steeper for products that rank higher on fit within their set. The Hypothesis~\ref{hyp:contest} estimand moves the moderator to the set,
\begin{equation}
\theta_3=\frac{\partial\theta_1}{\partial S_{s}},
\label{eq:contest-estimand}
\end{equation}
with prediction \(\theta_3<0\), where \(\bar T_s\) is the textual separation of the returned set and takes the place of \(S_s\) in the derivative. Section~\ref{subsec:deferral-model} adds a descriptive set-level analysis at a different unit altogether, relating the visual composition of a returned set to whether it produced any click; it uses a sample the choice models cannot use and carries no directional prediction.

Two forms of repeated variation identify these quantities. Products within a search differ in fit and in visual position, and the same product recurs across searches with different queries and competitors. Search-event fixed effects use the first comparison and hold constant the query, session state, returned-set size, time, and every other characteristic shared by the set. Product fixed effects use the second comparison and remove stable appeal, image quality, category, price position, popularity, and global visual uniqueness. Recorded-position indicators enter throughout. Because the platform's ranker may itself respond to fit and to visual similarity, conditioning on position isolates the consumer's response to a product's fit and visual position given where the platform placed it, which is the object of interest; the association that runs through placement is deliberately partialled out and would be a separate question.

These controls narrow the comparison without creating random assignment. Retrieval and position depend on platform signals absent from the release, the semantic representation may omit query-specific attributes that consumers value, and returned products need not have been visible. The coefficients are conditional associations within realized search environments. Editing an image, substituting a competitor, or moving a product's rank would define different causal estimands, and we do not claim to recover them.

\subsection{Fixed-effects linear probability model}
\label{subsec:hdfe-model}

The baseline is a linear probability model at the product-search level. Let \(\widetilde F_{is}\) and \(\widetilde D_{is}\) denote standardized relative fit and distinctiveness. Table~\ref{tab:main-models} builds the design in stages, with search-event and recorded-position effects present in every column so that the within-set comparison is fixed throughout. The first column adds fit and distinctiveness additively,
\begin{equation}
Y_{is}=\alpha_s+\lambda_{r(is)}
+\beta_F\widetilde F_{is}+\beta_D\widetilde D_{is}
+\varepsilon_{is},
\label{eq:additive-model}
\end{equation}
where \(\alpha_s\) are search-event effects and \(\lambda_{r(is)}\) indicators for recorded position, entered flexibly because click incidence need not fall linearly with rank. This is the additive relevance-diversity benchmark: it prices distinctiveness identically for every product. The second column adds product effects \(\mu_i\), which absorb every stable reason an unusual-looking product might attract clicks; the movement in \(\widehat\beta_D\) between the two columns measures how much of the raw association reflects enduring product heterogeneity rather than occasion-level position.

Returned sets differ in size, so we weight each observation by \(w_{is}=1/J_s\), which gives every search one unit of total weight; the unweighted, row-weighted estimate is a sensitivity analysis. Standard errors are two-way clustered by session and product \citep{camerongelbachmiller2011robust}, allowing dependence across searches within a shopping episode and across appearances of a product. The linear model accommodates two high-dimensional fixed effects and returns probability-point coefficients with a direct interpretation. It is not a claim about the choice process; the final column of Table~\ref{tab:main-models} re-estimates the premium model as a fixed-effects conditional logit, described below. Effect sizes are reported as probability contrasts over supported ranges, with confidence intervals, and relative to the mean random-choice probability \(1/J_s\), because a given probability-point change means different things in a set of six and a set of sixty.

\subsection{The preferred within-product models}
\label{subsec:within-between-model}

Even with product effects in the model, the level of distinctiveness \(D_{is}\) mixes two comparisons: whether a product does better when it is more distinctive than usual, and whether the set of occasions on which distinctive products appear differs in unmodeled ways. The theory's claim is the first one, so the confirmatory models are built on the decomposition
\begin{equation}
\bar D_i=\frac{1}{N_i}\sum_{s:\,i\in C_s}D_{is},
\qquad
D^W_{is}=D_{is}-\bar D_i,
\qquad
D^B_i=\bar D_i-\bar D,
\label{eq:between-differentiation}
\end{equation}
where \(N_i\) counts product \(i\)'s eligible appearances and \(D^B_i\) is the centered between-product component. After standardization, the preferred Hypothesis~\ref{hyp:distinctiveness} model is
\begin{equation}
Y_{is}=\alpha_s+\mu_i+\lambda_{r(is)}
+\beta_F\widetilde F_{is}+\beta_W\widetilde D^W_{is}
+\varepsilon_{is},
\label{eq:preferred-model}
\end{equation}
and the confirmatory coefficient is \(\beta_W\), the within-product distinctiveness slope. The between component's level is collinear with the product effects, which is the point: nothing in \(\widehat\beta_W\) can come from stable differences between unusual-looking and ordinary-looking products.

The Hypothesis~\ref{hyp:premium} model adds the interaction of fit with each component,
\begin{equation}
\begin{split}
Y_{is}=\;&\alpha_s+\mu_i+\lambda_{r(is)}
+\beta_F\widetilde F_{is}+\beta_W\widetilde D^W_{is}\\
&+\beta_{FW}\left(\widetilde F_{is}\times\widetilde D^W_{is}\right)
+\beta_{FB}\left(\widetilde F_{is}\times\widetilde D^B_i\right)
+\varepsilon_{is},
\end{split}
\label{eq:premium-model}
\end{equation}
and the confirmatory coefficient is \(\beta_{FW}\): is the within-product distinctiveness slope steeper when the product ranks higher on fit? The between interaction \(\beta_{FB}\) is identified although the between level is absorbed, and it serves as a diagnostic: a premium that loads on \(\beta_{FB}\) rather than \(\beta_{FW}\) would say that globally distinctive products have a steeper fit slope, a global-uniqueness pattern rather than a contextual one. We report the difference \(\widehat\beta_{FW}-\widehat\beta_{FB}\) with its interval. The marginal association of within-product distinctiveness implied by Equation~\ref{eq:premium-model} is
\begin{equation}
\frac{\partial\operatorname{E}[Y_{is}\mid\cdot]}{\partial\widetilde D^W_{is}}=\beta_W+\beta_{FW}\widetilde F_{is},
\label{eq:marginal-differentiation}
\end{equation}
which Figure~\ref{fig:marginal-effect} plots with 95\% confidence intervals over the central 90\% of fit, together with the fit-decile slopes from a flexible specification, so that the linear summary can be checked against the shape it summarizes. We also report the change in predicted probability from moving within-product distinctiveness between its 25th and 75th percentiles at the 10th, 50th, and 90th fit percentiles.

The two confirmatory tests are \(\beta_W>0\) in Equation~\ref{eq:preferred-model} and \(\beta_{FW}>0\) in Equation~\ref{eq:premium-model}, each reported with a two-sided 95\% interval and its p-value as estimated. The two are separate predictions about distinct coefficients in distinct models rather than repeated attempts at one claim, so no family-wise adjustment is applied; contrasts use the full covariance matrix. Identification rests on recurring products observed in different environments. We report the appearance distribution, the connectedness of the search-product graph, and estimates from its largest connected component; samples requiring at least 5, 10, and 20 appearances show whether sparse products drive the result; and product-by-week effects compare products over shorter periods to reduce sensitivity to changing popularity, promotion, or inventory. Because \(\bar D_i\) is estimated from a finite number of appearances, \(D^W_{is}\) contains estimation error for rarely observed products, which the recurrence thresholds also address.

\subsection{Channel moderation}
\label{subsec:contest-model}

Hypothesis~\ref{hyp:contest} asks whether the distinctiveness effect is larger in sets whose descriptions fail to separate the alternatives. The confirmatory moderator is the set's textual separation \(\bar T_s\), the mean contextual textual distance among its products, standardized; it enters the premium model with every lower-order term implied by the triple interaction and the corresponding between-product terms, exactly as below. Two variants probe the neighboring semantic margin, whose sign the model leaves unrestricted because it operates through the merit slot rather than a discrimination channel: the within-set standard deviation of query-product compatibility \(S_s\), which Table~\ref{tab:app-contest-diagnostics} reports and Figure~\ref{fig:boundaries} panel~(a) visualizes, and the gap between the set's best and second-best compatibility, in raw and range-normalized forms. Writing the moderator generically as \(\widetilde S_s\), the design is:
\begin{equation}
\begin{split}
Y_{is}=\;&\alpha_s+\mu_i+\lambda_{r(is)}
+\beta_F\widetilde F_{is}+\beta_W\widetilde D^W_{is}\\
&+\beta_{FW}\widetilde F_{is}\widetilde D^W_{is}
+\beta_{FS}\widetilde F_{is}\widetilde S_{s}
+\beta_{WS}\widetilde D^W_{is}\widetilde S_{s}
+\beta_{FWS}\widetilde F_{is}\widetilde D^W_{is}\widetilde S_{s}\\
&+\beta_{FB}\widetilde F_{is}\widetilde D^B_i
+\beta_{BS}\widetilde D^B_i\widetilde S_{s}
+\beta_{FBS}\widetilde F_{is}\widetilde D^B_i\widetilde S_{s}
+\varepsilon_{is}.
\end{split}
\label{eq:contest-model}
\end{equation}
For the confirmatory channel moderator the coefficient of interest is \(\beta_{WS}\) with \(\widetilde S_s\) replaced by standardized \(\bar T_s\), and the predicted sign is negative: a one-standard-deviation increase in the set's textual separation should flatten the visual-distance slope. Because \(S_s\) is constant within a search, its level is absorbed by the search effects while its interactions remain identified, and the between-product terms keep a dispersion-conditioned global-uniqueness pattern from loading on the contextual interaction. The marginal association of within-product distinctiveness is
\begin{equation}
\frac{\partial\operatorname{E}[Y_{is}\mid\cdot]}{\partial\widetilde D^W_{is}}
=\beta_W+\beta_{FW}\widetilde F_{is}+\beta_{WS}\widetilde S_{s}+\beta_{FWS}\widetilde F_{is}\widetilde S_{s},
\label{eq:contest-marginal}
\end{equation}
which Figure~\ref{fig:boundaries}, panel~(a), plots over relative fit for the dispersion variant. Because rank-based relative fit is uniform within every set by construction, it carries no information about how spread out the set is on either channel, so none of these moderators is mechanically linked to fit. All are constant within a search, so their levels are absorbed by the search effects while their interactions stay identified.

\subsection{Exploratory coherence moderation}
\label{subsec:coherence-model}

The exploratory analysis of Section~\ref{subsec:hypothesis} asks whether the premium varies with the categorical composition of the set. It adds standardized coherence \(\widetilde G_{is}\), every lower-order term implied by the triple interaction, and the corresponding interactions with between-product distinctiveness:
\begin{equation}
\begin{split}
Y_{is}=\;&\alpha_s+\mu_i+\lambda_{r(is)}
+\beta_F\widetilde F_{is}+\beta_W\widetilde D^W_{is}+\beta_G\widetilde G_{is}\\
&+\beta_{FW}\widetilde F_{is}\widetilde D^W_{is}
+\beta_{FG}\widetilde F_{is}\widetilde G_{is}
+\beta_{WG}\widetilde D^W_{is}\widetilde G_{is}
+\beta_{FWG}\widetilde F_{is}\widetilde D^W_{is}\widetilde G_{is}\\
&+\beta_{FB}\widetilde F_{is}\widetilde D^B_i
+\beta_{BG}\widetilde D^B_i\widetilde G_{is}
+\beta_{FBG}\widetilde F_{is}\widetilde D^B_i\widetilde G_{is}
+\varepsilon_{is}.
\end{split}
\label{eq:coherence-model}
\end{equation}
The between-product terms keep category-conditioned global uniqueness from loading on the contextual triple interaction. With centered regressors, \(\beta_{FW}\) is the premium at mean coherence and \(\beta_{FWG}\) is the moderation of interest; the cross-partial at coherence \(\widetilde G_{is}\) is
\begin{equation}
\frac{\partial^2\operatorname{E}[Y_{is}\mid\cdot]}{\partial\widetilde F_{is}\,\partial\widetilde D^W_{is}}
=\beta_{FW}+\beta_{FWG}\widetilde G_{is}.
\label{eq:coherence-cross-partial}
\end{equation}
Because the direction is theoretically ambiguous, \(\widehat\beta_{FWG}\) is reported two-sided, unadjusted, and outside the confirmatory family. Figure~\ref{fig:boundaries}, panel~(b), plots distinctiveness slopes over fit in coherence groups with the support of each curve, and we report interquartile distinctiveness contrasts at pre-specified fit and coherence percentiles, which are easier to read than the triple-interaction coefficient. Coherence uses the fine-grained second level of the hashed category hierarchy; the top level is nearly degenerate in this catalog (Section~\ref{subsec:category-coherence-measure}), and the top-level and search-level dominant-share versions appear as robustness variants.

Two category-restricted analyses accompany the moderation and also serve the category-mismatch diagnostic for the confirmatory hypotheses: the Hypothesis~\ref{hyp:distinctiveness} and~\ref{hyp:premium} models in fully category-homogeneous sets, and the same models with same-category distance (Equation~\ref{eq:same-category-differentiation}) in place of \(D_{is}\). Estimates that persist when every returned peer shares the focal product's category cannot reflect visual distance that merely marks a different product type.

\subsection{Set composition and click incidence: a descriptive companion}
\label{subsec:deferral-model}

The three hypotheses condition on a click having occurred and therefore describe how clicks are allocated among returned products. This section turns to their total, which needs a different sample and a different unit. The choice panel keeps only single-click searches; the data construction, however, evaluates every search that passes the structural rules (a valid query representation, at least five returned products, complete catalog coverage, a usable category for every product) and records its composition whether or not it produced a click. Those records, described in Section~\ref{subsec:analytic-sample}, form the estimation sample here, and the zero-click searches that the choice analysis excludes carry the identifying variation.

For search \(s\) let \(A_s\) indicate at least one click and \(\bar D_s\) the mean distance between the returned products and their nearest visual substitutes. The specification is
\begin{equation}
A_s=\gamma\,\widetilde{\bar D}_s+X_s^{\top}\delta+\omega_{w(s)}+\psi_{J_s}+\kappa_{c(s)}+\eta_s ,
\label{eq:deferral-model}
\end{equation}
where \(\omega_{w(s)}\) are week effects, \(\psi_{J_s}\) returned-set-size effects, and \(\kappa_{c(s)}\) effects for the modal category of the set, so that the comparison runs between sets returned in the same week, of the same size, and in the same product area that differ in how tightly their products cluster visually. The controls \(X_s\) hold constant the retrieval quality and composition that would otherwise be confounded with visual spread: the set's mean compatibility with the query, its compatibility dispersion, its mean textual distance, its dominant-category share, and log set size. Standard errors are clustered by session. We state no predicted sign. Interference implies \(\gamma>0\), since sets whose products are hard to tell apart should be resolved less often, while retrieval quality implies \(\gamma<0\), since a visually tight set is frequently what a well-targeted query returns and a dispersed set what a broad one returns. The two are not separately identified here, and the estimate is reported as a description of the search occasions rather than as a test. Two variants sharpen the reading. Replacing \(\bar D_s\) with the mean distance among all returned pairs is the set-level counterpart of the measurement contrasts, and entering both distinguishes near-duplication from general visual variety. A second outcome, whether another search follows in the same session within the attribution window, asks whether redundant sets send consumers back to the query box.

The design has limits worth stating in advance, and they are the reason the analysis is descriptive. A search without a click is not necessarily a failure: the consumer may have found what they needed on the page, been interrupted, or never intended to click. Retrieval is endogenous, and a query returning near-duplicates differs from one that does not in ways no observational control fully captures, which is why the specification conditions on compatibility level and dispersion rather than treating visual spread as randomly assigned. Nothing here identifies what deduplicating a ranker would deliver.

\subsection{Chronological ranking comparison}
\label{subsec:prediction}

A chronological holdout asks whether the distinctiveness terms carry ranking information. Sessions are ordered by the shifted timestamp of their first search; the first 80\% of searches form the development period and the final 20\% are held out until evaluation, with the last fifth of the development period serving as validation. Standardization constants, product histories, average distinctiveness, and tuning choices use development observations only.

Five rankings are compared within each held-out returned set. The recorded platform order is the historical benchmark. The remaining four are linear scores whose coefficients are estimated on the development period from a conditional logit with search effects, recorded-position indicators, the covariates below, and a smoothed product prior \(\widehat\pi_i\) (entered as a log-odds) in place of product fixed effects, so that every score is computable for products with any history. The relevance score is
\begin{equation}
S^{R}_{is}=\widehat\lambda_{r(is)}+\widehat\beta_F\widetilde F_{is}+\widehat\pi_i ,
\label{eq:relevance-score}
\end{equation}
the additive score adds distinctiveness,
\begin{equation}
S^{A}_{is}=S^{R}_{is}+\widehat\beta_D\widetilde D_{is},
\label{eq:additive-score}
\end{equation}
the goal-consistent score adds the Hypothesis~\ref{hyp:premium} interaction,
\begin{equation}
S^{I}_{is}=S^{A}_{is}+\widehat\beta_{FD}\left(\widetilde F_{is}\widetilde D_{is}\right),
\label{eq:interactive-score}
\end{equation}
and the coherence-conditioned score adds the exploratory coherence terms,
\begin{equation}
S^{G}_{is}=S^{I}_{is}
+\widehat\beta_G\widetilde G_{is}
+\widehat\beta_{FG}\widetilde F_{is}\widetilde G_{is}
+\widehat\beta_{DG}\widetilde D_{is}\widetilde G_{is}
+\widehat\beta_{FDG}\widetilde F_{is}\widetilde D_{is}\widetilde G_{is}.
\label{eq:coherence-score}
\end{equation}
Raw distinctiveness is used because it is available for every current set without product recurrence. Products first observed after the development period receive the common prior and are reported separately. The \(S^R\) versus \(S^A\) comparison measures the ranking value of distinctiveness itself, the \(S^A\) versus \(S^I\) comparison the increment from the goal-consistency premium, and \(S^I\) versus \(S^G\) the increment from the exploratory coherence terms. Mean reciprocal rank is primary; Hit@1 and Hit@3 are secondary. Confidence intervals resample sessions and keep each session's searches together. The metrics measure concordance with clicks logged under the historical display policy; they do not estimate click-through or revenue under a counterfactual ranking.

\subsection{Robustness, rival mechanisms, and falsification}
\label{subsec:robustness}

The diagnostic program is organized around threats to interpretation; Appendix Table~\ref{tab:app-diagnostic-map} maps each threat to its test. Position and incomplete exposure are examined by dropping position controls, restricting to the top 10 and top 5 positions, and letting position effects vary with set size. Category mismatch is addressed by homogeneous sets and same-category distance. Session path dependence is addressed with first-search and single-search-session samples and with indicators for whether the product was returned or clicked in an earlier retained search of the session. Product conditions are addressed with product-by-week effects, recurrence thresholds, and the largest connected component. Outcome and estimand sensitivity are addressed with the multiclick fractional outcome, unweighted estimation, one-percent tail trimming, and an add-to-cart or purchase outcome attributed to the clicked product within a short window (Appendix~\ref{app:downstream}). Fit curvature is addressed with fit-decile and restricted-cubic-spline specifications.

A distinct block of specifications estimates the aggregation questions stated in Section~\ref{subsec:hypothesis}. Replacing nearest-substitute distance with the average distance from all returned peers, or with a trimmed average that drops the single farthest peer, spreads the comparison over the whole set; replacing within-set fit rank with the globally standardized similarity level removes the cross-query calibration that makes fit comparable across occasions. The comparison is informative in both directions rather than a stability check with a preferred outcome. A strictly local field would concentrate the association in the nearest-substitute form; a crowded field, in which several close neighbors each contribute to conflation, lets broader aggregations carry it as well, and the minimum over a set of distances is in any case the noisiest of the constructions, so relative magnitudes across aggregations do not by themselves settle the localization question, which the pairwise analyses of Web Appendices~\ref{app:adjacency} and~\ref{app:twin-duel} address at the level where it is defined. Table~\ref{tab:robustness} groups these rows as measurement contrasts.

Three rival mechanisms receive dedicated controls. Visual distinctiveness could proxy for being an outlier in every representation, so we add contextual textual differentiation \(T_{is}\), its within-product component, and their interactions with fit. It could proxy for a price outlier, so we add the interaction of fit with the product's price decile relative to the set mean (the level itself is absorbed by the product and search effects). And the rank transform of fit could drive results through its functional form alone, so we re-estimate with fit centered on the set mean, an alternative relational form that preserves spacing information.

A fixed-effects conditional logit provides a choice-model scale. With one selected product per search,
\begin{equation}
\Pr(Y_{is}=1\mid C_s)=\frac{\exp(V_{is})}{\sum_{j\in C_s}\exp(V_{js})},
\label{eq:conditional-logit}
\end{equation}
where \(V_{is}\) contains product and position effects and the terms of Equation~\ref{eq:premium-model} \citep{mcfadden1974conditional}. We estimate it through its Poisson representation with high-dimensional fixed effects \citep{correia2020ppmlhdfe} and report latent-utility coefficients together with average predicted-probability contrasts, because coefficients in nonlinear models are not probability derivatives \citep{ainorton2003interaction}. Agreement with the linear model reduces concern that the conclusions depend on the additive probability scale; the logit also weights each returned product equally within the search, so it doubles as a check on the equal-event weighting.

Two analyses address unobserved confounding directly. A representation placebo permutes image vectors across products within top-level category, preserving categories, returned sets, clicks, queries, products, and positions while breaking the product-image assignment; nearest-substitute distance and its decomposition are recomputed and both confirmatory coefficients re-estimated over 200 draws (Appendix~\ref{app:placebo}). A coefficient-stability analysis reports how \(\widehat\beta_{W}\) and \(\widehat\beta_{FW}\) move as fixed effects and controls are added and the robustness value at which an unobserved confounder as strong as the included controls would eliminate the estimate \citep{oster2019unobservable,cinellihazlett2020sensitivity}. Figure~\ref{fig:specification-curve} collects the estimates from every specification in a single display \citep{simonsohn2020specification}.

The checks have asymmetric implications. Sensitivity to position restrictions points toward presentation. Attenuation with textual differentiation points toward representation outliers rather than visual position. Failure in category-comparable settings points toward category mismatch. Evidence carried by the between-product component supports global uniqueness rather than contextual distinctiveness. A placebo distribution centered near the observed estimates indicates that either representation geometry or a structural feature the permutation preserves can generate the pattern, and diagnosing which is what the placebo variants in Section~\ref{subsec:premium-diagnostics} are for. The measurement contrasts, by contrast, are estimates of how far the operative field extends rather than pass-fail checks. The paper's conclusions require support for Hypothesis~\ref{hyp:distinctiveness} in the within-product model with these controls, and for Hypothesis~\ref{hyp:premium} additionally the within-set interaction with the between-product diagnostic near zero.

%% file: sections/05_results.tex

\section{Main Results}
\label{sec:results}

\subsection{Sample, measurement validity, and identifying variation}
\label{subsec:sample-validity}

The choice panel contains 56,931 eligible searches from 53,334 sessions, 1,049,523 product-search observations, and 24,383 unique products; the search-event sample used in Section~\ref{subsec:deferral-results} contains 333,694 searches, 75.0\% of which received no click. The median returned set has 25 products (interquartile range 11 to 25). The 819,516 raw search records reduce to that panel in three main steps: requiring exactly one distinct clicked product retains 132,697 searches, requiring at least five returned products retains 93,549, and requiring complete catalog coverage with valid representations for every returned product retains 56,931 (Appendix Table~\ref{tab:app-attrition}). The single-click condition is by far the largest reduction, so the estimand applies to searches that produced exactly one click on a fully covered returned set of at least five products, not to search behavior in general.

\ifinlineexhibits\input{exhibits/blocks/table1_descriptives}\fi

Table~\ref{tab:sample-descriptives} reports four facts, each providing the foundation for a subsequent result. The first is that position behaves as a ranked list: conditional click incidence falls from 0.165 at position one to 0.040 at position ten (Appendix Figure~\ref{fig:app-position-gradient}), and Appendix~\ref{app:position-verification} records the documentary basis for treating the array as the ordered response returned to the client together with the specifications that bound the consequence if it is not. Without that gradient the position effects would be absorbing noise; with it they are absorbing the largest single influence on which product gets clicked.

\ifinlineexhibits\input{exhibits/blocks/figure2_model_free}\fi

The second is that the fit measure carries goal information rather than generic similarity. Within a search, the clicked product exceeds the mean compatibility of its nonclicked peers by 0.126 standard deviations (95\% CI[0.121, 0.131]), its median within-set fit percentile is 0.61, and click incidence rises monotonically across within-search fit deciles (Figure~\ref{fig:model-free-evidence}, panel~a). The unaligned cosine fails the same check, with a gap of \(-0.026\) standard deviations and a median percentile of 0.50. This contrast justifies our embedding alignment step before measuring similarity: the released query and description vectors live in spaces that are not comparable until the retrieval-estimated map aligns them, and before alignment, their cosine carries virtually no information about what the shopper asked for.

\ifinlineexhibits\input{exhibits/blocks/figure3_identifying_variation}\fi

The third fact is the engine of the design. Distinctiveness varies substantially for the same product across occasions: the Shapley decomposition attributes 65.2\% of its variance to products, 18.0\% to search environments, and 16.8\% to residual product-search variation, and the median within-product standard deviation is 0.028 (Figure~\ref{fig:identifying-variation}, panels (a) and (b)). The non-product portion variation is the main source of identification in the models below, and its substantive share (35\%) forms the premise of the paper: because retrieval assembles a different competitive field for each query, a fixed image is contextually distinctive on one occasion and locally duplicated on the next. Products appear in a median of 17 searches, 79.2\% appear in at least five and 64.5\% in at least ten, and the largest connected component of the search-product graph holds 99.99\% of observations, so the within-product comparison rests on repeated appearances rather than on a thin tail of recurring items.

Panel~(c) of Figure~\ref{fig:identifying-variation} further illustrates this. Each line is one product's nearest-substitute distinctiveness across the searches it appears in, with the product's own mean marked, and the lines wander: a fixed image sits well above its mean on some occasions and well below on others, purely because retrieval assembled a different set of competitors. The within-product component $D^W_{is}$ is the vertical deviation from each product's own line, and it is the only variation the confirmatory models use.

The final fact concerns the moderators, and it is the reason one of them is measured at a finer grain than the theory strictly requires. Because 74.4\% of searches are already category homogeneous at the top level, top-level coherence separates almost nothing, so the coherence analysis uses the second level, defined on 99.8\% of observations across 38 second-level categories; 99.0\% of observations have at least two same-category peers, which is what makes the category-restricted robustness checks feasible. The catalog has 8 top-level categories.

Figure~\ref{fig:model-free-evidence}, panel~(b), shows the raw surface over fit and distinctiveness before any of this structure is imposed. It is worth reading precisely because it is unpersuasive on its own: the surface adjusts for neither product appeal, position, nor platform selection, and the pattern that the models recover is not visible in it. What follows is an argument about which comparisons make that pattern appear, and why those are the right comparisons.

\subsection{Distinctiveness and selection}
\label{subsec:primary-results}

\ifinlineexhibits\input{exhibits/blocks/table2_main_models}\fi

Table~\ref{tab:main-models} builds the estimate one restriction at a time, and the sequence is the argument in miniature. Search-event and recorded-position effects are present in every column, so throughout the table a product is compared only against the alternatives returned beside it and at comparable ranks. Column~1 asks the question the marketplace literature asks, whether products that look unlike their neighbors are chosen more often, and answers it in the negative. Relative fit predicts selection strongly, 0.0110 (SE 0.0008) per standard deviation, while distinctiveness is indistinguishable from zero, 0.0012 (SE 0.0008). Read on its own, the results seem to indicate that visual distinctiveness has a null effect.

This, however, is not the case upon further decomposition. Column~2 controls for product fixed effects, and the distinctiveness coefficient moves to 0.0072 (SE 0.0010, \(p<0.001\)). This reversal from the null effect in Column~1 is pivotal. The pooled coefficient aggregates two effects that run in opposite directions. Products whose appearance is habitually unlike anything returned around them, i.e., globally visually unique, are on average somewhat less likely to be clicked, because habitual visual isolation in this catalog goes with being a niche item. Any given product, however, does better on the occasions when it happens to be retrieved without a close look-alike, that is, when it is \emph{contextually distinctive}. The first is a global effect, and the second is a within consideration set effect, and they offset each other until the product effects hold the first fixed. That reversal, rather than the size of any single coefficient, anchors the paper's central claim: the reference set against which visual difference pays is the returned page, not the full catalog, and therefore contextual distinctiveness is more important than global distinctiveness in modeling the search choices.

Columns~2 and~3 are the same regression written on two scales. Because product fixed effects already absorb each product's average distance, the level coefficient in column~2 is identified from exactly the occasion-to-occasion variation that column~3 names explicitly, and the two columns accordingly share a \(t\) statistic and a \(p\) value; their coefficients differ only because one standard deviation of the within-product deviation is about half a standard deviation of the level. Column~3 is the preferred form for two reasons. It represents the association in the units the theory is about, a movement of the same product \emph{relative to its own} norm rather than a movement across the catalog's distribution, and it frees the between-product component to enter as a regressor in its own right, which the interaction test of the next section needs.

On that scale the within-product coefficient \(\widehat\beta_{W}\) is 0.00348 (SE 0.00050; \(p<0.001\)). The same product, visually presented the same way, is more likely to be chosen on the occasions when the search returned set leaves it farther from its nearest visual substitute than it normally is. Against a mean random-choice probability of 0.0725, a one-standard-deviation movement is 4.8\% of a random-choice click and an interquartile movement is 0.0011 in probability. Two comparisons calibrate that. Within the same column, distinctiveness is about three fifths the size of relative fit, so the query remains the stronger cue, as it should be, but the visual field is not a rounding error beside it. Against the page itself the effect is small: raw click incidence spans roughly 0.125 between position ten and position one, so no amount of visual separation substitutes for a better slot. The quantity being described is correspondingly modest and specific. It is not whether the shopper clicks, but which of several similar products absorbs a click that one of them was going to receive, and on that margin a competitor's near-duplicate image is worth about as much as a meaningful move down the fit distribution. The conditional logit in column~6 puts the same term at 0.032 latent-utility units (SE 0.007, \(p<0.001\)), so the result is a property of the choice data rather than of the additive probability scale. Hypothesis~\ref{hyp:distinctiveness} is supported.

\subsection{The goal-consistency premium}
\label{subsec:premium-results}

If visual separation works by making a product easier to pick out as a distinct option, then what that separation is worth should depend on what the product has to offer once it has been picked out. Column~4 puts that to the test by letting the distinctiveness slope vary with relative fit, and it does so in a form that guards against the obvious alternative. The interaction with the occasion-specific component, \(\widehat\beta_{FW}\), is 0.00254 (SE 0.00049), while the parallel interaction with the product's enduring distinctiveness is essentially zero, 0.00020 (SE 0.00066), and the difference between them is 0.00233 (95\% CI 0.00078 to 0.00389). The contrast between those two numbers is the test. If steeper fit gradients belonged to products that always look unusual, the between-product term would carry them; it does not, so the premium belongs to the occasion rather than to the product.

\ifinlineexhibits\input{exhibits/blocks/figure4_marginal_effect}\fi

Figure~\ref{fig:marginal-effect} shows the shape behind the coefficient, and the shape is more informative than the linear summary. The decile-specific slopes are flat and indistinguishable from zero through the bottom four deciles of fit, lift through the middle, and rise steeply in the top two, so the linear interaction understates a gradient concentrated among the best answers to the query. Translated into probabilities, an interquartile movement in distinctiveness buys a product almost nothing at the tenth percentile of fit and roughly 3.3\% of a random-choice click at the ninetieth. The practical reading is that standing apart is not a substitute for being right for the query but a multiplier on it, and this is also where the framework parts company with the diversification objectives described in Section~\ref{sec:introduction}: a ranker that pays for visual novelty irrespective of relevance is paying most of its budget for the flat part of the curve.

The scale on which the premium appears tells a second story that the coefficient by itself does not. The same interaction is null in the conditional logit, 0.005 (SE 0.006), and a straightforward reading of that pair would be that one of the two estimates is wrong. Proposition~\ref{prop:scales} says otherwise: when visual distance operates on whether a product is individuated rather than on how much it is liked, a positive interaction in probability units and no interaction in log-odds is exactly the pattern to expect, because the linear-scale interaction is then an implication of the level effect rather than an additional force. A genuine taste for goal-consistent distinctiveness would show up on both scales; individuation shows up on one. This changes what the premium is evidence for. It supports the claim that the value of visual separation is concentrated among products that answer the query, which is the substantive content of Hypothesis~\ref{hyp:premium}, without supporting the stronger claim that consumers prefer goal-consistent distinctiveness as such. Two things could still undo even the weaker reading, since the same probability-scale pattern would arise mechanically if the interaction were returned-set-size structure, and the representation placebo of Section~\ref{subsec:robustness-results} reproduces an interaction of similar size. Section~\ref{subsec:premium-diagnostics} is where those two possibilities are separated from the individuation account.

\subsection{When visuals decide: channel substitution}
\label{subsec:contest-results}

\IfFileExists{exhibits/generated/tableA6_contest_diagnostics.tex}{%
  \input{exhibits/generated/tableA6_contest_diagnostics}
}{}

Hypothesis~\ref{hyp:contest} predicts that the distinctiveness association weakens as the textual separation of the returned set rises, because a set whose descriptions already stand apart gives the consumer a cheaper channel for telling the alternatives apart. Table~\ref{tab:app-contest-diagnostics} estimates Equation~\ref{eq:contest-model} with standardized \(\bar T_s\) as the moderator, carrying the complete interaction hierarchy and the between-product terms. The coefficient on within-product distinctiveness times textual separation, \(\widehat\beta_{WS}\), is \(-0.00230\) (SE 0.00064; \(p=0.0003\)). In slope terms, the within-product distinctiveness association is 0.0074 at the 25th percentile of textual separation and 0.0036 at the 75th: the visual channel does roughly twice the work where the descriptions sit close together, and roughly half where the text already discriminates. The tertile split in Web Appendix~\ref{app:heterogeneity} shows the same pattern nonparametrically (0.0082 in the tightest-text third of searches against about 0.0031 elsewhere). The triple interaction with fit is not significant, so the substitution operates on the distinctiveness level rather than reshaping the premium. This evidence supports Hypothesis~\ref{hyp:contest}: visual separation matters most where words fail, which is what Proposition~\ref{prop:channels} predicts and what a decorative account of imagery does not.

\ifinlineexhibits\input{exhibits/blocks/figure5_boundaries}\fi

The first row of Table~\ref{tab:app-contest-diagnostics} reports the neighboring variant in which the moderator is the within-set dispersion of query-product compatibility, \(S_s\). That interaction is not distinguishable from null at \(-0.00062\) (\(p=0.31\)); Figure~\ref{fig:boundaries}, panel~(a), converts it into slopes, 0.0048 at the 25th percentile of dispersion and 0.0038 at the 75th. The null is informative once the two moderators are separated. As Section~\ref{subsec:contest-model} notes, the compatibility structure of a returned set enters the model twice: how far apart the candidates stand determines whether the query can do the separating work, the discrimination margin on which Hypothesis~\ref{hyp:contest} operates, while how much merit separates the leaders determines what is at stake when individuation succeeds. Compatibility dispersion bundles the two, and it correlates 0.75 with textual separation, so it inherits part of the negative channel component while also carrying the merit component.

The top-two-gap variants unbundle them. The gap between the set's best and second-best compatibility carries an interaction of 0.00085 (SE 0.00049; \(p=0.08\)); measured as a share of the set's compatibility range, which removes the per-query calibration that the raw score lacks, it is 0.00190 (SE 0.00072; \(p=0.008\)), with distinctiveness slopes of 0.0019 at the 25th percentile of the normalized gap and 0.0028 at the 75th. Where the query separates a clear leader from the runner-up, distinctiveness is associated more strongly with selection, not less. That is the merit-margin sign the model leaves unrestricted: individuation lets merit act, so it is worth more where more merit rides on it (Section~\ref{subsec:model}). The correlation structure completes the account of the dispersion null: the normalized gap is essentially uncorrelated with textual separation (\(-0.07\)), while dispersion loads on it (0.75), so the dispersion interaction mixes a negative discrimination component with a positive merit component and lands near zero. The gap moderations carry no predicted sign and are unadjusted, so we read them as consistent with the model's merit margin rather than as confirmatory findings; the confirmatory statement of this section is the textual-separation moderation, and it holds.

\subsection{What visually tight sets signal}
\label{subsec:deferral-results}

\ifinlineexhibits\input{exhibits/blocks/table4_deferral}\fi

The models so far condition on a click and therefore describe how clicks are allocated. Table~\ref{tab:deferral} turns to whether a set receives one at all, using the 333,694 search events of the set-level sample, 75.0\% of which ended without a click. The association runs opposite to what interference alone would imply. With week, set-size, and modal-category effects and no further controls, a one-standard-deviation increase in the set's mean nearest-substitute distance is associated with a 4\% lower chance of any click, against a base rate of 25\%. Adding the retrieval-quality and composition controls halves it to -2.4\% but does not change the direction.

The comparison between distance measures locates the quantity. Entering the average distance among all returned pairs in place of nearest-substitute distance gives \(-0.0389\) (SE 0.0011), and entering both leaves the average at \(-0.0340\) (SE 0.0014) and collapses the nearest-substitute coefficient to \(-0.0064\) (SE 0.0012). What predicts a click at the set level is therefore overall visual spread rather than near-duplication specifically, which is the opposite of the pattern inside the choice model, where the nearest substitute is what matters and the average attenuates. Reformulation is flat: sets with closer look-alikes are no more likely to be followed by another search in the same session (0.0007, SE 0.0010, \(p=0.46\)).

We read this as a statement about what a returned set's visual spread signals rather than as evidence that redundancy helps consumers. A visually tight set is largely what good retrieval looks like: a query that picks out a coherent product family returns products that resemble one another, and such searches convert into clicks at a high rate. A visually dispersed set is more often what a broad or ambiguous query returns, and those searches end without a click more often. The controls for mean compatibility, compatibility dispersion, textual distance, and category concentration absorb part of that difference, which is why the coefficient halves when they enter, but nothing in observational retrieval logs separates the composition of a set from the quality of the query that produced it. The caveats of Section~\ref{subsec:deferral-model} bind with full force here: a search without a click is not necessarily a failed search, and this analysis identifies neither an interference effect nor what a deduplicating ranker would deliver. Its value in the paper is as a caution: the within-set result of Section~\ref{subsec:primary-results} is about which product wins a click, and it should not be read as implying that visually varied assortments perform better overall.
\section{Mechanisms, Robustness, and External Validity}
\label{sec:robustness}

Section~\ref{sec:results} established our main results and confirmed the three hypotheses. What remains is to ask what those conclusions survive further rigorous inspections, what is their potential mechanism, and whether they carry predictive content outside the sample that produced them. The order below runs from the widest question to the narrowest. We begin with the one moderator our framework could not predict: the categorical composition of the returned set. We then work through the rival mechanisms, measurement contrasts, and specification choices that could generate the confirmatory results without the framework being true, together with a placebo test that reassigns images across products and recomputes every visual construct. Additional robustness checks can be found in Appendix \ref{app:robustness}. Finally, we ask whether any of this impacts clicks in a held-out period, to ensure the validity of our results. 

\subsection{Exploratory coherence moderation}
\label{subsec:coherence-results}

The three confirmatory tests treat the returned set as a competitive field without asking what kinds of products populate it. This section asks whether the categorical composition of that field matters, and it is reported as an exploratory analysis rather than as a test because Section~\ref{subsec:hypothesis} could not sign the prediction in advance. Let \emph{category coherence} be the share of a product's returned peers that fall in its own fine-grained catalog category, so a value near one describes a page of same-type products and a lower value a product surrounded by other kinds of things. Two readings of the framework point in opposite directions. On the first, coherence should amplify the premium: among categorically comparable peers, visual distance cannot be marking a difference in product type, so a well-fitting product's separation is unambiguously separation from its substitutes. On the second, coherence should mute it: when every returned product is of the same type, visual separation already distinguishes substitutes at any level of fit, whereas in a categorically mixed set relative fit is precisely what tells a distinctive credible answer apart from a distinctive intruder. 

Appendix Table~\ref{tab:app-coherence} estimates Equation~\ref{eq:coherence-model}, which adds coherence to the premium model together with every lower-order term the triple interaction implies and the corresponding between-product terms, on the 99.8\% of observations whose returned sets carry a usable second-level category throughout. The coefficient of interest is the triple interaction of relative fit, within-product distinctiveness, and coherence, \(\widehat\beta_{FWG}\), which is \(-0.00017\) (SE 0.00039; 95\% CI $[-0.00093, 0.00059]$). Its sign is the one the muting account predicts and its magnitude is about a fifteenth of the premium it is supposed to moderate. The interval is the more informative object here: it rules out any moderation larger than roughly a third of the premium in the muting direction, or a quarter of it in the amplifying direction, so the flat result reflects a genuinely small quantity rather than an inability to measure one. The between-product analogue, \(\widehat\beta_{FBG}\), is 0.00054 (SE 0.00045), also indistinguishable from null, so nothing suggests the moderation is instead operating through enduring product uniqueness.

The probability contrasts say the same thing in units that are easier to read than a triple interaction. Among the best-fitting products, where the premium is concentrated and where any moderation of it has the most room to show, moving within-product distinctiveness across its interquartile range changes click probability by 0.00228 in the least coherent quarter of observations and by 0.00250 in the most coherent, a difference of 0.00023 (SE 0.00034; Figure~\ref{fig:boundaries}, panel~(b)). The two curves run close together across the whole supported range of fit, and the gap between them narrows slightly as fit rises, which is the shape a small negative triple interaction produces; at no level of fit does that gap separate from zero.

One feature of the measure bounds what this null can be asked to carry. Coherence is concentrated at the top of its range: the median and the 75th percentile of the standardized measure coincide, because most returned sets in this catalog are categorically homogeneous even at the second level, so the upper half of the distribution is a point mass rather than a gradient. What the design can identify is therefore a contrast between sets with some categorically unlike peers and sets with none, not a smooth comparison across degrees of mixing, and a moderation that operates only among heavily mixed sets would be difficult to detect here. Within that limit the reading is descriptive: the categorical composition of the returned set does not detectably change how visual distinctiveness relates to selection, and neither of the two mechanisms is supported over the other.

We still cannot tell from the results above whether category mismatch contaminates our main results. In the next section, we address this question via the category-homogeneous and same-category specifications, where the distinctiveness association persists at full strength when every returned peer shares the focal product's category.

\subsection{Rival mechanisms, measurement contrasts, and robustness}
\label{subsec:robustness-results}

\subsubsection*{Rival mechanisms, category structure, and specification}

\ifinlineexhibits\input{exhibits/blocks/table3_robustness}\fi

Table~\ref{tab:robustness} and Figure~\ref{fig:specification-curve} organize the checks around the threats in Section~\ref{subsec:robustness}; in every row H1 denotes the within-product distinctiveness coefficient and H2 the premium, both from the interaction specification of column~4, whose reference estimates on the robustness panel are 0.0040 (SE 0.0005) and 0.0025 (SE 0.0005). The rival-mechanism rows move little. Adding textual differentiation and its fit interactions lowers the distinctiveness coefficient to 0.0031 and leaves the premium at 0.0025, so about a fifth of the visual association overlaps the textual channel and the rest does not; the relative-price interaction (0.0039, 0.0026) and the prior-exposure indicators (0.0040, 0.0025) change nothing, and centering fit within the set rather than ranking it gives 0.0041 and 0.0019. The category rows address mismatch directly: in fully category-homogeneous sets the estimates are 0.0045 and 0.0025, same-category distance gives a larger level coefficient of 0.0068 with a smaller premium of 0.0014, and the dominant-share and top-level-coherence moderators leave both coefficients at their reference values. Visual distance that merely marks a different product type cannot produce estimates that persist when every returned peer shares the focal product's category.

The position rows separate the association from the page. Dropping position controls entirely moves the estimates to 0.0039 and 0.0025, and replacing them with position-by-set-size effects gives 0.0040 and 0.0025, so the result does not depend on how the rank gradient is absorbed. Restricting to recorded positions where products were plausibly visible amplifies rather than attenuates: 0.0140 (SE 0.0015) with a premium of 0.0034 in the top ten positions, and 0.0236 (SE 0.0028) with 0.0058 in the top five, roughly 3.5 and 6 times the full-panel coefficients. The full-panel estimate averages over returned products that likely never entered the viewport, so it is best read as a lower bound diluted by unseen items. The product, session, and outcome rows behave the same way: product-by-week effects, which discard all variation between calendar weeks, give 0.0028 and 0.0018; requiring at least 5, 10, or 20 appearances per product holds the level near 0.0040 while the premium eases from 0.0023 to 0.0018; the largest connected component is indistinguishable from the full panel; first searches in a session (0.0041, 0.0025) and single-search sessions (0.0043, 0.0025) rule out within-session exposure spillovers; the unweighted product-row estimand gives 0.0034 and 0.0015; trimming the top and bottom percent of the constructs gives 0.0039 and 0.0027; and relaxing the single-click restriction to a fractional multiclick outcome on 1.42 million observations gives 0.0024 and 0.0019. The conditional multinomial logit in its Poisson representation reproduces column~6: a clear level effect of 0.032 latent-utility units (SE 0.007) with a null interaction of 0.004 (SE 0.006), the two-scale pattern of Proposition~\ref{prop:scales} on a second multiplicative estimator.

\ifinlineexhibits\input{exhibits/blocks/figure6_specification_curve}\fi
The measurement contrasts speak to the aggregation question that Section~\ref{subsec:contextual-differentiation} left empirical: how far the operative visual field extends. They do not order the way a strictly local account would suggest. The average distance from all returned peers carries a level coefficient of 0.0093 (SE 0.0010), the trimmed average 0.0087, and the position-local construct 0.0043, all at least as large as the nearest-substitute coefficient of 0.0040. Two readings, not exclusive, fit this pattern. The minimum over a set of distances is the noisiest of these constructs, an order statistic built from a single pair, so attenuation from measurement error is largest exactly where the theory's preferred object is sharpest; and when several close neighbors crowd a product, each contributes to conflation, so field-wide averages aggregate real signal rather than diluting it. What the contrast does establish is that no particular construction manufactures the result: every aggregation carries the association. The premium, by contrast, is largest under the nearest-substitute construct (0.0025 against 0.0018, 0.0012, and 0.0015 under the alternatives), consistent with the interaction operating at the margin of the closest pair, and the pairwise analyses below, which need a specific nearest neighbor to exist, are where the localization claim earns its keep. Across all 29 completed specifications in Figure~\ref{fig:specification-curve}, both coefficients are positive in every one, and both are significant at the five percent level in every linear click-outcome specification, with the level estimates spanning 0.0024 to 0.0236; the exceptions are exactly the two rows the framework predicts should differ, the latent-utility model, where the premium is null by Proposition~\ref{prop:scales}, and the downstream-action outcome, which the funnel analysis below takes up.
\subsubsection*{Falsification and sensitivity}
The placebo and the sensitivity analysis address confounding directly. In the category-within representation placebo, which reassigns images and recomputes nearest-substitute distance and its decomposition on every draw, the two coefficients behave very differently. The observed \(\widehat\beta_{W}\) exceeds all 200 permuted estimates (randomization \(p=0.005\)), whose mean is 0.00022 with a central interval of \(-0.00075\) to 0.00129: the distinctiveness effect is not something arbitrary image assignments produce. The observed \(\widehat\beta_{FW}\), by contrast, lies at the 37.5th percentile of its permutation distribution, whose mean is 0.00269 with a central interval of 0.00161 to 0.00358 (randomization \(p=0.63\)). The premium is therefore reproduced, in magnitude, by permuted images (Appendix Figure~\ref{fig:app-placebo}). The coefficient-stability analysis gives robustness values of 0.007 for \(\widehat\beta_{W}\) and 0.005 for \(\widehat\beta_{FW}\), so neither survives a confounder explaining even one percent of the residual variance in both the regressor and the outcome, which is a reminder that these are conditional associations of modest size.

Three results therefore point the same way about the premium: it is null in the conditional logit, it is indistinguishable from the representation placebo, and it shrinks as returned sets grow (Web Appendix~\ref{app:heterogeneity}). All three are consistent with Proposition~\ref{prop:scales} and Corollary~\ref{cor:setsize}: an individuation effect implies a probability-scale interaction with no log-odds counterpart, and implies that both the slope and the interaction shrink as merits spread over more products. The same facts are also consistent with a blunter reading, that the linear interaction is set-size structure with no behavioral content at all, since nearest-substitute distance falls mechanically as returned sets grow while the random-choice probability falls as \(1/J_s\). The two readings differ in what should survive within narrow set-size bands, which is what Section~\ref{subsec:premium-diagnostics} estimates. The distinctiveness effect itself is unaffected by this concern: it is estimated off the level of the within-product deviation rather than its interaction with a within-set rank, and it clears the placebo decisively.
\subsubsection*{Mechanism probes}
Four exploratory probes in Web Appendix~\ref{app:mechanism} ask how the pattern arises; each states its prediction before the estimate, none is adjusted for multiplicity, and none is a confirmatory test. The most informative is the \emph{near-twin duel}. A near-twin duel is the cleanest natural quasi-experiment the page offers: two listings so similar that a shopper would struggle to say what distinguishes them, competing in the same set, for the same query, at different heights on the page. Among 4,321 returned pairs whose members are each other's nearest visual substitute, whose distance falls in the lowest decile, and exactly one of whose members was clicked, display position dominates: a one-standard-deviation position advantage raises the probability of winning the pair by 0.138 (SE 0.007), and the higher-placed twin wins 63.9\% of the time. Relative fit matters but modestly (0.024, SE 0.007, \(p=0.002\); the better-fitting twin wins 55.8\% of duels), and the price advantage is null (0.008, SE 0.008). When two products are visually interchangeable, in other words, the page decides and the query breaks a small part of the remaining tie. Across the heterogeneity moderators (Appendix Figure~\ref{fig:app-heterogeneity}), the distinctiveness association is largest where the returned descriptions are closest together (0.0082 against about 0.0031 elsewhere), is flat in the display separation between a product and its nearest substitute (0.0032 when adjacent, 0.0056 nearby, 0.0035 distant), and is near zero for products returned earlier in the same session but not clicked (0.0001). The adjacency null is worth stating plainly: if visual interference operated through competition for a single glance, separation on the page should have mattered, and it does not. The restricted cubic spline in within-product distinctiveness stays positive through the top of the range (0.0076 at high fit at the center of the range,  \(p<0.001\)), so the linear specification is not hiding a reversal, and the relevant-peer split is directionally as predicted and null (distance from high-fit peers 0.0011, from low-fit peers 0.0003, difference \(p=0.43\)).
\subsubsection*{The purchase funnel}
The click-quality analysis adds the most consequential qualification in the paper. Among clicked products, within-product distinctiveness predicts a lower probability that the click is followed by an add-to-cart or purchase: \(-0.0077\) (SE 0.0015, \(p<0.001\)) against a base conversion of 11.6\%, so the clicks that distinctiveness recruits convert about seven percent less often than the average click. Two readings fit this. Under the individuation model the marginal encounter converted into a click by easier findability carries lower purchase intent than the average click, so per-click conversion falls even if total downstream demand rises; under a curiosity reading, distinctive presentation attracts clicks that were never purchase-directed and total demand does not rise. The two are separated by the funnel rather than by the pooled indicator, and Appendix Table~\ref{tab:app-funnel} decomposes it on the 51,656 clicked searches whose product had no matching action earlier in the session. The intuition for what follows is that the marginal click has a specific identity. It belongs to a shopper who would otherwise have scrolled past this product, and a click won by being easier to pick out of the page carries no more purchase intent than the browsing that preceded it, so its conversion rate should sit below the average click's without implying that anything was lost. The declines are close to proportional at every stage: relative to each outcome's base, a one-standard-deviation increase in within-product distinctiveness predicts 6.7\% fewer adds (\(-0.0077\), SE 0.0015, on a base of 0.116), 5.3\% fewer purchases within the attribution window (\(-0.0018\), SE 0.0008, base 0.033), 6.0\% fewer purchases in the session (\(-0.0023\), SE 0.0009, base 0.039), and 6.7\% less cart abandonment (\(-0.0052\), SE 0.0013, base 0.078). A curiosity account concentrated in idle clicks would thin the deep end of the funnel faster than the shallow end and pile up abandoned carts; instead every stage, abandonment included, thins by roughly the same five to seven percent, which is what marginal-click selection implies: the recruited encounter is an ordinary shopper's shallower one, not a different species of click. The one interaction that moves is consistent with the goal-consistency reading, though only at \(p=0.04\) and unadjusted: the fit-by-distinctiveness term is positive for window purchases (0.0013, SE 0.0006), so the conversion penalty concentrates in clicks on distinctive but poorly fitting products. The within-set ledger closes the account. If the clicks that distinctiveness recruits converted like average clicks, the add-or-purchase outcome in Table~\ref{tab:robustness} would show roughly \(0.0040\times0.116\approx0.00046\); pure reallocation of attention with no new demand would show zero. The estimate is 0.00017 (SE 0.00016), between the two benchmarks and distinguishable from neither: the click-side gain and the per-click conversion decline are of about the same relative size and cancel. Distinctiveness reallocates attention within the returned set; the data cannot show that it creates demand, and they rule out both a demand gain beyond the proportional benchmark and any material demand loss. Decision latency is uninformative here: only 7.7\% of clicks match a detail view under the timing rules, and the estimates are null with wide intervals.

\subsection{Is the premium a returned-set-size artifact?}
\label{subsec:premium-diagnostics}

The results so far leave one rival account of the premium standing: that the linear interaction is returned-set-size structure, since nearest-substitute distance falls mechanically as sets grow while every within-set slope scales inversely with set size. Table~\ref{tab:app-premium-diagnostics} separates the two with three diagnostics. The concern has a simple form: on a page of eight products, each one carries an eighth of the stakes, while on a page of thirty, each carries a thirtieth, and nearest-substitute distance falls as pages grow, so the two move together for reasons that have nothing to do with consumers. Panel~A estimates the premium inside narrow bands of returned-set size, where that mechanical covariation has little room to operate. The three small-set bands hold few searches and are individually uninformative (\(-0.005\), SE 0.005, in sets of five to seven; 0.006, SE 0.004, in eight to ten; 0.002, SE 0.003, in eleven to fourteen). The two bands that hold most of the data are both positive and significant: 0.0044 (SE 0.0020, \(p=0.025\)) in sets of fifteen to twenty and 0.0007 (SE 0.0003, \(p=0.024\)) in sets of twenty-one to thirty, the band with three quarters of the observations. A pure artifact predicts zero inside every band; the estimates instead stay positive where the data are dense and shrink with set size, which is the pattern Corollary~\ref{cor:setsize} predicts for a real effect whose stakes spread over more products.

Panel~B is the sharper test. It returns to the full panel and interacts both relative fit and within-product distinctiveness with set-size bin indicators, so the premium is identified only from variation within bins and anything operating through set size, the \(1/J_s\) scaling of the fit slope included, is absorbed by construction. The premium survives at 0.00213 (SE 0.00045; \(p<0.001\)), 84\% of its unrestricted size, with the distinctiveness level at 0.0061 (SE 0.0024). The linear interaction is not set-size structure.

Panel~C explains why the representation placebo of Section~\ref{subsec:robustness-results} nonetheless reproduces the interaction. That placebo permutes images within top-level category, and because 74\% of searches are category homogeneous, a permuted set largely keeps its own distance structure; a catalog-wide permutation breaks it. Five draws are feasible, since each requires recomputing every visual construct and re-estimating the full model, so Panel~C is a magnitude comparison rather than a formal test (with five draws the sharpest attainable one-sided randomization \(p\) is one in six). The level effect again clears the bar: the observed 0.0035 exceeds all five draws, whose mean is \(-0.0001\) with a central interval of \(-0.0010\) to 0.0008. The interaction does not: the placebo mean is 0.0029 (central interval 0.0024 to 0.0033) and the observed 0.0025 sits at the 20th percentile. Permuted images reproduce the interaction even when the permutation destroys all within-category structure, and that pins down the channel, because a placebo draw preserves exactly one thing: the returned sets and their sizes. A minimum over more arbitrary distances is smaller, so permuted nearest-substitute distance still carries set-size structure, and a regression without fit-by-set-size terms loads that structure onto the fit-by-distance interaction whether or not the images mean anything. The placebo percentile therefore measures the vulnerability of the unbinned design, an interaction of about 0.003 from set-size structure alone, not the content of the observed coefficient; Panel~B removes that channel and is where the confirmatory reading rests.

Taken together, the diagnostics resolve the tension that Sections~\ref{subsec:premium-results} and~\ref{subsec:robustness-results} set up. Within-bin identification preserves the premium; the band estimates shrink with set size as Corollary~\ref{cor:setsize} predicts while staying positive where the data are dense; and the placebo reproduction is accounted for by a set-size channel that the placebo design shares with the unbinned model. The premium is a regularity of the data on the probability scale, not an artifact of returned-set size. Its absence in the conditional logit and the Poisson representation then does the interpretive work of Proposition~\ref{prop:scales}: visual distance operates on whether a product is individuated, and the linear interaction is the level effect expressed on the probability scale rather than a second force.

\subsection{Out-of-time ranking}
\label{subsec:ranking-results}

\ifinlineexhibits\input{exhibits/blocks/table5_ranking}\fi

The held-out period contains 11,387 searches from the final 20\% of the observation window, 96.6\% of whose clicked products were seen during development. Table~\ref{tab:ranking} compares the recorded platform order with the relevance, additive, goal-consistent, and coherence-conditioned scores, all estimated from earlier observations. The platform order attains a mean reciprocal rank of 0.364, Hit@1 of 0.180, and Hit@3 of 0.429; the relevance score attains 0.398, 0.220, and 0.467. Every subsequent contrast is null. Adding distinctiveness moves mean reciprocal rank by \(-0.00005\), adding the fit interaction by 0.0003 relative to the additive score (session-bootstrap 95\% CI \(-0.0008\) to 0.0012), and adding the coherence terms by a further \(-0.0007\) (95\% CI \(-0.0019\) to 0.0004). None of the intervals excludes zero. The within-set associations documented above are real but small relative to relevance and position, which is what a null here should be taken to mean: at the margin of an already relevance-ordered list, visual distinctiveness does not reorder held-out clicks. Because the test-period clicks were generated under the historical display policy, these figures measure concordance with logged choices, not the response to a deployed ordering.

%% file: exhibits/blocks/table1_descriptives.tex
\IfFileExists{exhibits/generated/table1_sample_descriptives.tex}{%
  \input{exhibits/generated/table1_sample_descriptives}
}{%
\begin{table}[!htbp]
\centering
\caption{Descriptive statistics, measurement validity, and identifying support}
\label{tab:sample-descriptives}
\small
\begin{threeparttable}
\begin{tabularx}{\textwidth}{@{}Yrrrr@{}}
\toprule
& Mean & SD & P25 & P75 \\
\midrule
\multicolumn{5}{@{}l}{\textit{Panel A: Product-search variables}}\\
Clicked indicator & \TBD & \TBD & \TBD & \TBD \\
Within-set relative fit $F_{is}$ (rank) & \TBD & \TBD & \TBD & \TBD \\
Aligned semantic fit (global scale) & \TBD & \TBD & \TBD & \TBD \\
Nearest-substitute distinctiveness $D_{is}$ & \TBD & \TBD & \TBD & \TBD \\
Within-product distinctiveness $D^W_{is}$ & \TBD & \TBD & \TBD & \TBD \\
Average-peer visual distance (contrast measure) & \TBD & \TBD & \TBD & \TBD \\
Contextual textual differentiation $T_{is}$ & \TBD & \TBD & \TBD & \TBD \\
Second-level category coherence $G_{is}$ & \TBD & \TBD & \TBD & \TBD \\
Top-level category coherence & \TBD & \TBD & \TBD & \TBD \\
Relative price position & \TBD & \TBD & \TBD & \TBD \\
Recorded position & \TBD & \TBD & \TBD & \TBD \\
\midrule
\multicolumn{5}{@{}l}{\textit{Panel B: Search-level variables}}\\
Returned-set size $J_s$ & \TBD & \TBD & \TBD & \TBD \\
Dominant-category share & \TBD & \TBD & \TBD & \TBD \\
Category-homogeneous set (share of searches) & \multicolumn{4}{r}{\TBD} \\
\midrule
\multicolumn{5}{@{}l}{\textit{Panel C: Measurement validity}}\\
Clicked-minus-nonclicked fit gap (SD units; 95\% CI) & \multicolumn{4}{r}{\TBD} \\
Median within-set fit percentile of clicked product & \multicolumn{4}{r}{\TBD} \\
Click incidence, position 1 vs.\ position 10 & \multicolumn{4}{r}{\TBD} \\
\midrule
\multicolumn{5}{@{}l}{\textit{Panel D: Identifying support}}\\
Variance share of $D_{is}$: product / search / residual & \multicolumn{4}{r}{\TBD} \\
Median within-product SD of $D_{is}$ & \multicolumn{4}{r}{\TBD} \\
Products appearing in $\geq 5$ / $\geq 10$ searches (share) & \multicolumn{4}{r}{\TBD} \\
Share of rows in largest connected component & \multicolumn{4}{r}{\TBD} \\
Rows with at least two same-category peers (share) & \multicolumn{4}{r}{\TBD} \\
Rows in supported fit-differentiation-coherence cells (share) & \multicolumn{4}{r}{\TBD} \\
\midrule
Searches / sessions / products / product-search rows & \multicolumn{4}{r}{\TBD} \\
\bottomrule
\end{tabularx}
\begin{tablenotes}[flushleft]
\footnotesize
\item Notes: Statistics use the frozen analytic sample of single-click searches. Panel A statistics are equal-event weighted. Relative price position is the product's price decile minus the returned-set mean decile. Variance shares come from the two-way Shapley decomposition in Appendix~\ref{app:variance-support}. Sequential attrition from the raw search records appears in Appendix Table~\ref{tab:app-attrition}.
\end{tablenotes}
\end{threeparttable}
\end{table}
}

%% file: exhibits/generated/table1_sample_descriptives.tex
\begin{table}[tbp]
\centering
\caption{Descriptive statistics, measurement validity, and identifying support}
\label{tab:sample-descriptives}
\small
\begin{threeparttable}
\begin{tabularx}{\textwidth}{@{}Yrrrr@{}}
\toprule
& Mean & SD & P25 & P75 \\
\midrule
\multicolumn{5}{@{}l}{\textit{Panel A: Product-search variables}}\\
Clicked indicator & 0.073 & 0.259 & 0.000 & 0.000 \\
Within-set relative fit $F_{is}$ (rank) & 0.536 & 0.285 & 0.286 & 0.800 \\
Nearest-substitute distinctiveness $D_{is}$ & 0.171 & 0.136 & 0.084 & 0.218 \\
Within-product distinctiveness $D^W_{is}$ & 0.001 & 0.066 & -0.016 & 0.006 \\
Between-product distinctiveness $D^B_i$ & 0.000 & 0.117 & -0.079 & 0.049 \\
Near-twin indicator & 0.100 & 0.300 & 0.000 & 0.000 \\
Positions to nearest substitute & 5.305 & 4.941 & 1.000 & 8.000 \\
Average-peer visual distance (contrast measure) & 0.439 & 0.233 & 0.258 & 0.580 \\
Contextual textual differentiation $T_{is}$ & 0.406 & 0.276 & 0.180 & 0.607 \\
Set fit dispersion (SD within set) & 0.167 & 0.116 & 0.064 & 0.250 \\
Set mean textual distance & 0.406 & 0.238 & 0.207 & 0.603 \\
Second-level category coherence $G_{is}$ & 0.790 & 0.303 & 0.625 & 1.000 \\
Top-level category coherence & 0.915 & 0.200 & 0.958 & 1.000 \\
Relative price position & 0.000 & 1.390 & -0.562 & 0.640 \\
Recorded position & 9.718 & 6.905 & 4.000 & 15.000 \\
\midrule
\multicolumn{5}{@{}l}{\textit{Panel B: Search-level variables}}\\
Returned-set size $J_s$ & 18.435 & 7.638 & 11.000 & 25.000 \\
Dominant-category share & 0.942 & 0.124 & 0.960 & 1.000 \\
Category-homogeneous set (share of searches) & \multicolumn{4}{r}{0.744} \\
\midrule
\multicolumn{5}{@{}l}{\textit{Panel C: Measurement validity}}\\
Clicked-minus-nonclicked fit gap (SD units; 95\% CI) & \multicolumn{4}{r}{0.126 [0.121, 0.131]} \\
Median within-set fit percentile of clicked product & \multicolumn{4}{r}{0.611} \\
Click incidence, position 1 vs.\ position 10 & \multicolumn{4}{r}{0.165 vs.\ 0.040} \\
\midrule
\multicolumn{5}{@{}l}{\textit{Panel D: Identifying support}}\\
Variance share of $D_{is}$: product / search / residual & \multicolumn{4}{r}{0.652 / 0.180 / 0.168} \\
Median within-product SD of $D_{is}$ & \multicolumn{4}{r}{0.028} \\
Products appearing in $\geq 5$ / $\geq 10$ searches (share) & \multicolumn{4}{r}{0.792 / 0.645} \\
Share of rows in largest connected component & \multicolumn{4}{r}{1.000} \\
Rows with at least two same-category peers (share) & \multicolumn{4}{r}{0.990} \\
Rows in supported fit-differentiation-coherence cells (share) & \multicolumn{4}{r}{1.000} \\
\midrule
Searches / sessions / products / product-search rows & \multicolumn{4}{r}{56,931 / 53,334 / 24,383 / 1,049,523} \\
\bottomrule
\end{tabularx}
\begin{tablenotes}[flushleft]
\footnotesize
\item Notes: Statistics use the frozen analytic sample of single-click searches. Panel A statistics are equal-event weighted; Panel B statistics are unweighted across searches. Relative price position is the product's price decile minus the returned-set mean decile. Variance shares come from the two-way Shapley decomposition in Appendix B.4. The fit-gap interval uses session-clustered standard errors. Sequential attrition appears in Appendix Table A1.
\end{tablenotes}
\end{threeparttable}
\end{table}

%% file: exhibits/blocks/figure2_model_free.tex
\begin{figure}[!htbp]
\centering
\IfFileExists{exhibits/generated/figure2_model_free_evidence.pdf}{%
  \includegraphics[width=0.95\textwidth]{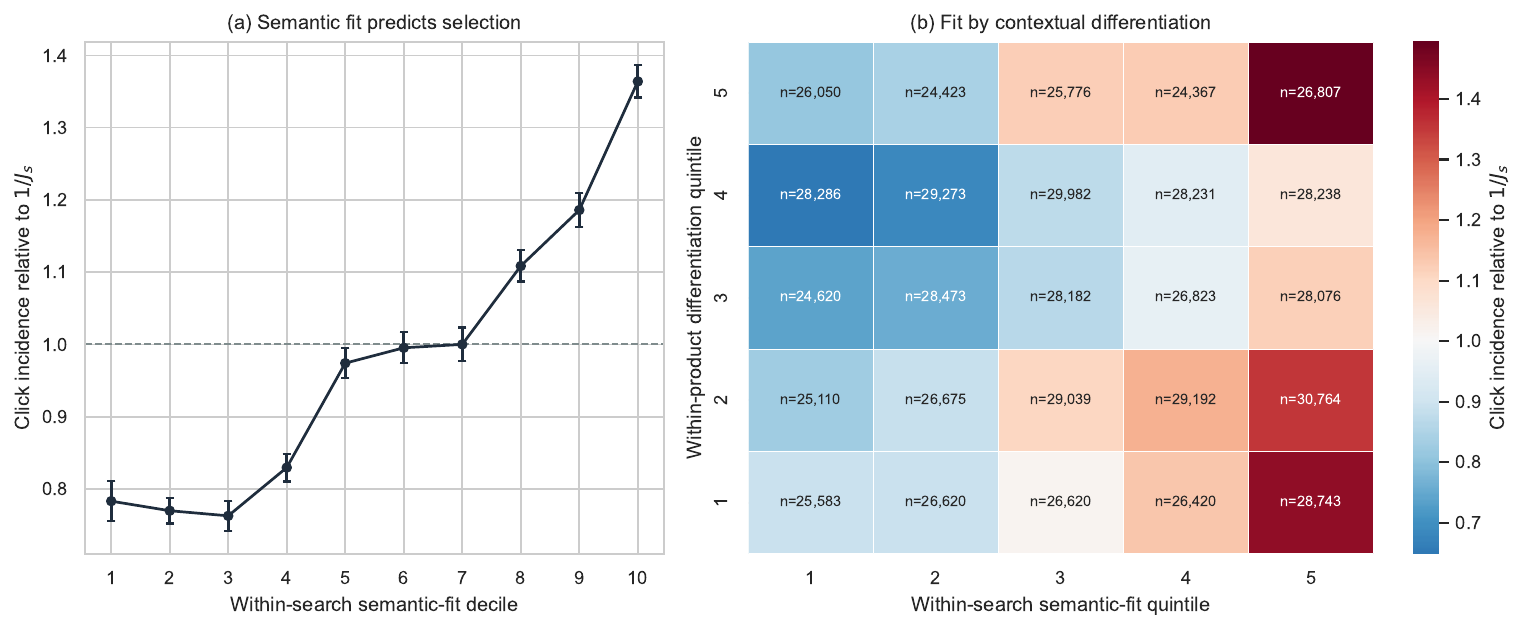}
}{%
  \ExhibitPlaceholder{3.5in}{Analysis output pending. Panel (a): click incidence relative to $1/J_s$ by within-search fit decile, with 95 percent intervals. Panel (b): click incidence relative to $1/J_s$ over within-search fit and distinctiveness bins, with cell counts printed and unsupported cells masked.}
}
\caption{Model-free evidence. Panel (a) validates the fit measure: click incidence relative to the random-choice benchmark $1/J_s$ by within-search fit decile. Panel (b) plots the same quantity over within-search fit and nearest-substitute distinctiveness bins; the surface adjusts for neither product appeal nor position.}
\label{fig:model-free-evidence}
\end{figure}

%% file: exhibits/blocks/figure3_identifying_variation.tex
\begin{figure}[!htbp]
\centering
\IfFileExists{exhibits/generated/figure3_identifying_variation.pdf}{%
  \includegraphics[width=0.95\textwidth]{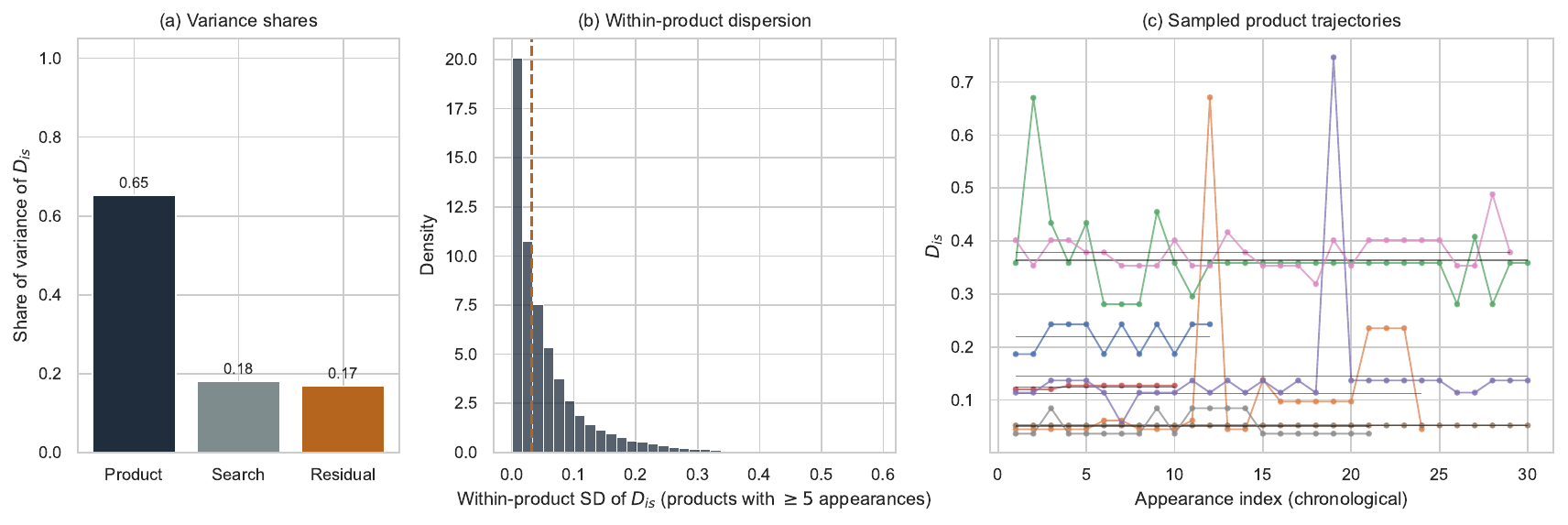}
}{%
  \ExhibitPlaceholder{3.5in}{Analysis output pending. Panel (a): Shapley variance shares of $D_{is}$ attributed to product effects, search effects, and residual product-search variation. Panel (b): distribution of within-product standard deviations of $D_{is}$ among products with at least five appearances. Panel (c): $D_{is}$ across appearances for a random sample of recurring products, one line per product, with the product mean marked.}
}
\caption{Identifying variation in nearest-substitute distinctiveness. Panel (a) decomposes the variance of $D_{is}$ into product, search, and residual components. Panel (b) shows how much the same product's distinctiveness varies across its appearances. Panel (c) traces $D_{is}$ across appearances for sampled products; the within-product component $D^W_{is}$ is the deviation from each product's mean.}
\label{fig:identifying-variation}
\end{figure}

%% file: exhibits/blocks/table2_main_models.tex
\IfFileExists{exhibits/generated/table2_main_models.tex}{%
  \input{exhibits/generated/table2_main_models}
}{%
\begin{table}[!htbp]
\centering
\caption{Distinctiveness, the goal-consistency premium, and set-level moderation}
\label{tab:main-models}
\scriptsize
\begin{threeparttable}
\begin{tabularx}{\textwidth}{@{}Ycccccc@{}}
\toprule
& \multicolumn{5}{c}{Linear probability model, change in click probability} & Logit \\
\cmidrule(lr){2-6}\cmidrule(lr){7-7}
& (1) & (2) & (3) & (4) & (5) & (6) \\
\midrule
Relative fit & \TBD & \TBD & \TBD & \TBD & \TBD & \TBD \\
Nearest-substitute distinctiveness & \TBD & \TBD &  &  &  &  \\
Within-product distinctiveness (H1) &  &  & \TBD & \TBD & \TBD & \TBD \\
Fit $\times$ within-product distinctiveness (H2) &  &  &  & \TBD & \TBD & \TBD \\
Fit $\times$ between-product distinctiveness &  &  &  & \TBD & \TBD & \TBD \\
Set fit dispersion &  &  &  &  & \TBD &  \\
Fit $\times$ dispersion &  &  &  &  & \TBD &  \\
Within distinctiveness $\times$ dispersion (H3) &  &  &  &  & \TBD &  \\
Fit $\times$ within distinctiveness $\times$ dispersion &  &  &  &  & \TBD &  \\
Between distinctiveness $\times$ dispersion &  &  &  &  & \TBD &  \\
Fit $\times$ between distinctiveness $\times$ dispersion &  &  &  &  & \TBD &  \\
\midrule
Search-event effects & Yes & Yes & Yes & Yes & Yes & Yes \\
Recorded-position effects & Yes & Yes & Yes & Yes & Yes & Yes \\
Product effects & No & Yes & Yes & Yes & Yes & Yes \\
Equal-event weights & Yes & Yes & Yes & Yes & Yes & n/a \\
$\widehat\beta_{FW}-\widehat\beta_{FB}$ (95\% CI) &  &  &  & \TBD & \TBD &  \\
IQR distinctiveness contrast at P10 / P50 / P90 fit &  &  &  & \TBD & \TBD & \TBD \\
Observations & \TBD & \TBD & \TBD & \TBD & \TBD & \TBD \\
Search events & \TBD & \TBD & \TBD & \TBD & \TBD & \TBD \\
\bottomrule
\end{tabularx}
\begin{tablenotes}[flushleft]
\footnotesize
\item Notes: The dependent variable is the product-level click indicator in single-click returned sets. Relative fit is the product's within-set fit rank and distinctiveness its visual distance from the most similar returned competitor, decomposed within and between products; continuous regressors are centered and standardized with equal-event weights. Standard errors in parentheses are two-way clustered by session and product. Column 3 is the preferred Hypothesis~1 model, column 4 the Hypothesis~2 model, and column 5 the Hypothesis~3 model, which interacts the distinctiveness slope with the dispersion of query-product compatibility inside the returned set and carries the complete lower-order hierarchy; column 6 is the fixed-effects conditional logit corresponding to column 4, in latent-utility units, with average predicted-probability changes in its interquartile row. The exploratory category-coherence moderation appears in the web appendix. The restriction row reports the within-minus-between interaction difference with its interval. Interquartile contrasts move within-product distinctiveness from its 25th to its 75th percentile at the stated fit percentiles (column 5 at median dispersion).
\end{tablenotes}
\end{threeparttable}
\end{table}
}

%% file: exhibits/generated/table2_main_models.tex
\begin{table}[tbp]
\centering
\caption{Distinctiveness, the goal-consistency premium, and set-level moderation}
\label{tab:main-models}
\small\setlength{\tabcolsep}{4pt}
\begin{threeparttable}
\begin{tabular}{@{}>{\raggedright\arraybackslash}p{5cm}ccccc@{}}
\toprule
& \multicolumn{4}{c}{Linear probability model, change in click probability} & Logit \\
\cmidrule(lr){2-5}\cmidrule(lr){6-6}
& (1) & (2) & (3) & (4) & (5) \\
\midrule
Relative fit & $0.0110^{***}$ & $0.0059^{***}$ & $0.0059^{***}$ & $0.0059^{***}$ & $0.0935^{***}$ \\
 & (0.0008) & (0.0007) & (0.0007) & (0.0007) & (0.0094) \\
Distinctiveness, level & $0.0012$ & $0.0072^{***}$ &  &  &  \\
 & (0.0008) & (0.0010) &  &  &  \\
Distinctiveness, within (H1) &  &  & $0.0035^{***}$ & $0.0040^{***}$ & $0.0316^{***}$ \\
 &  &  & (0.0005) & (0.0005) & (0.0067) \\
Fit $\times$ within (H2) &  &  &  & $0.0025^{***}$ & $0.0050$ \\
 &  &  &  & (0.0005) & (0.0059) \\
Fit $\times$ between &  &  &  & $0.0002$ & $0.0011$ \\
 &  &  &  & (0.0007) & (0.0095) \\
\midrule
Search-event FEs & Yes & Yes & Yes & Yes & Yes \\
Recorded-position FEs & Yes & Yes & Yes & Yes & Yes \\
Product FEs & No & Yes & Yes & Yes & Yes \\
$\widehat\beta_{FW}-\widehat\beta_{FB}$ &  &  &  & 0.0023 &  \\
\quad 95\% CI &  &  &  & [0.0008, 0.0039] &  \\
IQR contrast at P10 fit &  &  &  & 0.0002 & 0.0005 \\
\quad at P50 fit &  &  &  & 0.0013 & 0.0007 \\
\quad at P90 fit &  &  &  & 0.0024 & 0.0008 \\
Observations & 1,049,523 & 1,047,903 & 1,047,903 & 1,047,903 & 868,423 \\
Search events & 56,931 & 56,930 & 56,930 & 56,930 & 56,931 \\
\bottomrule
\end{tabular}
\begin{tablenotes}[flushleft]
\scriptsize
\item \textbf{Notes}: The dependent variable is click. Columns 1 to 4 are linear probability models, on a mean random-choice probability of 0.0725. Relative fit is the product's within-set fit rank and distinctiveness its visual distance from the most similar returned competitor; \emph{within} and \emph{between} denote the occasion-specific deviation and the product mean of that distance. Continuous regressors are standardized, and standard errors in parentheses are two-way clustered by session and product. Interquartile contrasts move within-product distinctiveness from its 25th to its 75th percentile at the stated fit percentiles. $^{*}p<0.10$, $^{**}p<0.05$, $^{***}p<0.01$.
\end{tablenotes}
\end{threeparttable}
\end{table}

%% file: exhibits/blocks/figure4_marginal_effect.tex
\begin{figure}[!htbp]
\centering
\IfFileExists{exhibits/generated/figure4_marginal_effect.pdf}{%
  \includegraphics[width=0.88\textwidth]{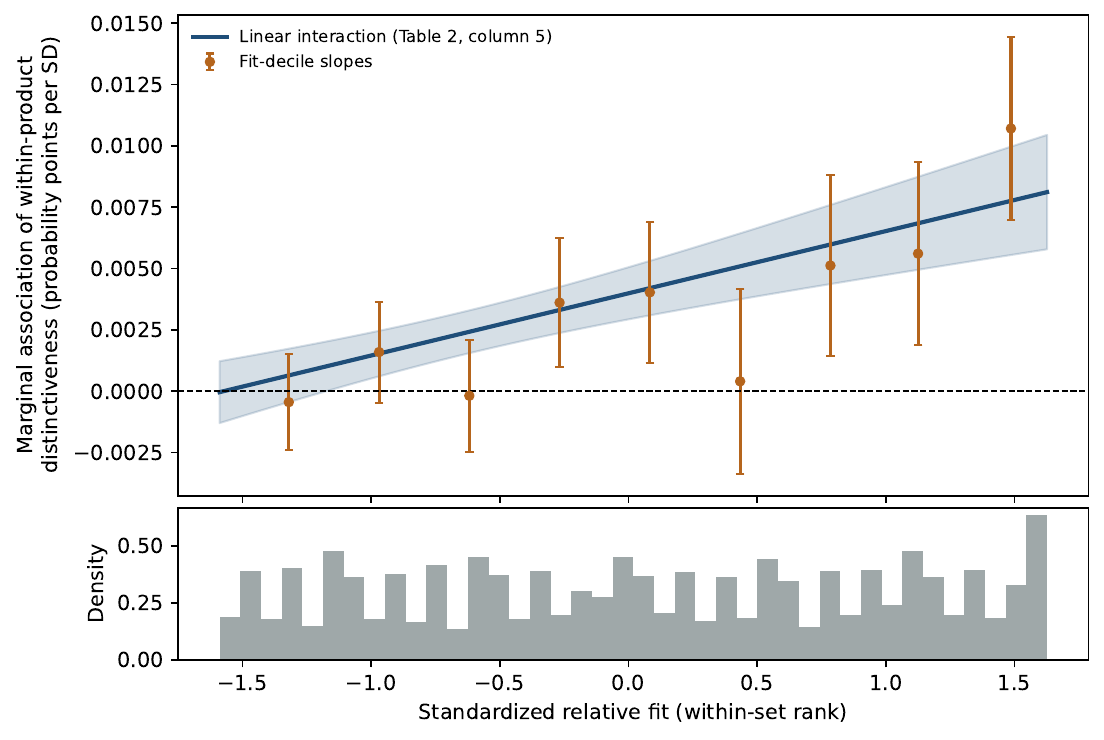}
}{%
  \ExhibitPlaceholder{3.5in}{Analysis output pending. Plot the marginal association of within-product distinctiveness (Equation~\ref{eq:marginal-differentiation}) over the central 90 percent of relative fit with 95 percent confidence bands, overlay the fit-decile slopes from the flexible specification as points with intervals, and add a support density or rug for fit along the horizontal axis.}
}
\caption{Marginal association of within-product distinctiveness across relative fit. The line is the estimate implied by column 4 of Table~\ref{tab:main-models}, whose level at mean fit is the Hypothesis~1 effect and whose slope is the Hypothesis~2 premium; points are decile-specific slopes from the flexible specification; the rug shows the support of fit.}
\label{fig:marginal-effect}
\end{figure}

%% file: exhibits/generated/tableA6_contest_diagnostics.tex
\begin{table}[!htp]
\centering
\caption{Channel moderation of the distinctiveness association: confirmatory moderator and variants}
\label{tab:app-contest-diagnostics}
\small
\begin{threeparttable}
\begin{tabularx}{\textwidth}{@{}Yrrrr@{}}
\toprule
Moderator & Interaction & SE & Slope at P25 & Slope at P75 \\
\midrule
Within-set SD of compatibility ($S_s$) & $-0.001$ & (0.001) & 0.005 & 0.004 \\
Best minus second-best compatibility & $0.001^{*}$ & (0.000) & 0.002 & 0.003 \\
Top-two gap as a share of the set's range & $0.002^{***}$ & (0.001) & 0.002 & 0.003 \\
Set mean textual distance & $-0.002^{***}$ & (0.001) & 0.007 & 0.004 \\
\bottomrule
\end{tabularx}
\begin{tablenotes}[flushleft]
\scriptsize
\item Notes: Each row estimates the moderation design of Equation~\ref{eq:contest-model} with a different measure of how far the returned products stand apart, carrying the complete lower-order hierarchy and the between-product terms. The reported interaction is distinctiveness by moderator. All models carry search, product, and recorded-position fixed effects, and standard errors reported are clustered by two-way session and product. $^{*}p<0.10$, $^{**}p<0.05$, $^{***}p<0.01$.
\end{tablenotes}
\end{threeparttable}
\end{table}

%% file: exhibits/blocks/figure5_boundaries.tex
\begin{figure}[!htbp]
\centering
\IfFileExists{exhibits/generated/figure5_boundaries.pdf}{%
  \includegraphics[width=0.98\textwidth]{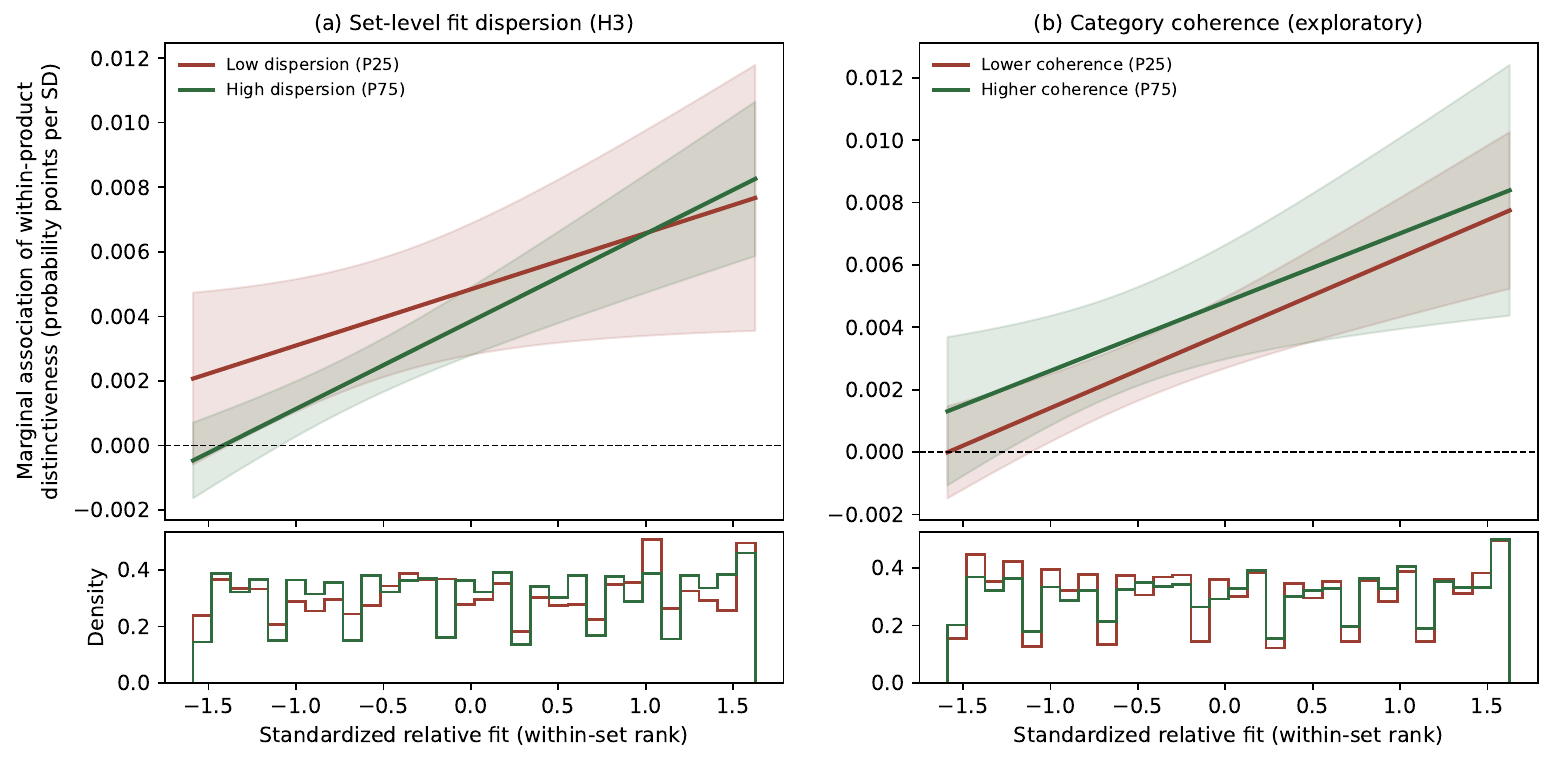}
}{%
  \ExhibitPlaceholder{3.5in}{Analysis output pending. Two panels of within-product distinctiveness slopes over relative fit with 95 percent confidence bands and support displays. Panel (a): low versus high dispersion of query-product compatibility within the returned set (Hypothesis 3). Panel (b): lower versus higher second-level category coherence (exploratory).}
}
\caption{Boundaries of the distinctiveness effect. Panel (a) plots within-product distinctiveness slopes over relative fit in sets whose products are close together and far apart in query-product compatibility, from column 5 of Table~\ref{tab:main-models}, the compatibility-dispersion variant of the channel moderation; the confirmatory textual-separation moderator of Hypothesis~\ref{hyp:contest} appears in Table~\ref{tab:app-contest-diagnostics}. Panel (b) repeats the display for category coherence, an exploratory moderator reported two-sided. Shaded regions are 95 percent confidence bands and the strips below show the support of each curve.}
\label{fig:boundaries}
\end{figure}

%% file: exhibits/blocks/table4_deferral.tex
\IfFileExists{exhibits/generated/table4_deferral.tex}{%
  \input{exhibits/generated/table4_deferral}
}{%
\begin{table}[!htbp]
\centering
\caption{Set-level visual redundancy and click incidence}
\label{tab:deferral}
\small
\begin{threeparttable}
\begin{tabularx}{\textwidth}{@{}Yccccc@{}}
\toprule
& \multicolumn{4}{c}{Any click on the returned set} & Reformulation \\
\cmidrule(lr){2-5}\cmidrule(lr){6-6}
& (1) & (2) & (3) & (4) & (5) \\
\midrule
Mean nearest-substitute distance & \TBD & \TBD &  & \TBD & \TBD \\
Mean average-peer distance (contrast) &  &  & \TBD & \TBD &  \\
Mean query-product compatibility &  & \TBD & \TBD & \TBD & \TBD \\
Compatibility dispersion &  & \TBD & \TBD & \TBD & \TBD \\
Mean textual distance &  & \TBD & \TBD & \TBD & \TBD \\
Dominant-category share &  & \TBD & \TBD & \TBD & \TBD \\
Log returned-set size &  & \TBD & \TBD & \TBD & \TBD \\
\midrule
Week, set-size, and modal-category effects & Yes & Yes & Yes & Yes & Yes \\
Outcome mean & \TBD & \TBD & \TBD & \TBD & \TBD \\
Search events & \TBD & \TBD & \TBD & \TBD & \TBD \\
\bottomrule
\end{tabularx}
\begin{tablenotes}[flushleft]
\footnotesize
\item Notes: The unit is a structurally eligible search event, including the zero-click searches the choice panel excludes. Columns 1 to 4 take an indicator that the search received at least one click; column 5 takes an indicator that another search followed in the same session within the attribution window. Regressors are standardized across search events and standard errors are clustered by session. Column 3 is the set-level counterpart of the measurement contrasts in Table~\ref{tab:robustness}, and column 4 enters both distance measures.
\end{tablenotes}
\end{threeparttable}
\end{table}
}

%% file: exhibits/generated/table4_deferral.tex
\begin{table}[tbp]
\centering
\caption{Set-level visual redundancy and click incidence}
\label{tab:deferral}
\small
\begin{threeparttable}
\begin{tabularx}{\textwidth}{@{}Yccccc@{}}
\toprule
& \multicolumn{4}{c}{Any click on the returned set} & Reformulation \\
\cmidrule(lr){2-5}\cmidrule(lr){6-6}
& (1) & (2) & (3) & (4) & (5) \\
\midrule
Mean nearest-substitute distance & $-0.040^{***}$ & $-0.024^{***}$ &  & $-0.006^{***}$ & $0.001$ \\
 & (0.001) & (0.001) &  & (0.001) & (0.001) \\
Mean average-peer distance (contrast) &  &  & $-0.039^{***}$ & $-0.034^{***}$ &  \\
 &  &  & (0.001) & (0.001) &  \\
Mean query-product compatibility &  & $0.025^{***}$ & $0.019^{***}$ & $0.020^{***}$ & $-0.010^{***}$ \\
 &  & (0.001) & (0.001) & (0.001) & (0.001) \\
Compatibility dispersion &  & $0.017^{***}$ & $0.018^{***}$ & $0.017^{***}$ & $-0.012^{***}$ \\
 &  & (0.001) & (0.001) & (0.001) & (0.001) \\
Mean textual distance &  & $-0.011^{***}$ & $-0.005^{***}$ & $-0.004^{**}$ & $0.015^{***}$ \\
 &  & (0.002) & (0.002) & (0.002) & (0.002) \\
Dominant-category share &  & $0.022^{***}$ & $0.019^{***}$ & $0.019^{***}$ & $-0.006^{***}$ \\
 &  & (0.001) & (0.001) & (0.001) & (0.001) \\
\midrule
Week, set-size, and modal-category effects & Yes & Yes & Yes & Yes & Yes \\
Outcome mean & 0.250 & 0.250 & 0.250 & 0.250 & 0.233 \\
Search events & 333,694 & 333,694 & 333,694 & 333,694 & 333,694 \\
\bottomrule
\end{tabularx}
\begin{tablenotes}[flushleft]
\footnotesize
\item Notes: The unit is a structurally eligible search event, including the zero-click searches that the choice panel excludes. The dependent variable in columns 1 to 4 is an indicator that the search received at least one click and in column 5 an indicator that another search followed in the same session within the attribution window. Regressors are standardized across search events. All columns absorb week, returned-set-size, and modal-category effects; standard errors in parentheses are clustered by session. Column 3 replaces nearest-substitute distance with the average distance among all returned pairs, the set-level counterpart of the measurement contrasts in Table 3, and column 4 enters both. $^{*}p<0.10$, $^{**}p<0.05$, $^{***}p<0.01$.
\end{tablenotes}
\end{threeparttable}
\end{table}

%% file: exhibits/blocks/table3_robustness.tex
\IfFileExists{exhibits/generated/table3_robustness.tex}{%
  \input{exhibits/generated/table3_robustness}
}{%
\begin{table}[!ht]
\centering
\caption{Rival mechanisms, corroborating evidence, and robustness}
\label{tab:robustness}
\scriptsize
\begin{threeparttable}
\begin{tabularx}{\textwidth}{@{}Yrrrrr@{}}
\toprule
Specification & H1 & SE & H2 & SE & Observations \\
\midrule
Preferred fixed-effects model (nearest-substitute distinctiveness, relative fit) & \TBD & \TBD & \TBD & \TBD & \TBD \\
\multicolumn{6}{@{}l}{\textit{Rival mechanisms}}\\
Add textual differentiation and its fit interactions & \TBD & \TBD & \TBD & \TBD & \TBD \\
Add relative price position $\times$ fit interaction & \TBD & \TBD & \TBD & \TBD & \TBD \\
Add prior-exposure indicators (returned or clicked earlier in session) & \TBD & \TBD & \TBD & \TBD & \TBD \\
Fit centered within set & \TBD & \TBD & \TBD & \TBD & \TBD \\
\multicolumn{6}{@{}l}{\textit{Measurement contrasts (theory predicts attenuation)}}\\
Average distance from all returned peers & \TBD & \TBD & \TBD & \TBD & \TBD \\
Trimmed-average distance (drops the farthest peer) & \TBD & \TBD & \TBD & \TBD & \TBD \\
Globally standardized (aligned) fit instead of within-set rank & \TBD & \TBD & \TBD & \TBD & \TBD \\
\multicolumn{6}{@{}l}{\textit{Category structure}}\\
Category-homogeneous sets & \TBD & \TBD & \TBD & \TBD & \TBD \\
Same-category average distance & \TBD & \TBD & \TBD & \TBD & \TBD \\
Search-level dominant-category share as moderator & \TBD & \TBD & \TBD & \TBD & \TBD \\
Top-level category coherence as moderator & \TBD & \TBD & \TBD & \TBD & \TBD \\
\multicolumn{6}{@{}l}{\textit{Position and exposure}}\\
No position controls & \TBD & \TBD & \TBD & \TBD & \TBD \\
Top 10 recorded positions & \TBD & \TBD & \TBD & \TBD & \TBD \\
Top 5 recorded positions & \TBD & \TBD & \TBD & \TBD & \TBD \\
Position-by-set-size effects & \TBD & \TBD & \TBD & \TBD & \TBD \\
\multicolumn{6}{@{}l}{\textit{Neighborhood, product, session, and outcome}}\\
Position-local distinctiveness ($\pm k$ positions) & \TBD & \TBD & \TBD & \TBD & \TBD \\
Product-by-week effects & \TBD & \TBD & \TBD & \TBD & \TBD \\
At least 10 appearances per product & \TBD & \TBD & \TBD & \TBD & \TBD \\
First search in session & \TBD & \TBD & \TBD & \TBD & \TBD \\
Multiclick fractional outcome & \TBD & \TBD & \TBD & \TBD & \TBD \\
Add-to-cart or purchase outcome & \TBD & \TBD & \TBD & \TBD & \TBD \\
Conditional multinomial logit (Poisson representation) & \TBD & \TBD & \TBD & \TBD & \TBD \\
\multicolumn{6}{@{}l}{\textit{Falsification and sensitivity}}\\
Representation placebo: observed percentile & \TBD &  & \TBD &  & \TBD \\
Robustness value (share of residual variance) & \TBD &  & \TBD &  & \TBD \\
\bottomrule
\end{tabularx}
\begin{tablenotes}[flushleft]
\footnotesize
\item Notes: H1 denotes the within-product distinctiveness coefficient and H2 its interaction with within-set relative fit, each from the model in the row. Unless stated otherwise, estimates use nearest-substitute distinctiveness and relative fit with search, product, and flexible recorded-position effects, equal-event weights, and two-way session/product clustering. The measurement-contrast rows replace the construct with a theoretically weaker alternative and are expected to attenuate. The conditional multinomial-logit row is in latent-utility units. Figure~\ref{fig:specification-curve} displays these estimates with confidence intervals. The appendix reports the remaining recurrence thresholds, flexible functions, unweighted estimates, tail trimming, single-search sessions, the largest connected component, and both placebo distributions.
\end{tablenotes}
\end{threeparttable}
\end{table}
}

%% file: exhibits/generated/table3_robustness.tex
\begin{table}[tbp]
\centering
\caption{Rival mechanisms, corroborating evidence, and robustness}
\label{tab:robustness}
\footnotesize
\begin{threeparttable}
\begin{tabularx}{\textwidth}{@{}Yrrrrr@{}}
\toprule
Specification & H1 & SE & H2 & SE & Observations \\
\midrule
Preferred fixed-effects model (nearest-substitute distinctiveness, relative fit) & $0.0040^{***}$ & 0.0005 & $0.0025^{***}$ & 0.0005 & 1,046,032 \\
\multicolumn{6}{@{}l}{\textit{Rival mechanisms}}\\
Add textual differentiation and its fit interactions & $0.0031^{***}$ & 0.0005 & $0.0025^{***}$ & 0.0005 & 1,046,032 \\
Add relative price position $\times$ fit interaction & $0.0039^{***}$ & 0.0005 & $0.0026^{***}$ & 0.0005 & 1,045,938 \\
Add prior-exposure indicators (returned or clicked earlier in session) & $0.0040^{***}$ & 0.0005 & $0.0025^{***}$ & 0.0005 & 1,046,032 \\
Fit centered within set & $0.0041^{***}$ & 0.0005 & $0.0019^{***}$ & 0.0003 & 1,046,032 \\
\multicolumn{6}{@{}l}{\textit{Measurement contrasts (alternative constructions)}}\\
Average distance from all returned peers & $0.0093^{***}$ & 0.0010 & $0.0018^{***}$ & 0.0005 & 1,046,032 \\
Trimmed-average distance (drops the farthest peer) & $0.0087^{***}$ & 0.0010 & $0.0012^{**}$ & 0.0005 & 1,046,032 \\
Globally standardized (aligned) fit instead of within-set rank & $0.0051^{***}$ & 0.0007 & $0.0026^{***}$ & 0.0005 & 1,046,032 \\
\multicolumn{6}{@{}l}{\textit{Category structure}}\\
Category-homogeneous sets & $0.0045^{***}$ & 0.0007 & $0.0025^{***}$ & 0.0007 & 720,860 \\
Same-category average distance & $0.0068^{***}$ & 0.0009 & $0.0014^{***}$ & 0.0005 & 898,605 \\
Search-level dominant-category share as moderator & $0.0040^{***}$ & 0.0005 & $0.0025^{***}$ & 0.0005 & 1,047,903 \\
Top-level category coherence as moderator & $0.0040^{***}$ & 0.0005 & $0.0025^{***}$ & 0.0005 & 1,047,903 \\
\multicolumn{6}{@{}l}{\textit{Position and exposure}}\\
No position controls & $0.0039^{***}$ & 0.0005 & $0.0025^{***}$ & 0.0005 & 1,046,032 \\
Top 10 recorded positions & $0.0140^{***}$ & 0.0015 & $0.0034^{***}$ & 0.0008 & 380,536 \\
Top 5 recorded positions & $0.0236^{***}$ & 0.0028 & $0.0058^{***}$ & 0.0017 & 146,441 \\
Position-by-set-size effects & $0.0040^{***}$ & 0.0005 & $0.0025^{***}$ & 0.0005 & 1,046,032 \\
\multicolumn{6}{@{}l}{\textit{Neighborhood, product, session, and outcome}}\\
Position-local distinctiveness ($\pm k$ positions) & $0.0043^{***}$ & 0.0005 & $0.0015^{***}$ & 0.0004 & 1,046,032 \\
Product-by-week effects & $0.0028^{***}$ & 0.0006 & $0.0018^{***}$ & 0.0005 & 996,036 \\
At least 10 appearances per product (complete sets) & $0.0040^{***}$ & 0.0006 & $0.0020^{***}$ & 0.0006 & 837,470 \\
First search in session & $0.0041^{***}$ & 0.0006 & $0.0025^{***}$ & 0.0005 & 978,539 \\
Multiclick fractional outcome & $0.0024^{***}$ & 0.0004 & $0.0019^{***}$ & 0.0004 & 1,424,703 \\
Add-to-cart or purchase outcome & $0.0002$ & 0.0002 & $0.0002^{*}$ & 0.0001 & 1,028,396 \\
Conditional multinomial logit (Poisson representation) & $0.0317^{***}$ & 0.0067 & $0.0041$ & 0.0059 & 866,956 \\
\multicolumn{6}{@{}l}{\textit{Falsification and sensitivity}}\\
Representation placebo: observed percentile & 1.000 & & 0.375 & & 200 draws \\
Robustness value (share of residual variance) & 0.007 & & 0.005 & & \\
\bottomrule
\end{tabularx}
\begin{tablenotes}[flushleft]
\scriptsize
\item Notes: H1 denotes the within-product distinctiveness coefficient and H2 its interaction with within-set relative fit, each from the model in the row. Unless stated otherwise, estimates use nearest-substitute distinctiveness and relative fit with search, and product FEs, flexible recorded-position effects, and SEs are two-way session - product clustered. The measurement-contrast rows replace each primary construct with an alternative construction; Section~\ref{subsec:robustness-results} discusses their reading. A blank H2 cell indicates that the moderator is fixed by the sample restriction. The conditional multinomial-logit row is in latent-utility units. Figure~\ref{fig:specification-curve} displays these estimates with confidence intervals; the appendix reports the remaining specifications, flexible functions, and both placebo distributions.
\end{tablenotes}
\end{threeparttable}
\end{table}

%% file: exhibits/blocks/figure6_specification_curve.tex
\begin{figure}[!htbp]
\centering
\IfFileExists{exhibits/generated/figure6_specification_curve.pdf}{%
  \includegraphics[width=\textwidth]{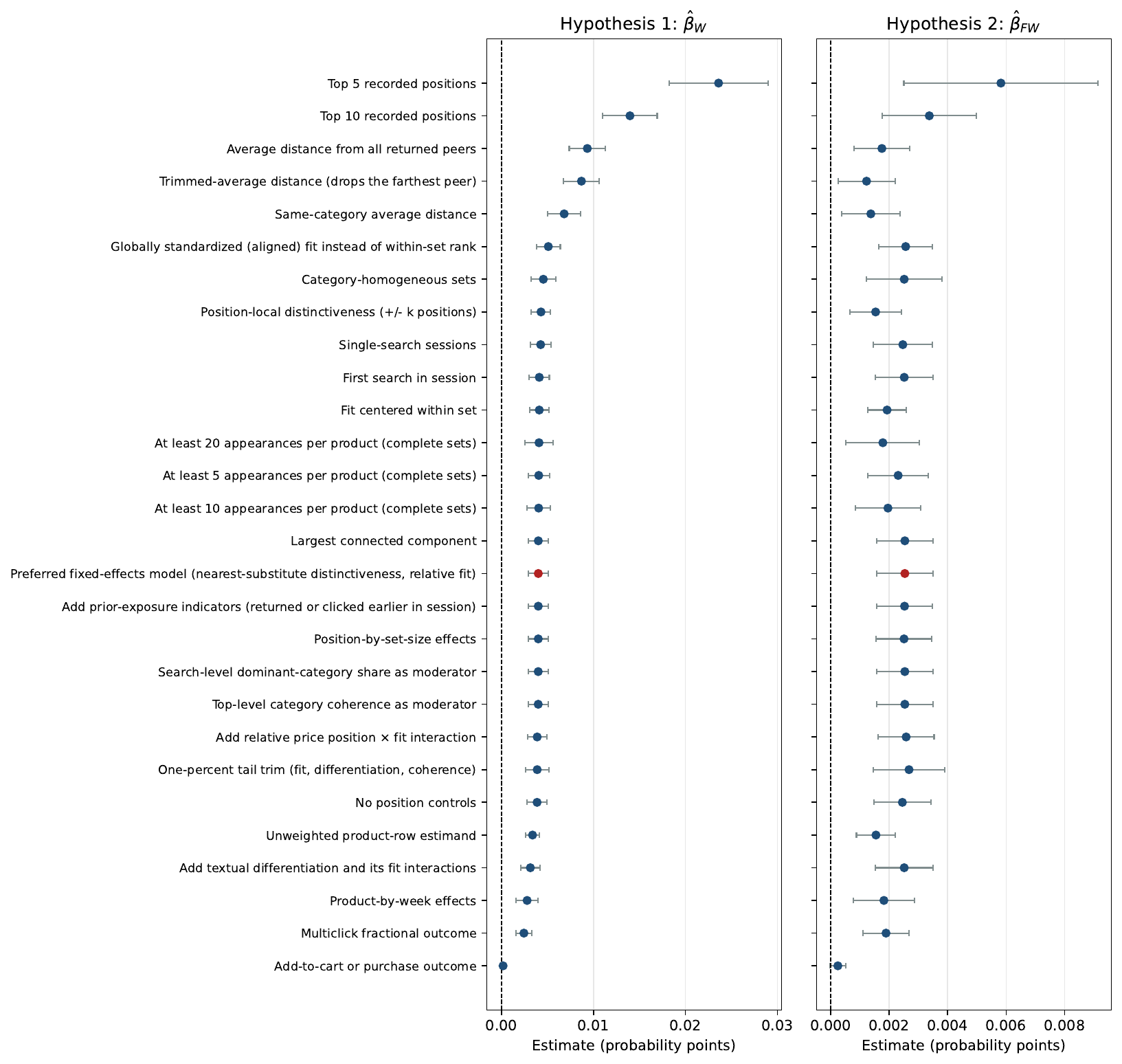}
}{%
  \ExhibitPlaceholder{3.5in}{Analysis output pending. Forest plot with one row per specification in Table~\ref{tab:robustness} and the appendix, sorted by the Hypothesis 1 estimate: left panel $\widehat\beta_{W}$, right panel $\widehat\beta_{FW}$, each with a 95 percent interval, the preferred model highlighted, and a zero reference line.}
}
\caption{Specification curve for the confirmatory coefficients. Each row is one specification from Table~\ref{tab:robustness} and the appendix, sorted by the Hypothesis~1 estimate; the left panel plots the within-product distinctiveness coefficient $\widehat\beta_{W}$ and the right panel the goal-consistency premium $\widehat\beta_{FW}$ with 95 percent confidence intervals. The preferred model is highlighted; the conditional logit is omitted because it is on a different scale. Measurement-contrast rows replace a primary construct with an alternative construction, and Section~\ref{subsec:robustness-results} discusses their reading.}
\label{fig:specification-curve}
\end{figure}

%% file: exhibits/blocks/table5_ranking.tex
\IfFileExists{exhibits/generated/table5_ranking.tex}{%
  \input{exhibits/generated/table5_ranking}
}{%
\begin{table}[!htbp]
\centering
\caption{Chronological holdout ranking comparison}
\label{tab:ranking}
\small
\begin{threeparttable}
\begin{tabular}{@{}lccc@{}}
\toprule
Ranking & Mean reciprocal rank & Hit@1 & Hit@3 \\
\midrule
Recorded platform order & \TBD & \TBD & \TBD \\
Position, relevance, and product prior ($S^R$) & \TBD & \TBD & \TBD \\
Additive relevance and distinctiveness ($S^A$) & \TBD & \TBD & \TBD \\
Goal-consistency premium, H2 ($S^I$) & \TBD & \TBD & \TBD \\
Coherence-conditioned score, exploratory ($S^G$) & \TBD & \TBD & \TBD \\
\midrule
$S^I$ minus $S^A$ & \TBD & \TBD & \TBD \\
\quad Session-bootstrap 95\% CI & \TBD & \TBD & \TBD \\
$S^G$ minus $S^I$ & \TBD & \TBD & \TBD \\
\quad Session-bootstrap 95\% CI & \TBD & \TBD & \TBD \\
\midrule
Held-out searches / warm-product share & \multicolumn{3}{c}{\TBD} \\
\bottomrule
\end{tabular}
\begin{tablenotes}[flushleft]
\footnotesize
\item Notes: Sessions are ordered by the shifted timestamp of their first search; the first 80 percent of searches form the development period and the final 20 percent the held-out period. Standardization constants, coefficients, and product priors use development observations only. Warm and cold products and strata by set size and coherence appear in Appendix Figure~\ref{fig:app-ranking-strata}.
\end{tablenotes}
\end{threeparttable}
\end{table}
}

%% file: exhibits/generated/table5_ranking.tex
\begin{table}[tbp]
\centering
\caption{Chronological holdout ranking comparison}
\label{tab:ranking}
\small\setlength{\tabcolsep}{2pt}
\begin{threeparttable}
\begin{tabular}{@{}lccc@{}}
\toprule
Ranking & Mean reciprocal rank & Hit@1 & Hit@3 \\
\midrule
Recorded platform order & 0.364 & 0.180 & 0.429 \\
Position, relevance, and product prior ($S^R$) & 0.398 & 0.220 & 0.467 \\
Additive relevance and distinctiveness ($S^A$) & 0.398 & 0.219 & 0.467 \\
Goal-consistency premium, H2 ($S^I$) & 0.398 & 0.220 & 0.466 \\
Coherence-conditioned score, exploratory ($S^G$) & 0.397 & 0.219 & 0.467 \\
\midrule
$S^I$ minus $S^A$ & 0.000 & 0.001 & -0.001 \\
\quad 95\% CI & [-0.001, 0.001] & [-0.001, 0.002] & [-0.003, 0.002] \\
$S^G$ minus $S^I$ & -0.001 & -0.001 & 0.001 \\
\quad 95\% CI & [-0.002, 0.000] & [-0.003, 0.001] & [-0.002, 0.003] \\
\midrule
Held-out searches / warm-product share & \multicolumn{3}{c}{11,387 / 0.966} \\
\bottomrule
\end{tabular}
\begin{tablenotes}[flushleft]
\footnotesize
\item Notes: Sessions are ordered by the shifted timestamp of their first retained search; the first 80 percent of searches form the development period and the final 20 percent the held-out period. Standardization constants, coefficients, and product priors use development observations only. Warm and cold products and strata by set size and coherence appear in Appendix Figure A3.
\end{tablenotes}
\end{threeparttable}
\end{table}

%% file: sections/06_discussion.tex

\section{Discussion}
\label{sec:discussion}

Online search places each returned product in relation to an expressed goal and to a contemporaneous set of competitors, and the closest competitor is the one that matters. The preferred specification estimates a within-product distinctiveness coefficient of 0.0035 (SE 0.0005), about 4.8\% of a random-choice click, concentrated entirely in the within-product component: the between-product interaction is 0.0002 (SE 0.0007). It exceeds all 200 draws of a representation placebo and holds in a conditional logit. The premium is 0.0025 (SE 0.0005) on the linear scale and 0.005 (SE 0.006) in the logit, the two-scale pattern Proposition~\ref{prop:scales} implies when distance operates on individuation rather than utility; it survives identification within narrow bands of returned-set size, and the set-size structure that lets a permuted-image placebo reproduce it is a property of the unbinned design rather than of the estimate. The distinctiveness association weakens as the set's textual separation rises, doing about twice the work where the returned descriptions sit closest together, while the compatibility-dispersion variant of the same moderation is null. The clicks distinctiveness recruits convert five to seven percent less often at every stage of the purchase funnel, so the within-set demand ledger closes flat. At the level of the search event, visually tight sets receive clicks more often, which we read as retrieval quality rather than a benefit of redundancy, and a chronological holdout shows no reordering gain over a relevance-only score.

\subsection{Contributions}
\label{subsec:theoretical-contributions}

The framework places visual differentiation inside an active goal and localizes it at the nearest substitute. Research on prototypicality and unity explains why coherent designs are attractive, and research on salience and uniqueness explains why standing apart can help \citep{veryzerhutchinson1998unity,landwehr2011gut,milosavljevic2012saliency,feng2025visual}. Our evidence concerns a different quantity: not whether a product looks unusual, but whether the assortment returned on a particular occasion contains a close look-alike. Treating the nearest substitute as the operative reference imports the central regularity of visual search, that interference comes from the most similar distractors \citep{duncan1989visual}, into a demand setting. The evidence that the association is contextual is strongest in the nested models, where it is invisible until product effects absorb enduring appearance, and in the placebo, where reassigning images destroys it; the evidence that it is local rests on the pairwise analyses and on where the premium concentrates, since broader aggregations of the visual field carry the level association as well. The goal-consistency premium would extend context-dependent choice to an environment with an observable statement of the consumer's objective, since the query supplies a second reference point beside the surrounding options \citep{tverskysimonson1993context,dharsherman1996common,rooderkerk2011context}. The individuation model reframes what that premium is. The gradient is clear on the linear scale and in the fit deciles, vanishes in the conditional logit, weakens sharply as returned sets grow yet survives when identification is confined within narrow set-size bins, and Proposition~\ref{prop:scales} with Corollary~\ref{cor:setsize} implies exactly this configuration when visual separation works by letting merits act rather than by adding utility. The honest summary is that the level of the distinctiveness effect is established, and that the premium is best read as its implication on the probability scale rather than as evidence that consumers value goal-consistent distinctiveness per se.

The third prediction, that visual separation matters most where words fail, is confirmed in the moderation design: the distinctiveness slope falls from 0.0074 at the 25th percentile of the set's textual separation to 0.0036 at the 75th, which is what a cue-diagnosticity reading predicts and a decorative reading does not, since a decorative attribute has no reason to matter more exactly where the substantive channel stops discriminating \citep{feldmanlynch1988self,chernev2005feature}. The neighboring compatibility-dispersion moderation is null, and the top-two-gap variant carries the opposite, merit-margin sign, which is what the model expects of a moderator that bundles how well the query discriminates with how much merit is at stake; the dispersion null is the mixture, not a failure of the channel logic. The set-level companion analysis, meanwhile, cautions against reading any of this as an assortment-design prescription. Visually tight returned sets receive clicks more often, not less, and the most plausible account is that visual tightness is what a well-targeted query produces \citep{iyengarlepper2000choice,townsendkahn2014visual}. Which product wins a click and whether the page produces one are different questions with different answers, and the funnel behind the click is a third: the clicks distinctiveness recruits convert five to seven percent less often at every stage, abandonment included, a proportional thinning that matches marginal-click selection rather than the abandonment-heavy signature of curiosity clicking, and the within-set downstream total sits between pure attention reallocation and fully proportional demand, distinguishable from neither. The within-set evidence is therefore about where attention lands, not about how much demand the page generates.

The paper also changes the level at which visual differentiation is measured. Marketplace research characterizes an image as unusual relative to a category or population \citep{feng2025visual}. Nearest-substitute distinctiveness varies across occasions while the image is fixed, because ranking algorithms continuously place the same product beside new competitors. The within-product design uses that variation to separate local contrast from enduring uniqueness, so the estimates cannot be produced by stable differences between unusual-looking and ordinary-looking products. This is a measurement contribution in the tradition of using image representations for demand analysis \citep{zhang2022image,dew2022logos,burnap2023aesthetic}, applied to a relational rather than an absolute quantity. Relative fit makes the parallel move on the semantic side, replacing an uncalibrated similarity level with the product's standing among the candidates actually scored against the same query.

For search and recommendation, the results bear on the additive treatment of relevance and nonredundancy \citep{carbonellgoldstein1998mmr,ziegler2005diversification}. The holdout comparison provides no evidence that distinctiveness or its fit interaction adds ranking information beyond a relevance score: every contrast is within its bootstrap interval of zero. That null is informative about magnitude rather than about existence. The within-set associations are real but small next to relevance and position, so at the margin of an already relevance-ordered list they do not reorder held-out clicks. The exploratory coherence analysis is likewise flat, and click logs cannot say whether the operative process is earlier attention, reduced confusion, easier comparison, or a clearer reason for choice.

\subsection{Implications for platforms and practitioners}
\label{subsec:managerial-implications}

The results locate where visual differentiation matters, and the location is specific enough to act on. Distinctiveness earns most where four conditions meet: the product already answers the query, since the within-product contrast rises from 0.2\% of a random-choice click at the 10th fit percentile to 3.3\% at the 90th; the set's descriptions fail to separate the alternatives, since the slope roughly doubles in textually tight sets; the product is actually seen, since the association is about six times larger in the top five recorded positions than in the full panel; and the returned set is small enough that individual products still carry stakes, since both the slope and its fit interaction shrink as sets grow. And what distinctiveness wins under these conditions is attention rather than demand: the funnel evidence says the recruited clicks convert proportionally less often and the within-set downstream total stays flat, so every rule below is a rule about which product wins the click, not about how many clicks or purchases the page produces.

For a platform this implies a constrained form of visual diversification aimed at near-duplication rather than a uniform diversity bonus. A ranker can preserve a region of strong relevance and reduce visual redundancy inside it, penalizing candidates whose near twin already occupies a slot, letting the weight on visual separation rise with a candidate's standing on the query, and targeting the queries whose returned descriptions sit closest together, which a platform can identify from its own text representations without any of the measurement in this paper. The cheapest version of the rule needs no reranking at all: when a near-twin pair is returned, place the better-fitting member first. The duel estimates say the higher slot decides 63.9\% of near-twin pairs while relative fit alone decides 55.8\%, and the incumbent ordering placed the better-fitting twin below its partner in 44\% of those pairs, where it went on to win the click only 44\% of the time. Those two descriptive facts are the whole case for the rule, and they do not depend on any coefficient. Web Appendix~\ref{app:counterfactual} adds the arithmetic of exchanging the two slots, worth about nineteen percentage points to the better-fitting member in the pairs the rule would touch, but that figure is an upper bound rather than a forecast: the platform chose the original order using signals the release does not contain, and any of them that predict clicks are absorbed into the estimated value of a slot. The set-level evidence disciplines expectations about what such a rule buys the platform as a whole. If redundancy mainly reallocates clicks, deduplication changes which seller wins and little else, and the share model in Web Appendix~\ref{app:counterfactual} makes the accounting explicit: the clicks a separated pair gains are the clicks the rest of the returned set loses, roughly eight per thousand twin-containing searches in each direction. The estimated coefficients are diagnostics for such a rule, not deployment parameters: position and candidate composition arise from the incumbent system, and the public data do not identify which returned products were visible. A prospective A/B test should randomize a pre-specified reranking within narrow relevance bands while holding the candidate pool fixed and should measure clicks, purchases, reformulation, return to search, abandonment, and latency together.

For practitioners, the relevant visual benchmark is query-specific and pairwise, so the audit unit is not the image but the pair of a high-volume query and the adjacent competitor it returns. An image is redundant for one search and distinctive for another because retrieval changes its neighbors, which is why a catalog-wide uniqueness score misses the operative comparison: the question to ask of each traffic-driving query is whether it returns the product beside a near twin, and the most actionable cases are strongly fitting products whose presentation is nearly interchangeable with one specific competitor's in exactly those environments. The same evidence says where restyling is wasted: on products with weak standing on their core queries, in sets whose descriptions already separate the alternatives, and at positions the shopper never reaches. Because the payoff is attention, the right success metric is the share of the pair's clicks against the adjacent substitute, and a mechanical dip in measured conversion per click should be expected rather than read as failure, since the recruited clicks are marginal ones.

The counterfactual simulations in Appendix~\ref{app:counterfactual} demonstrate the magnitude of the restyling returns. Separating a pair just far enough to leave the \emph{near-twin} range is worth about 2 clicks per thousand impressions of those products, and a larger move to the median distance is worth about 10. The spread of simulation outcomes is also informative. The same move is worth about 3 times as much for products in the top third of relative fit as in the bottom third, and about 2.5 times as much in sets whose descriptions sit close together in comparison to sets where they spread apart. Those two gradients are the premium and the channel result restated in units a merchandiser can act on, and together they say the restyling budget belongs to products that already answer their queries, in categories where descriptions are not enough to tell products apart. A seller who cannot restyle is left with the position lever, which the duel says is the stronger one. Sellers rarely control the returned set, competitor images change, and extreme styling can reduce fit, or create expectations the product does not meet, so controlled changes to crop, background, composition, or view angle should be evaluated within high-volume query environments and against purchase and post-purchase outcomes, since a click alone does not establish durable value.

\subsection{Limitations and future research}
\label{subsec:limitations}

Identification is the primary limitation. Search-event and product effects remove conditions shared within a set and stable product attributes, position effects absorb the average rank gradient, and product-by-week effects provide a shorter-run comparison, but the platform still selects candidates and positions with signals absent from the release, and consumers can respond to query-specific attributes that the semantic representation misses. The estimates are conditional associations within realized returned sets, not causal effects of changing an image or its competitors. Exposure and outcome measurement form a second boundary. Returned products need not enter the viewport or the consideration set; top-position restrictions reduce this concern without observing exposure \citep{abaluckadamsprassl2021consider,ursu2018rankings}. A click can express information gathering, comparison, curiosity, or purchase intent; the downstream check and the exploratory click-quality, latency, relevant-peer, and heterogeneity probes in Web Appendix~\ref{app:mechanism} leave the process open, and the null on display separation between a product and its nearest substitute counts against the simplest attentional reading, and offline ranking metrics inherit the historical display policy. The representations impose a third constraint. Raw queries, descriptions, and images are unavailable, so the visual attributes that cosine distance encodes cannot be inspected; the released query and description vectors required a retrieval-estimated linear alignment before their cosine carried goal information (Appendix~\ref{app:fit-validation}), so fit is measured through the platform's own query understanding; image vectors may mix product design, color, background, crop, and photographic style, and hashed categories may be too coarse or too fine for some comparisons. Same-category distance, alternative neighborhoods, textual controls, and the representation placebo diagnose these concerns without replacing perceptual validation.

A specific limit applies to the set-level analysis. A search without a click is not a failed search: consumers abandon, get interrupted, or find what they needed on the results page. Retrieval is not randomly assigned either, so a query returning near-duplicates may differ from one that does not in ways the compatibility and composition controls do not capture. The set-level estimates therefore speak to whether the total number of clicks is fixed, not to what a deduplicating ranker would deliver.

The most informative next study would manipulate the competitive environment while holding the query and focal product constant, varying the similarity of the nearest peer and the composition of the remaining set orthogonally to fit and recording fixation, time to first attention, decision time, perceived similarity, comparison difficulty, choice, and confidence. A field experiment could randomize order within tight relevance bands and compare an additive diversification rule with the goal-consistent rule on purchases, reformulations, abandonment, returns, and longer-run satisfaction. Further measurement work can connect embedding distance to interpretable image attributes and human similarity judgments, separate product design from photography, and test whether nearest-substitute contrast aids early screening, detailed comparison, or both. Cross-platform evidence would show whether the relations generalize across retailers, categories, interfaces, devices, and cultures.

Product search creates a dual comparison, against the query and against the products returned beside it. The framework's principle is that the second comparison is local and the first gives it meaning: a product gains when the current set leaves it without a close visual substitute, and it gains most when it is also among the best answers to the question the consumer asked. The results establish the first of these and give the second its precise form: the same product is chosen more often when its returned set contains no close look-alike, most of all where the set's descriptions cannot be told apart, and the advantage grows with the product's standing on the query in probability units while vanishing in log-odds, the signature of a mechanism that works by letting merits act rather than by adding to them. Within the boundary of conditional clicks under the historical retrieval and display policy, the claim the evidence supports is more specific than what the literature currently records: the reference against which visual difference matters is not a category norm or a catalog-wide uniqueness score but the competitive neighborhood of the page itself, at its sharpest the single most similar product returned beside the focal one, and what visual difference buys there is being seen as a distinct option.

%% file: exhibits/main_exhibits.tex

\input{exhibits/blocks/table1_descriptives}
\input{exhibits/blocks/figure2_model_free}
\input{exhibits/blocks/figure3_identifying_variation}
\input{exhibits/blocks/table2_main_models}
\input{exhibits/blocks/figure4_marginal_effect}
\input{exhibits/blocks/figure5_boundaries}
\input{exhibits/blocks/table4_deferral}
\input{exhibits/blocks/table3_robustness}
\input{exhibits/blocks/figure6_specification_curve}
\input{exhibits/blocks/table5_ranking}

%% file: appendices/web_appendix.tex

\section{Data construction and audit}
\label{app:data-audit}

This appendix records the construction decisions that determine the empirical sample. The guiding rule is event integrity: each eligible search retains the exact product array the platform returned, and every exclusion triggered by a candidate-level defect is applied to the whole search event, so that the realized choice environment used to compute differentiation and coherence is never modified by the researcher.

\subsection{Source files and observation keys}
\label{app:source-files}

The release contains separate search, catalog, and browsing tables. Table~\ref{tab:app-source-contract} lists the fields each analysis uses. The raw files are kept immutable; all derived data are written elsewhere, and the replication archive records file size, cryptographic digest, row count, and parsing-error count for every source.

\begin{table}[htbp]
\centering
\caption{Raw-data contract}
\label{tab:app-source-contract}
\small
\begin{tabularx}{\textwidth}{@{}p{0.19\textwidth}p{0.17\textwidth}YY@{}}
\toprule
Source & Raw unit & Required fields & Analytic role \\
\midrule
Search table & Search event & Session hash, shifted timestamp, query vector, returned-product array, clicked-product array & Returned-set reconstruction, click outcome, semantic fit, recorded position \\
Catalog table & Product & Product hash, category hierarchy, price decile, description vector, image vector & Product representations, category coherence, textual differentiation, relative price \\
Browsing table & Session event & Session hash, shifted timestamp, event type, product action, product hash & Session sequence, prior-exposure indicators, downstream action check, decision latency \\
\bottomrule
\end{tabularx}
\end{table}

Each search receives a deterministic identifier built from the source-file version and physical row number, which resolves sessions containing two searches with the same shifted timestamp. The product-search key combines that identifier with the returned product hash and must be unique after expansion. The returned-product field is parsed as an ordered array; parsing preserves the recorded sequence and never sorts or deduplicates. A repeated product within an array triggers event-level exclusion because it would assign two positions to one product and change the neighborhood. The clicked-product field is parsed independently, and a search passes the click-integrity check when every clicked product occurs among its returned candidates. The primary sample then requires exactly one distinct clicked product.

\subsection{Position verification}
\label{app:position-verification}

\begin{figure}[htbp]
\centering
\IfFileExists{exhibits/generated/figureA1_position_gradient.pdf}{%
  \includegraphics[width=0.85\textwidth]{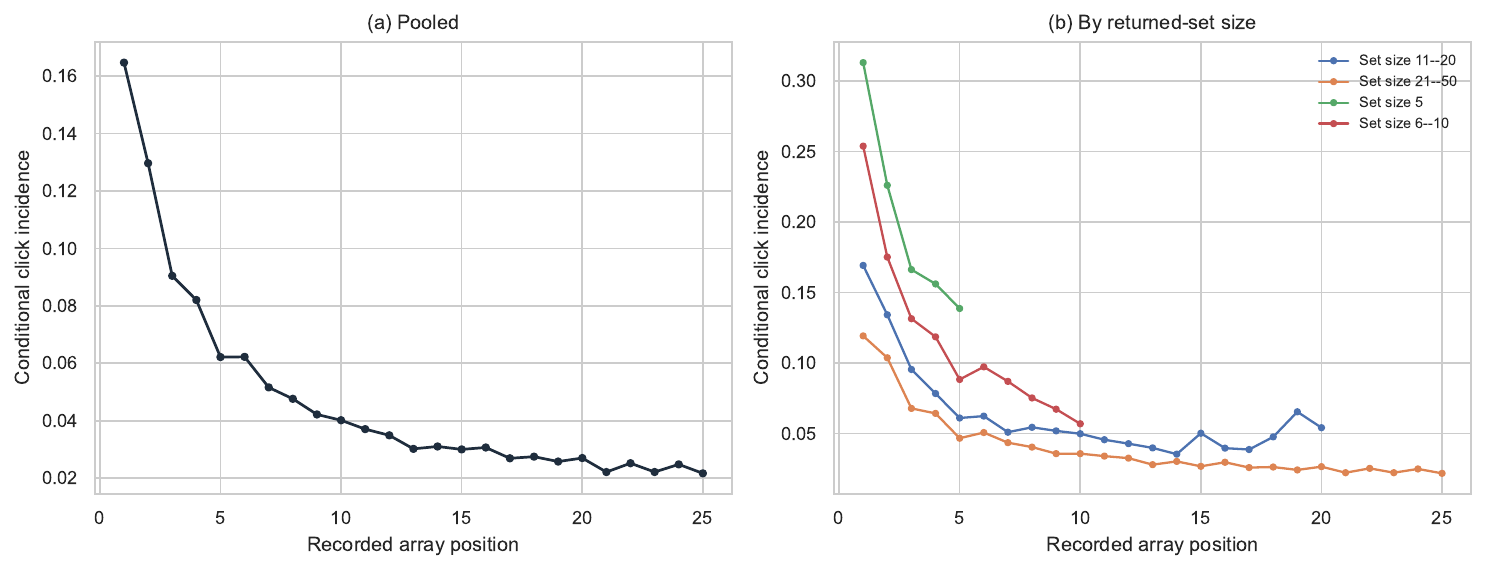}
}{%
  \ExhibitPlaceholder{2.8in}{Analysis output pending. Plot conditional click incidence by recorded position for the pooled sample and separately for returned-set-size strata (for example 5 to 10, 11 to 20, 21 or more), with 95 percent intervals and the random-choice benchmark for each stratum.}
}
\caption{Conditional click incidence by recorded position, pooled and by returned-set-size stratum.}
\label{fig:app-position-gradient}
\end{figure}

The recorded sequence supports position adjustment only if it is the order returned to the client. The dataset documentation describes the field as the products in the search response without stating that the array is ranked, so the audit assembles what the release does establish and bounds what it does not. The steep decline of click incidence with array position (Figure~\ref{fig:app-position-gradient}) is a consistency check, not proof, because other array-generating processes could correlate with relevance. Table~\ref{tab:app-position-audit} records the evidence and its conclusion. Because ordering rests on the documented meaning of the field and the gradient rather than on an explicit ranking statement, the design carries specifications whose conclusions do not depend on it: the no-position-controls row and the top-position restrictions of Table~\ref{tab:robustness}, and the placebo, which permutes images while holding whatever the array records fixed.

\begin{table}[htbp]
\centering
\caption{Returned-array position audit}
\label{tab:app-position-audit}
\small
\begin{tabularx}{\textwidth}{@{}p{0.23\textwidth}Yp{0.28\textwidth}@{}}
\toprule
Audit item & Evidence & Status \\
\midrule
Dataset documentation & Release README describes \path{product_skus_hash} as the hashed identifiers of the products in the search response; no explicit ordering statement & Recorded; ordering not stated \\
Parsing & The loader preserves the recorded sequence and never sorts or deduplicates (Appendix~\ref{app:data-audit}) & Verified by construction \\
Position-specific click incidence & Conditional click incidence 0.165 at position 1, declining monotonically to 0.040 at position 10 (Figure~\ref{fig:app-position-gradient}), in every set-size stratum & Consistent with a ranked list \\
Author confirmation & Dated correspondence or issue response & Not obtained; not required given the bounding specifications \\
Final conclusion & Ordered response, unordered set, or unresolved & Treated as the ordered response; consequence of mis-ordering bounded by the no-position and top-\(k\) rows \\
\bottomrule
\end{tabularx}
\end{table}

\subsection{Sequential eligibility rules}
\label{app:eligibility}

Sample construction proceeds in a fixed order so that every attrition count is reproducible (Table~\ref{tab:app-attrition}). The pipeline first requires a parseable, nonempty returned array with distinct product identifiers; then a valid query representation, a single distinct clicked product contained in the returned array, and at least five returned products; then a catalog join in which every returned product has one unambiguous catalog row, valid description and image representations, and a usable top-level category. Suppose one candidate lacked an image representation: dropping that candidate would change the average distance of every remaining product and replace the platform-returned set with a researcher-modified one, so the whole event is removed instead. Zero-click searches are retained in an audit population that characterizes deferral and representation coverage but contribute no selected alternative to the conditional model. Multiclick searches enter the fractional-outcome analysis (Appendix~\ref{app:multiclick}). Searches with a clicked product outside the returned array are integrity failures and are excluded from all item-level analyses.

\IfFileExists{exhibits/generated/tableA1_attrition.tex}{%
  \input{exhibits/generated/tableA1_attrition}
}{%
\begin{table}[htbp]
\centering
\caption{Sequential sample construction and comparison of retained with excluded searches}
\label{tab:app-attrition}
\small
\begin{threeparttable}
\begin{tabularx}{\textwidth}{@{}Yrrrr@{}}
\toprule
& Searches & Product-search rows & Products & Sessions \\
\midrule
Raw search records & \TBD & \TBD & \TBD & \TBD \\
Parseable, nonempty, duplicate-free returned arrays & \TBD & \TBD & \TBD & \TBD \\
Valid query representation & \TBD & \TBD & \TBD & \TBD \\
Exactly one returned product clicked & \TBD & \TBD & \TBD & \TBD \\
At least five candidates & \TBD & \TBD & \TBD & \TBD \\
Complete catalog and representation data & \TBD & \TBD & \TBD & \TBD \\
Final fixed-effects sample & \TBD & \TBD & \TBD & \TBD \\
\midrule
\multicolumn{5}{@{}l}{\textit{Retained versus excluded searches}}\\
& Retained & Excluded & Difference & \\
Returned-set size & \TBD & \TBD & \TBD & \\
Number of clicks & \TBD & \TBD & \TBD & \\
Search order within session & \TBD & \TBD & \TBD & \\
Shifted week & \TBD & \TBD & \TBD & \\
Category-homogeneous share & \TBD & \TBD & \TBD & \\
\bottomrule
\end{tabularx}
\begin{tablenotes}[flushleft]
\footnotesize
\item Notes: Counts at each step are cumulative. Excluded searches in the lower panel are those with at least one click that fail a later eligibility rule.
\end{tablenotes}
\end{threeparttable}
\end{table}
}

\subsection{Representation integrity}
\label{app:representation-integrity}

Representation checks precede normalization. Each vector family has a single documented dimension; every vector is parsed into finite numeric values, and vectors with a nonfinite element or zero Euclidean norm are invalid. The catalog table must map each product hash to one description vector and one image vector; exact duplicate rows are collapsed after verification, and conflicting rows exclude the product and every returned set containing it. Valid vectors are normalized once and stored in a product-indexed matrix; the audit verifies unit norm within tolerance, and a hand-computed fixture plus a random production sample verify the cosine implementation. Representation coverage can select the sample, so Table~\ref{tab:app-attrition} compares retained and excluded searches on shifted time, set size, click multiplicity, session search order, and category coverage.

\section{Measurement construction and validation}
\label{app:measurement}

\subsection{Efficient distance computation and alternative neighborhoods}
\label{app:efficient-differentiation}

Nearest-substitute distinctiveness (Equation~\ref{eq:nearest-neighbor}) is computed from blockwise similarity products within each returned set, taking the maximum off-diagonal similarity per product; block size is a configuration setting and the computation is exact, not approximate. For the average-form contrast measures, unit-normalized representations admit the identity
\begin{equation}
\sum_{\substack{j\in C_s\\ j\neq i}}v_i^{\top}v_j=v_i^{\top}\Bigl(\sum_{j\in C_s}v_j-v_i\Bigr),
\label{eq:app-distance-identity}
\end{equation}
so \(D^{avg}_{is}\) requires one vector sum per returned set and one dot product per product rather than a pairwise matrix. The production audit compares the identity-based and explicit pairwise computations on returned sets sampled across the set-size distribution and requires a maximum absolute discrepancy below \(10^{-10}\). The same identity yields textual differentiation \(T_{is}\) when description vectors replace image vectors.

Beside the all-peer average of Equation~\ref{eq:visual-differentiation}, two further aggregations are computed. Trimmed-average distance removes the single most distant peer,
\begin{equation}
D^{TR}_{is}=1-\frac{1}{J_s-2}\sum_{j\in C_s\setminus\{i,\,j^{\min}_{is}\}}v_i^{\top}v_j,
\qquad
j^{\min}_{is}=\arg\min_{j\neq i}v_i^{\top}v_j,
\label{eq:app-trimmed-distance}
\end{equation}
which limits the influence of one visually remote candidate; like the all-peer average it is a measurement contrast whose comparison with the nearest-substitute form speaks to how far the operative visual field extends (Section~\ref{subsec:robustness-results} reports the reading). Position-local distinctiveness restricts the peer set to products whose recorded position lies within \(k\) of the focal product's position, with \(k=3\) as a configuration setting; it approximates the products displayed nearest the focal one, probes the reference group rather than the aggregation, and is undefined for products with fewer than two peers in the window.

\subsection{Category coherence and category-restricted neighborhoods}
\label{app:category-coherence}

The top-level category is parsed as a product attribute before expansion; an event is ineligible if any returned product lacks a usable top-level category. Coherence is the count of returned peers sharing the focal category divided by \(J_s-1\) (Equation~\ref{eq:category-coherence}), computed from a returned-set frequency table in time linear in \(J_s\). Three invariants audit the construction: \(G_{is}\in[0,1]\), every row in a homogeneous set equals one, and a focal product with no same-category peer equals zero. The primary moderator uses the second element of the hashed hierarchy and is defined only for returned sets in which every product carries a usable second-level code, so that the peer count is unambiguous; the moderation analysis re-standardizes on that subsample and reports its coverage. The top-level version and the search-level dominant-category share, the largest category frequency in the set divided by \(J_s\), are moderation variants. Same-category distance uses \(M_{is}\) in Equation~\ref{eq:same-category-differentiation} and is retained only when \(|M_{is}|\geq2\); the same-category peer count is stored so that support can be reported separately from estimates.

\subsection{Within-between decomposition, controls, and scaling}
\label{app:within-between}

The product mean \(\bar D_i\) is the arithmetic mean across eligible appearances of product \(i\), each appearance contributing once. The within component has mean zero within product before singleton pruning, and the between component \(D^B_i=\bar D_i-\bar D\) captures persistent visual location; each distance measure exercised in the robustness suite is decomposed the same way on its own sample. Relative price position is the product's price decile minus the mean decile of its returned set; its level is absorbed by the product and search effects, so only its interaction with fit is identified. Prior-exposure indicators equal one when the product was returned, or clicked, in an earlier retained search of the same session. The primary fit measure is the within-set rank of aligned compatibility \(\phi_{is}\) (Equation~\ref{eq:relative-fit}); within-set-centered fit is \(\phi_{is}\) minus the set mean, and the globally standardized level of \(\phi_{is}\) is the measurement-contrast variant.

Fit, both differentiation components, textual differentiation, coherence, and relative price are standardized with the equal-event weights used in the primary regression. For a variable \(X_{is}\), the weighted mean is
\begin{equation}
\bar X_w=\frac{\sum_s J_s^{-1}\sum_{i\in C_s}X_{is}}{\sum_s J_s^{-1}\sum_{i\in C_s}1},
\label{eq:app-weighted-mean}
\end{equation}
with the corresponding weighted standard deviation. Every variable is centered before interactions are formed, and the coherence moderation model (Equation~\ref{eq:coherence-model}) builds every lower-order term from the same standardized components. In the ranking exercise all scaling constants and product means come from the development period.

\subsection{Variance decomposition and support}
\label{app:variance-support}

Product and search effects can explain overlapping variation in \(D_{is}\), so a sequential attribution would depend on entry order. We use a two-order Shapley decomposition. With \(R_P^2\), \(R_S^2\), and \(R_{PS}^2\) the weighted coefficients of determination from the product-only, search-only, and joint projections, the shares are
\begin{align}
\phi_P&=\tfrac{1}{2}\bigl[R_P^2+(R_{PS}^2-R_S^2)\bigr],\nonumber\\
\phi_S&=\tfrac{1}{2}\bigl[R_S^2+(R_{PS}^2-R_P^2)\bigr],\qquad
\phi_U=1-R_{PS}^2,
\label{eq:app-shapley-variance}
\end{align}
which sum to one. The residual share \(\phi_U\), the within-product standard-deviation distribution, and recurrence frequencies (Figure~\ref{fig:identifying-variation}) together establish whether the same image occupies materially different neighborhoods.

Common support for Hypotheses~\ref{hyp:distinctiveness} and~\ref{hyp:premium} is evaluated in the joint distribution of within-search fit rank and within-product distinctiveness; the exploratory coherence moderation adds coherence bins. Cells below the minimum event count are suppressed, and no contrast is reported outside the intersection of supported ranges. Figure~\ref{fig:marginal-effect} spans the central 90 percent of fit, and Figure~\ref{fig:boundaries} shows slopes in low- and high-moderator groups with separate support displays for each panel.

\subsection{Query-description alignment and behavioral validation of semantic fit}
\label{app:fit-validation}

\paragraph{Why alignment is necessary.} Three diagnostics, computed on the first 150{,}000 raw search records before any sample restriction (11{,}785 usable single-click searches with complete catalog coverage; the replication archive includes the script, seed, and output), establish that the released query vectors and description vectors do not share a coordinate system. First, a retrieval check: the mean raw cosine between the query and the products the engine actually returned is \(-0.039\), against \(-0.004\) for random catalog products drawn per search (paired \(t=-20.9\)); the returned products are, if anything, farther from the query than chance. Second, a category check: assigning each query to the nearest top-level-category description centroid recovers the returned set's dominant category in 27.7 percent of searches, below the 50.2 percent majority-class baseline over eight categories. Third, the behavioral check of this section fails for the raw cosine: the clicked-minus-nonclicked gap is \(-0.010\) standard deviations (standard error \(0.006\)), the clicked product sits at the median of its set, and click incidence by within-search decile is flat. The information is nonetheless present: a linear map from query space to description space, estimated from retrieval as in Equation~\ref{eq:query-alignment} on half of these searches, produces mapped queries whose mean cosine with returned products is \(0.490\) against \(0.009\) for random products (\(t=124.9\)) on the half never used to fit the map, with a clicked-minus-nonclicked gap of \(+0.096\) standard deviations and a nearly monotone decile gradient rising to \(1.80\) times the random-choice benchmark.

\paragraph{Estimation of the map.} \(\widehat{W}\) solves the ridge normal equations on all searches with a valid query vector and at least two returned products with catalog coverage whose timestamps lie in the first three quarters of the usable-search timestamp distribution (415{,}158 of 553{,}544 usable searches); the penalty is \(10^{-6}\) per training search and results are insensitive to it. The mean cosine between the mapped query and the returned-set centroid is \(0.575\) in the training window and \(0.549\) in the untouched final quarter, so the map generalizes across time with little decay. The cutoff is written to the replication archive and lies strictly before the chronological holdout period, so no holdout retrieval informs the measurement of any observation. Because the supervision is the returned-set centroid, the map encodes the platform's own query understanding; it uses no click and no image information, so neither the outcome nor the treatment side of the hypotheses can leak into the fit measure. Own-search influence on \(\widehat{W}\) is of order one in several hundred thousand and affects primarily the set-level component of fit, which the search fixed effects absorb.

\paragraph{Behavioral validation.} The clicked-versus-nonclicked compatibility gap is computed within each search,
\begin{equation}
G_s^{\phi}=\phi_{c(s),s}-\frac{1}{J_s-1}\sum_{j\neq c(s)}\phi_{js},
\label{eq:app-fit-gap}
\end{equation}
where \(c(s)\) is the clicked product and \(\phi\) is the aligned compatibility of Equation~\ref{eq:semantic-fit}. We report its mean and median, a session-clustered interval for the mean, the clicked product's within-set percentile, and click incidence by within-search fit decile (Figure~\ref{fig:model-free-evidence}, panel~a). A search-fixed-effects linear probability model with flexible position indicators provides the covariate-adjusted validation coefficient. This establishes behavioral convergence between the supplied representation and selection; it cannot reveal which semantic attributes drive the similarity score. The rank transform of Equation~\ref{eq:relative-fit} inherits this validation because it is a monotone within-set function of \(\phi\).

\section{Estimation and inference details}
\label{app:estimation}

\subsection{Equal-event weighting}
\label{app:event-weighting}

The primary linear model solves
\begin{equation}
\min_{\beta,\alpha,\mu,\lambda}\sum_s\frac{1}{J_s}\sum_{i\in C_s}\bigl(Y_{is}-X_{is}^{\top}\beta-\alpha_s-\mu_i-\lambda_{r(is)}\bigr)^2 ,
\label{eq:app-weighted-objective}
\end{equation}
so that each search has total weight one and the typical search occasion, rather than the typical returned row, drives the estimates. The unweighted model targets the row-weighted population and is a sensitivity analysis. Neither model uses negative sampling.

\subsection{Coherence moderation model and contrasts}
\label{app:coherence-estimation}

The exploratory moderation regressor matrix contains standardized fit, within-product distinctiveness, coherence, their three pairwise interactions and triple interaction, and fit by between-product distinctiveness, between-product distinctiveness by coherence, and their triple interaction, as in Equation~\ref{eq:coherence-model}; no lower-order term is dropped after inspection. At coherence \(g\) the estimated premium is
\begin{equation}
\widehat\theta_2(g)=\widehat\beta_{FW}+\widehat\beta_{FWG}\,g,
\label{eq:app-coherence-cross-partial}
\end{equation}
and at fit \(f\) and coherence \(g\) the marginal association of within-product distinctiveness is
\begin{equation}
\widehat m_D(f,g)=\widehat\beta_W+\widehat\beta_{FW}f+\widehat\beta_{WG}g+\widehat\beta_{FWG}fg .
\label{eq:app-coherence-marginal}
\end{equation}
The model is estimated on the subsample with defined second-level coherence and reported in Table~\ref{tab:app-coherence}; the main text carries only its summary. We report \(\widehat\theta_2(g)\) at the 25th, 50th, and 75th coherence percentiles and \(\widehat m_D(f,g)\) at the 10th, 50th, and 90th fit percentiles; interquartile contrasts multiply these slopes by the supported distinctiveness range. Delta-method standard errors use the full covariance matrix, and session-block bootstrap intervals check the plotted contrasts. The triple interaction is reported two-sided and unadjusted, consistent with its exploratory status in Section~\ref{subsec:hypothesis}.

\IfFileExists{exhibits/generated/tableA4_coherence.tex}{%
  \input{exhibits/generated/tableA4_coherence}
}{%
\begin{table}[htbp]
\centering
\caption{Exploratory category-coherence moderation}
\label{tab:app-coherence}
\small
\begin{threeparttable}
\begin{tabularx}{\textwidth}{@{}Yrr@{}}
\toprule
& Estimate & SE \\
\midrule
Relative fit & \TBD & \TBD \\
Within-product distinctiveness & \TBD & \TBD \\
Category coherence & \TBD & \TBD \\
Fit $\times$ within-product distinctiveness & \TBD & \TBD \\
Fit $\times$ coherence & \TBD & \TBD \\
Within distinctiveness $\times$ coherence & \TBD & \TBD \\
Fit $\times$ within distinctiveness $\times$ coherence & \TBD & \TBD \\
Fit $\times$ between distinctiveness & \TBD & \TBD \\
Between distinctiveness $\times$ coherence & \TBD & \TBD \\
Fit $\times$ between distinctiveness $\times$ coherence & \TBD & \TBD \\
\midrule
Observations / search events & \multicolumn{2}{r}{\TBD} \\
\bottomrule
\end{tabularx}
\begin{tablenotes}[flushleft]
\footnotesize
\item Notes: Estimated on the subsample with defined second-level category coherence, with search, product, and recorded-position effects, equal-event weights, and two-way session and product clustering. Reported two-sided and unadjusted; the moderation is exploratory and its direction was not predicted.
\end{tablenotes}
\end{threeparttable}
\end{table}
}

\subsection{Fixed-effect support and connectedness}
\label{app:connectedness}

The search-product incidence structure is a bipartite graph in which each returned row is an edge. We report the number of connected components, the row share of the largest, product degree, and component-specific support for the interaction. Estimation routines that absorb search and product effects prune singleton groups iteratively, and the reported sample for each specification is the post-pruning sample for that fixed-effect structure. Position indicators are supported only at ranks observed in the estimation sample; rare tail positions are pooled into a terminal bin under a rule fixed before estimation. Product-by-week effects use weeks defined from the shifted timestamps, whose relative order the privacy shift preserves; the specification requires products to recur within a week and therefore uses a smaller sample, which we report alongside the coefficient.

\subsection{Clustered inference}
\label{app:clustered-inference}

All fixed-effects models are estimated with the \texttt{fixest} package \citep{berge2018fixest}, whose alternating-projection demeaning handles the search, product, and position effects; the primary covariance matrix clusters by session and product \citep{camerongelbachmiller2011robust}, permitting arbitrary dependence among searches within a shopping episode and across appearances of a product; search clustering is redundant with session clustering. We report cluster counts on both dimensions and use the finite-sample corrections of the estimation package. Effect magnitudes are reported as probability-point coefficients and interquartile contrasts, the differentiation slope at pre-specified fit percentiles, and the fit value at which the slope crosses zero when that crossing lies within support. Two-sided 95 percent intervals accompany both confirmatory coefficients.

\subsection{Conditional choice representation}
\label{app:conditional-choice}

The conditional logit is estimated as a Poisson pseudo-likelihood with search fixed effects and one selected outcome per search; conditioning on that total yields the multinomial-logit probability in Equation~\ref{eq:conditional-logit}. Product and position effects enter the systematic utility with the terms of the premium model (Equation~\ref{eq:premium-model}), so that expanded rows are not treated as independent count processes. Because latent-utility coefficients are not probability derivatives, we report average predicted-probability contrasts from moving within-product distinctiveness between its supported quartiles at fixed fit percentiles, with delta-method or session-block bootstrap standard errors.

\subsection{Flexible functional form}
\label{app:flexible-form}

The first flexible model replaces standardized fit with within-sample decile indicators and interacts every nonreference bin with within-product distinctiveness; the second uses a restricted cubic spline with knots at fixed empirical quantiles. Coherence versions let the bin-specific or spline slopes differ between low- and high-coherence groups. All retain search, product, and position effects and the same weighting and clustering. Stable evidence for the confirmatory pair is a distinctiveness slope that is positive over most of the supported fit range and becomes more favorable as fit rises; an isolated tail coefficient is weak evidence. The coherence versions describe the exploratory moderation and carry no directional prediction.

\subsection{Coefficient stability and sensitivity to unobserved confounding}
\label{app:sensitivity}

We report the movement of \(\widehat\beta_{W}\) and \(\widehat\beta_{FW}\) as search effects, position effects, product effects, and the rival-mechanism controls are added, together with the change in \(R^2\). The movement of the coefficients as controls are added is the coefficient-stability evidence in the spirit of \citet{oster2019unobservable}; following \citet{cinellihazlett2020sensitivity} we also report the robustness value, the share of residual variance in both the regressor and the outcome that an unobserved confounder would need to explain to eliminate the estimate, computed from the clustered \(t\) statistic and the residual degrees of freedom. Both are reported for the preferred model, the premium model, and the controlled specifications.

\section{Robustness, falsification, and auxiliary outcomes}
\label{app:robustness}

Table~\ref{tab:app-diagnostic-map} maps each threat to the comparison that bears on it. For most rows the framework predicts stability and material attenuation weakens it; the measurement-contrast rows are estimates of how far the operative field extends rather than pass-fail checks. The homogeneous-set and same-category analyses diagnose category mismatch for the confirmatory pair; the continuous coherence interaction is the exploratory moderation.

\begin{table}[htbp]
\centering
\caption{Diagnostic map and interpretation}
\label{tab:app-diagnostic-map}
\footnotesize
\begin{tabularx}{\textwidth}{@{}>{\raggedright\arraybackslash}p{0.19\textwidth}YY@{}}
\toprule
Threat & Principal diagnostic & Interpretation of material attenuation \\
\midrule
Recorded position and incomplete exposure & Flexible position effects; no position controls; top 10; top 5; position by set-size effects & Relation is concentrated in presentation or in low-visibility candidates \\
Representation outlier rather than visual position & Textual differentiation and its fit interactions & Focal product is an outlier in every representation space, not visually in particular \\
Price outlier & Relative price position and its fit interaction & Visual distance proxies for a price position \\
Relational fit form & Within-set-centered fit & Interpretation depends on the rank transform rather than relational standing \\
Aggregation of the constructs & Measurement contrasts: all-peer average, trimmed average, globally standardized fit level & Locates how far the operative field extends; larger broad-aggregate estimates indicate crowding or order-statistic noise rather than refuting locality \\
Coherence moderation (exploratory) & Continuous fit by distinctiveness by coherence interaction; top-level and dominant-share variants & Describes where the premium operates; direction unrestricted \\
Category mismatch & Homogeneous sets; same-category distance & Visual distance mainly marks a different product type \\
Global product uniqueness & Product effects; within-between decomposition; between-product interaction diagnostic & Stable distinctiveness carries the estimates \\
Changing product conditions & Product-by-week effects; recurrence thresholds & Promotions, popularity, inventory, or sparse support contribute materially \\
Reference group & Position-local distinctiveness & Result depends on the display neighborhood rather than the returned set \\
Fit curvature & Fit bins and restricted cubic spline & Linear specification summarizes an omitted nonlinear pattern poorly \\
Session path dependence & First search; single-search sessions; prior-exposure indicators & Reformulation or within-session learning contributes materially \\
Outcome definition & Fractional multiclick outcome; add-to-cart or purchase outcome & Relation is specific to single-click selection \\
Representation geometry & Category-within vector permutation placebo & Similar interactions arise from arbitrary assignments \\
Unobserved query-product appeal & Coefficient stability; robustness value & Modest unobserved selection would eliminate the estimate \\
\bottomrule
\end{tabularx}
\end{table}

\subsection{Representation placebo}
\label{app:placebo}

\begin{figure}[htbp]
\centering
\IfFileExists{exhibits/generated/figureA2_placebo.pdf}{%
  \includegraphics[width=0.9\textwidth]{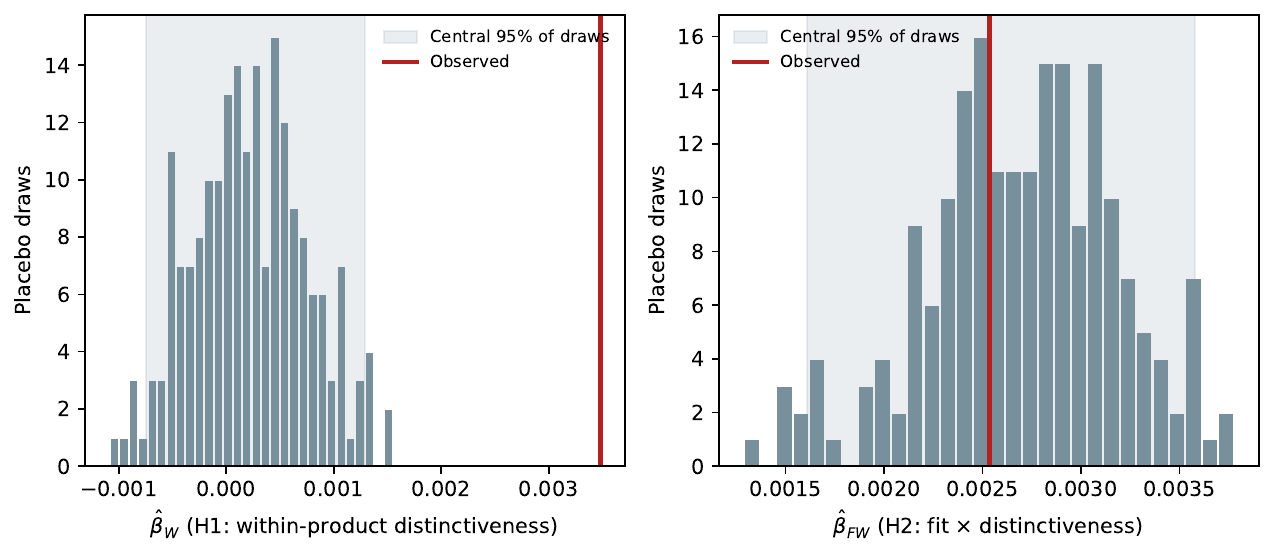}
}{%
  \ExhibitPlaceholder{2.6in}{Analysis output pending. Two histograms of the 200 placebo estimates, one for $\widehat\beta_{W}$ and one for $\widehat\beta_{FW}$, each with the observed estimate marked by a vertical line and the central 95 percent interval shaded.}
}
\caption{Representation placebo distributions for the two confirmatory coefficients. The two panels have different nulls, and the difference is the point. In panel (a) the permuted distribution is centered on zero, so the observed $\widehat\beta_{W}$ lying outside all 200 draws is a falsification pass. In panel (b) the permuted distribution is centered on 0.0027 and excludes zero: because a placebo draw preserves the returned sets and their sizes, permuted nearest-substitute distance still carries set-size structure, and a specification without fit-by-set-size terms loads that structure onto the fit-by-distance interaction whether or not the images mean anything. Panel (b) therefore measures the vulnerability of the unbinned design rather than the content of $\widehat\beta_{FW}$, which Section~\ref{subsec:premium-diagnostics} isolates by identifying the premium within set-size bins.}
\label{fig:app-placebo}
\end{figure}

The placebo permutes the mapping from image vectors to product identifiers within top-level category. Each replication draws one one-to-one mapping and applies it to every appearance of a product, preserving the vector distribution, category composition, returned sets, product recurrence, clicks, queries, and positions while breaking the observed product-image link. For each permutation the pipeline recomputes nearest-substitute distance within every returned set, product means, within-product deviations, and both confirmatory coefficients, the within-product distinctiveness slope and the fit interaction; coherence is unchanged because permutation occurs within category. On the observed mapping the placebo estimator must reproduce the stage-3 confirmatory estimates within a fixed tolerance before any draw is accepted. With \(k\in\{W,FW\}\), observed estimate \(\widehat\beta_k^{(0)}\), draw-\(b\) estimate \(\widehat\beta_k^{(b)}\), and \(B=200\), the one-sided randomization \(p\)-value is
\begin{equation}
p_{RI,k}=\frac{1+\sum_{b=1}^{B}\mathbb{1}\{\widehat\beta_k^{(b)}\geq\widehat\beta_k^{(0)}\}}{B+1}.
\label{eq:app-placebo-p}
\end{equation}
Figure~\ref{fig:app-placebo} reports each distribution with its mean, central 95 percent interval, and the observed estimate marked. The placebo diagnoses generic representation geometry; it does not address omitted query-specific product appeal.

\subsection{Premium and moderator diagnostics}
\label{app:premium-diagnostics}

Two diagnostic tables, generated by \path{analysis/diagnose_premium.py} and \path{analysis/diagnose_contest.py}, support Sections~\ref{subsec:contest-results} and~\ref{subsec:premium-diagnostics}. The first separates the individuation reading of the probability-scale interaction from mechanical set-size structure. The second estimates the channel moderation of Hypothesis~\ref{hyp:contest} with the set's textual separation as moderator, alongside the compatibility-based variants, whose top-gap forms do not inherit the calibration problem of the raw compatibility scale. Both run on the frozen panel without re-running any pipeline stage.

\IfFileExists{exhibits/generated/tableA5_premium_diagnostics.tex}{%
  \input{exhibits/generated/tableA5_premium_diagnostics}
}{}

\subsection{Set-level redundancy sample and controls}
\label{app:deferral}

The set-level sample of Section~\ref{subsec:deferral-model} is the set of searches that pass the structural rules of Appendix~\ref{app:eligibility} without the single-click condition: a parseable returned array with distinct identifiers, a valid query representation, at least five returned products, complete catalog coverage with valid vectors, and a usable top-level category for every returned product. Stage 1 writes one row per such search with the composition of its returned set, so the zero-click searches never enter the choice panel yet remain available here. For each search we store the mean, minimum, and lower quartile of the nearest-substitute distances of its products, the mean average-peer distance, the mean textual distance, the mean and standard deviation of query-product compatibility, the gap between its best and second-best compatibility, the dominant-category share, the modal category, and the share of products whose nearest substitute lies within a fixed cosine similarity of them.

Equation~\ref{eq:deferral-model} absorbs week, returned-set-size, and modal-category effects, so identification comes from sets returned in the same period, of the same size, and in the same product area. The controls are the set's mean compatibility, its compatibility dispersion, its mean textual distance, its dominant-category share, and log set size; the first two matter most, because a query whose retrieval is poor returns both weaker answers and, often, more homogeneous ones. Standard errors cluster by session. Two auxiliary readings accompany the primary coefficient: the average-peer contrast, which asks whether the operative quantity is near-duplication or overall visual variety, and the reformulation outcome, which asks whether redundant sets return the consumer to the query box. A regressor that is constant in the estimation sample is dropped and reported, and fixed-effect factors with a single realized level are omitted rather than allowed to fail.

\subsection{Multiclick outcome}
\label{app:multiclick}

For a search with \(K_s\geq1\) distinct clicked products, the fractional outcome is
\begin{equation}
Y^{MC}_{is}=\frac{\mathbb{1}\{i\text{ is clicked in }s\}}{K_s},\qquad\sum_{i\in C_s}Y^{MC}_{is}=1,
\label{eq:app-multiclick}
\end{equation}
which preserves one unit of outcome mass per search and is compatible with equal-event weighting. The sample includes all otherwise eligible searches with at least one click. We do not use the first recorded clicked product alone because the clicked array may record a set rather than a verified temporal sequence.

\subsection{Downstream action check}
\label{app:downstream}

The browsing table permits an auxiliary outcome when product-specific add-to-cart or purchase events can be linked. A downstream action is attributed to a clicked product when the matching event occurs after the search and before the earlier of the next search in the session or 30 minutes; alternative windows of 10 and 30 minutes are reported, products with matching actions before the focal search are excluded, and searches with multiple prior clicks to the eventual product are omitted. The outcome asks whether clicks associated with goal-consistent differentiation also receive a proximate lower-funnel action. It remains supplementary because sessions contain multiple searches and latent intentions, and it does not create a causal purchase estimand.

\section{Chronological ranking protocol}
\label{app:ranking-protocol}

\subsection{Split construction and leakage prevention}
\label{app:split-construction}

Sessions are ordered by the shifted timestamp of their first search, so that all searches from a session fall on one side of any boundary. The first 80 percent of searches form the development period and the last 20 percent the held-out period; the final fifth of the development period is validation, giving approximate train, validation, and test shares of 64, 16, and 20 percent. Training data determine scaling constants, product priors, product-average differentiation, recurrence status, and tuning parameters; the query-alignment map behind the fit measure is likewise estimated from retrieval strictly before the holdout period (Appendix~\ref{app:fit-validation}); validation data select smoothing and regularization; the held-out period is evaluated once after those choices are fixed. Products first observed in validation or test receive the cold-item prior and never inform training. The product prior is a smoothed training click rate: with \(c_i\) clicks in \(n_i\) returned appearances,
\begin{equation}
\widehat\pi_i=\frac{c_i+a\pi_0}{n_i+a},
\label{eq:app-product-prior}
\end{equation}
where \(\pi_0\) is the training click rate and \(a\) is chosen on validation data. The prior enters the scores as the log-odds of \(\widehat\pi_i\). The score coefficients \(\widehat\lambda\), \(\widehat\beta_F\), \(\widehat\beta_D\), \(\widehat\beta_{FD}\), and the coherence terms are estimated on the development period from a conditional logit with search effects (Poisson representation), position indicators capped at a terminal rank, and the prior as a covariate in place of product effects, using raw distinctiveness; each score is the estimated systematic utility.

\subsection{Ranking metrics and uncertainty}
\label{app:ranking-metrics}

\begin{figure}[htbp]
\centering
\IfFileExists{exhibits/generated/figureA3_ranking_strata.pdf}{%
  \includegraphics[width=0.9\textwidth]{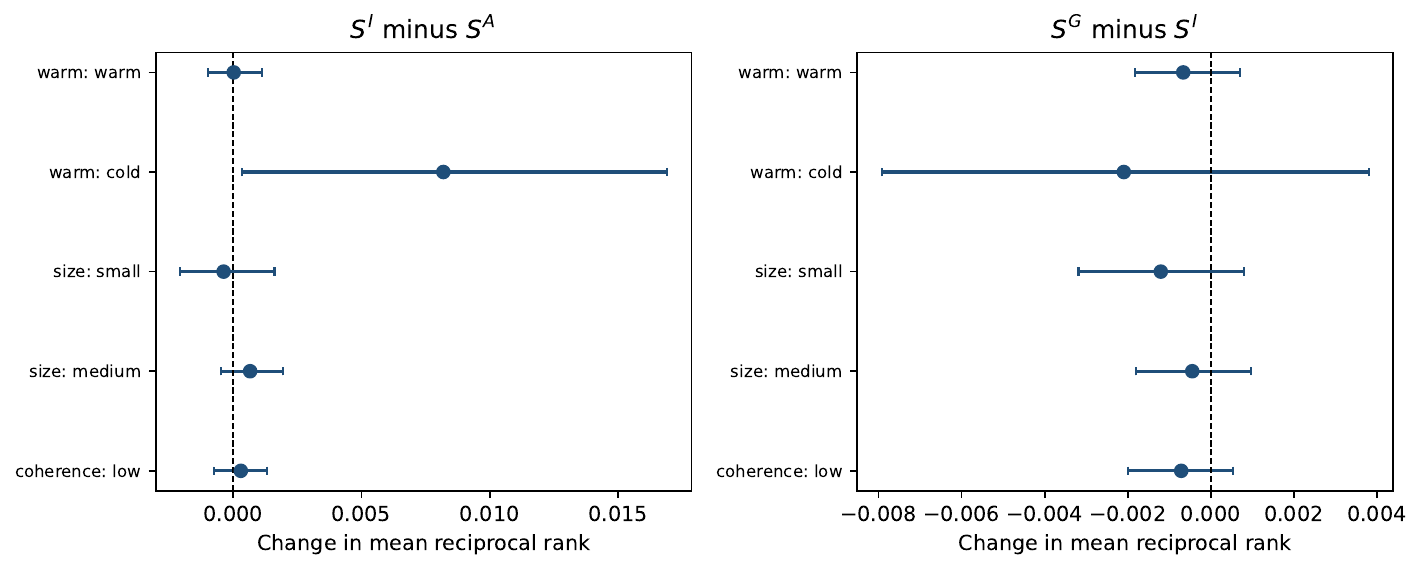}
}{%
  \ExhibitPlaceholder{2.6in}{Analysis output pending. Plot the $S^I$ minus $S^A$ and $S^G$ minus $S^I$ mean-reciprocal-rank contrasts with session-bootstrap intervals for warm versus cold products, returned-set-size terciles, and coherence terciles.}
}
\caption{Holdout ranking contrasts by product history, returned-set size, and category coherence.}
\label{fig:app-ranking-strata}
\end{figure}

For a single-click search, let \(r_s^{(m)}\) be the rank that model \(m\) assigns to the clicked product. Then
\begin{equation}
\operatorname{MRR}^{(m)}=\frac{1}{S}\sum_{s=1}^{S}\frac{1}{r_s^{(m)}},\qquad
\operatorname{Hit@k}^{(m)}=\frac{1}{S}\sum_{s=1}^{S}\mathbb{1}\{r_s^{(m)}\leq k\}.
\label{eq:app-ranking-metrics}
\end{equation}
Ties receive the average rank implied by a deterministic secondary ordering on the product hash, applied identically to every model. The two bootstrapped contrasts are \(S^I\) minus \(S^A\), the increment from the goal-consistency premium, and \(S^G\) minus \(S^I\), the increment from the exploratory coherence terms, each with 1,000 paired session-block bootstrap replications; the \(S^A\) minus \(S^R\) difference, the ranking counterpart of the distinctiveness effect, is read from the metric rows. Figure~\ref{fig:app-ranking-strata} reports the bootstrapped contrasts for warm and cold products and by set-size and coherence strata. These metrics evaluate agreement with choices logged under the historical display policy; a randomized deployment is required to estimate click-through, purchase, or revenue effects under a new ranking.

\section{Mechanism and exploratory evidence}
\label{app:mechanism}

The confirmatory tests establish a conditional association and the robustness suite rules out the rival accounts named in Section~\ref{subsec:robustness}. They do not say how the pattern arises. This section reports four exploratory probes of process. Each is tied to a directional prediction stated before the estimate, none was pre-registered, none is adjusted for multiplicity, and none changes the confirmatory conclusions; they are evidence about mechanism, to be read jointly with the theory in Section~\ref{sec:theory}. Unless stated otherwise every model uses the frozen analytic panel, the stage-3 standardization of Table~\ref{tab:main-models}, equal-event weights, search, product, and recorded-position effects, and two-way session and product clustering. The analysis code writes every estimate to \path{results/production/appendix/mechanism/} with its own checkpoint.

\subsection{Relevant-peer decomposition of visual distance}
\label{app:peer-split}

\begin{figure}[htbp]
\centering
\IfFileExists{exhibits/generated/figureA4_mechanism.pdf}{%
  \includegraphics[width=0.98\textwidth]{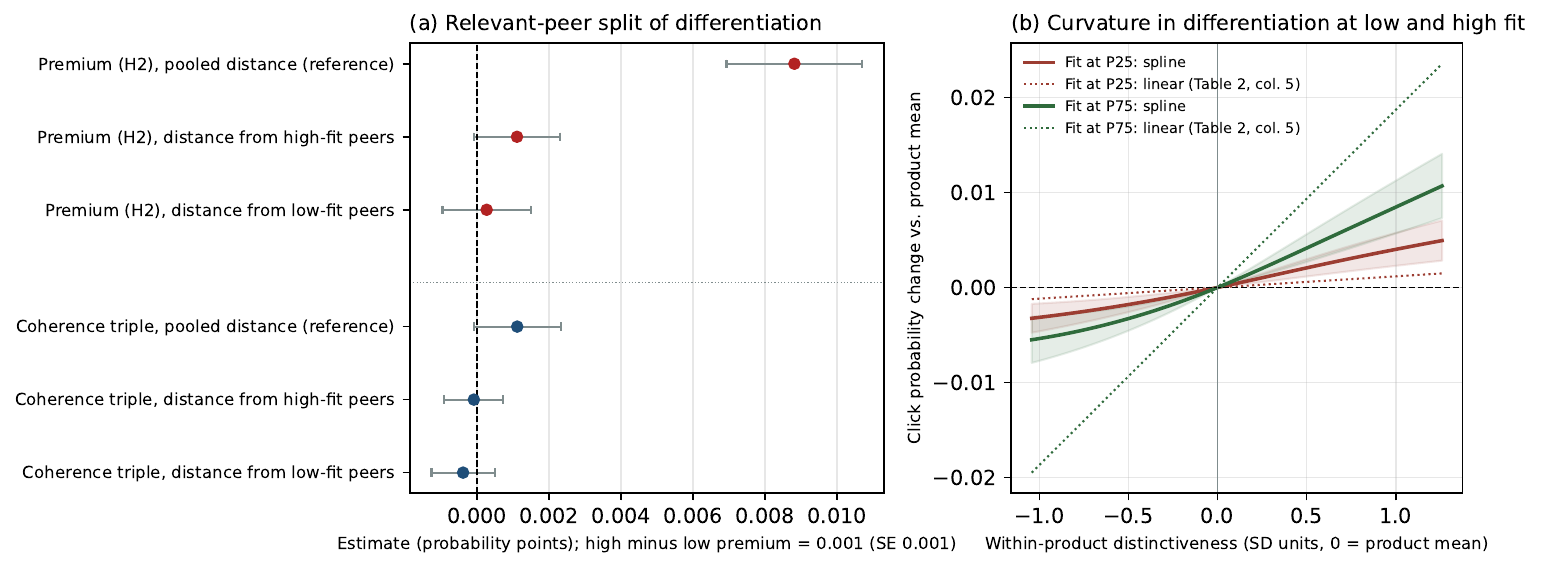}
}{%
  \ExhibitPlaceholder{2.8in}{Analysis output pending. Panel (a): dot-and-whisker estimates of the premium coefficient and the coherence triple for pooled distance, distance from high-fit peers, and distance from low-fit peers on the complete-set sample. Panel (b): restricted-cubic-spline change in click probability over within-product distinctiveness at the 25th and 75th fit percentiles, with the linear estimate from column 4 of Table~\ref{tab:main-models} dotted for reference.}
}
\caption{Mechanism probes. Panel (a) splits visual distance into distance from the high-fit and the low-fit halves of the returned set (Appendix~\ref{app:peer-split}). Panel (b) plots the change in click probability relative to the product mean over within-product distinctiveness at low and high fit from a restricted cubic spline, with the linear model dotted (Appendix~\ref{app:differentiation-curvature}). Whiskers and bands are 95 percent intervals with two-way session and product clustering.}
\label{fig:app-mechanism}
\end{figure}

The theory treats distinctiveness as separation among the alternatives the consumer is weighing, which raises a question the nearest-substitute measure does not answer by itself: does separation from the plausible answers matter more than separation from the implausible ones? This probe splits the peer group by fit and measures average distance from each half. Split each returned set at its median compatibility (ties to the upper half) and define
\begin{equation}
D^{H}_{is}=1-\frac{1}{|H_s\setminus\{i\}|}\sum_{j\in H_s\setminus\{i\}}\cos(v_i,v_j),\qquad
D^{L}_{is}=1-\frac{1}{|L_s\setminus\{i\}|}\sum_{j\in L_s\setminus\{i\}}\cos(v_i,v_j),
\label{eq:app-peer-split}
\end{equation}
where \(H_s\) and \(L_s\) are the upper and lower fit halves and \(v\) denotes the normalized image vector. A component is defined when the product has at least two peers in that half, so both are defined for every returned product once \(J_s\geq7\); the sample is the complete sets on which both components exist, reported with its minimum set size and coverage. Each component is decomposed within and between products and standardized on that sample exactly as in Appendix~\ref{app:within-between}, and the premium model (Equation~\ref{eq:premium-model}) is re-estimated on the same sample as the pooled reference. The split model replaces the single within-product term with both components:
\begin{equation}
\begin{split}
Y_{is}={}&\beta_F F_{is}+\beta_{HW}D^{HW}_{is}+\beta_{LW}D^{LW}_{is}+\beta_{FHW}F_{is}D^{HW}_{is}+\beta_{FLW}F_{is}D^{LW}_{is}\\
&+\beta_{FHB}F_{is}D^{HB}_{i}+\beta_{FLB}F_{is}D^{LB}_{i}+\alpha_s+\gamma_i+\delta_{p(i,s)}+\varepsilon_{is},
\end{split}
\label{eq:app-peer-split-model}
\end{equation}
with the coherence model extended in the same way for the triple interactions. A consideration-set contrast predicts \(\beta_{FHW}>\beta_{FLW}\): standing apart from the relevant competitors is what carries the premium, while distance from irrelevant products mainly marks a different product type. A general salience account predicts no difference. Figure~\ref{fig:app-mechanism}(a) reports the premium coefficient and the coherence triple for the pooled reference and for each component, and the difference \(\widehat\beta_{FHW}-\widehat\beta_{FLW}\) with its delta-method standard error. The two within-product components are correlated because both move with the focal image, so the contrast is informative about relative loading rather than a clean decomposition of variance.

\subsection{Attention and familiarity heterogeneity}
\label{app:heterogeneity}

\begin{figure}[htbp]
\centering
\IfFileExists{exhibits/generated/figureA5_heterogeneity.pdf}{%
  \includegraphics[width=0.98\textwidth]{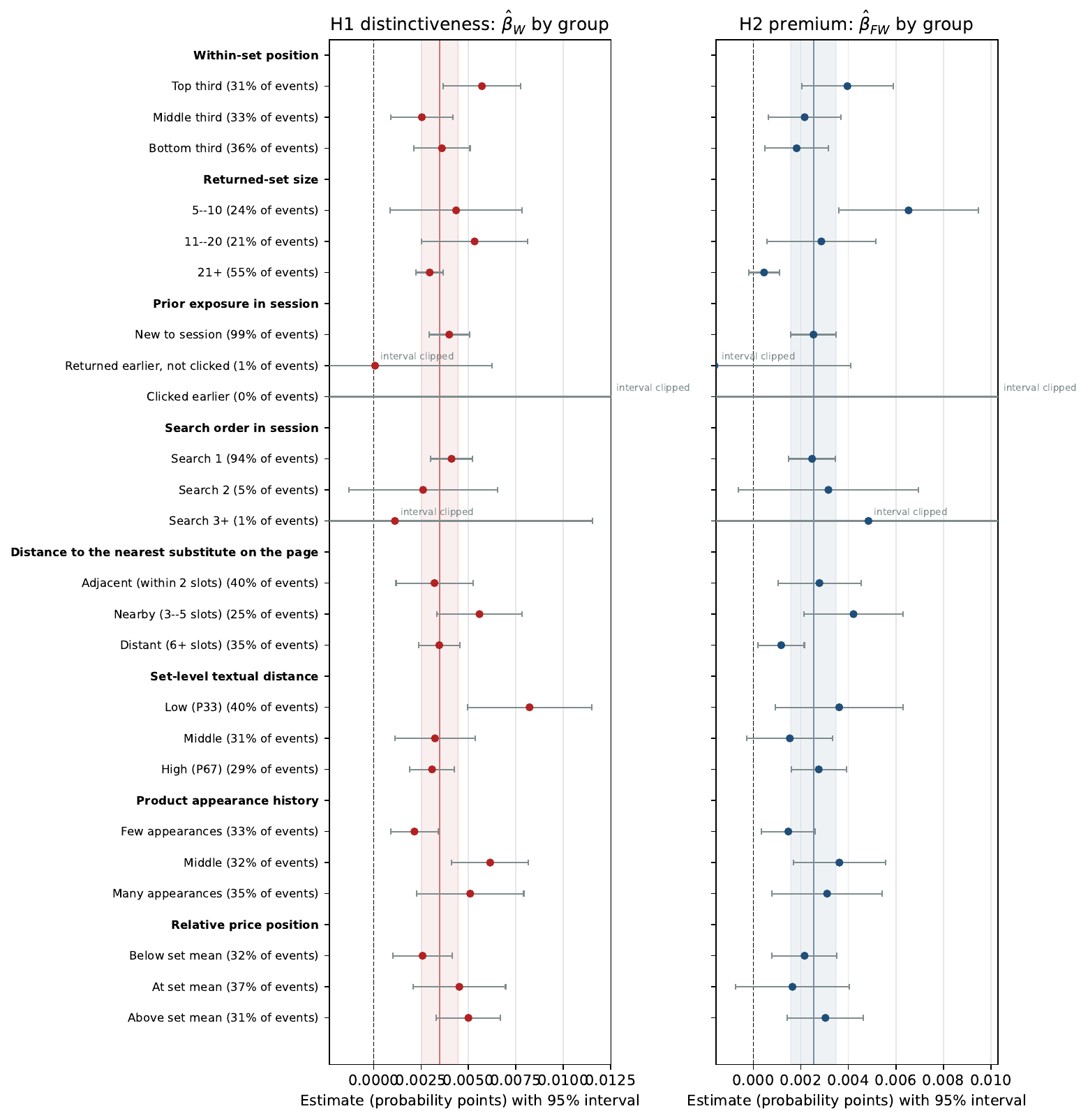}
}{%
  \ExhibitPlaceholder{3.2in}{Analysis output pending. Forest plot of the premium $\widehat\beta_{FW}$ (left) and the exploratory coherence triple $\widehat\beta_{FWG}$ (right) by within-set position tercile, returned-set size, prior exposure in the session, and search order in the session, each from a fully interacted model, with the pooled estimate and its interval shaded for reference.}
}
\caption{The distinctiveness coefficient and the goal-consistency premium by within-set position, returned-set size, prior exposure, search order, display distance to the nearest substitute, set-level textual separation, product appearance history, and relative price position, from fully interacted models. The shaded band is the pooled estimate from Table~\ref{tab:main-models} with its 95 percent interval; whiskers are 95 percent intervals with two-way session and product clustering; percentages are weighted event shares.}
\label{fig:app-heterogeneity}
\end{figure}

If the goal-consistency premium works by drawing inspection to a fitting product during a visual scan, it should be stronger where inspection is scarcer and weaker for products the consumer has already examined. Eight moderators are defined on the frozen panel: within-set position tercile (top, middle, bottom third of the returned array), returned-set size (5 to 10, 11 to 20, 21 or more products), prior exposure in the session (new to the session; returned earlier but not clicked; clicked earlier), search order in the session (first, second, third or later), the display distance between a product and its nearest visual substitute (Appendix~\ref{app:adjacency}), terciles of the set's textual separation \(\bar T_s\), which proxy the channel term of Hypothesis~\ref{hyp:contest} and complement the continuous moderation of Table~\ref{tab:app-contest-diagnostics}, product appearance history, and price position relative to the set mean. Each moderator enters a fully interacted version of the premium model (Equation~\ref{eq:premium-model}), with the group indicators interacting every slope term, so that the group-specific premium is \(\widehat\beta_{FW}+\widehat\beta_{FW,g}\) with a delta-method interval; the coherence model (Equation~\ref{eq:coherence-model}) is interacted in the same way for \(\widehat\beta_{FWG}\). Group main effects are included for the product-level moderators and omitted for the set-level moderators, whose levels are absorbed by the search effects. Position terciles are defined relative to set size and sit alongside the flexible recorded-position effects, so they capture where in the array a product sits rather than absolute rank. An attention-allocation process predicts larger premiums in the bottom terciles and in larger sets; an inference-from-a-cue process predicts attenuation for products returned or clicked earlier in the session; a pure evaluation process predicts no position dependence; the channel account of Hypothesis~\ref{hyp:contest} predicts the largest distinctiveness slope in the tightest-description tercile. Search order has no directional prediction and is reported as a description of how the pattern evolves across reformulations. Figure~\ref{fig:app-heterogeneity} plots the group estimates against the pooled estimate and its interval; weighted event shares are printed with each group, small groups can carry wide intervals, and the four moderators are not independent tests.

\subsection{Display adjacency of the nearest substitute}
\label{app:adjacency}

If distinctiveness matters because a near-identical neighbor interferes with identifying the focal product, the interference should depend on where that neighbor sits. Stage 1 records the recorded position of each product's nearest visual substitute, so the display separation \(\lvert r(i,s)-r(n(i,s),s)\rvert\) is observed for every row, where \(n(i,s)\) is the nearest substitute. Rows are grouped into adjacent (within two slots), nearby (three to five), and distant (six or more) and entered as a moderator in the fully interacted design of Appendix~\ref{app:heterogeneity}. A visual-interference account predicts a larger distinctiveness effect where the look-alike is displayed close by, because that is where the two compete for the same glance; an account in which the returned set is evaluated as a whole predicts no dependence on display separation. The moderator is a property of the platform's layout rather than of the product, and the flexible recorded-position effects remain in the model, so the comparison is between products at similar ranks whose closest look-alike happens to fall nearer or farther on the page.

\subsection{The near-twin duel}
\label{app:twin-duel}

\IfFileExists{exhibits/generated/tableA3_twin_duel.tex}{%
  \input{exhibits/generated/tableA3_twin_duel}
}{%
\begin{table}[htbp]
\centering
\caption{The near-twin duel: which member of a near-identical returned pair is chosen}
\label{tab:app-twin-duel}
\small
\begin{threeparttable}
\begin{tabularx}{\textwidth}{@{}Yccc@{}}
\toprule
& (1) & (2) & (3) \\
\midrule
Compatibility advantage over the twin & \TBD & \TBD & \TBD \\
Display-position advantage over the twin &  & \TBD & \TBD \\
Price advantage over the twin (cheaper) &  &  & \TBD \\
\midrule
Near-twin duels & \TBD & \TBD & \TBD \\
Share won by the better-fitting twin & \multicolumn{3}{c}{\TBD} \\
Share won by the higher-placed twin & \multicolumn{3}{c}{\TBD} \\
\bottomrule
\end{tabularx}
\begin{tablenotes}[flushleft]
\footnotesize
\item Notes: One observation per returned pair whose members are each other's nearest visual substitute, whose distance falls in the lowest decile, and exactly one of whose members was clicked. Advantages are within-pair differences standardized across pairs; the query, session, and remaining returned set are identical within a pair. Standard errors clustered by session. Exploratory.
\end{tablenotes}
\end{threeparttable}
\end{table}
}

The sharpest version of the substitution question uses pairs. A pair qualifies when the two products are each other's nearest visual substitute, their distance falls in the lowest decile of the estimation sample, and exactly one of them was clicked. Within such a pair the query, the session, the returned set, and the retrieval policy are identical by construction, so they cannot explain which member won; what remains are the differences between the twins. We estimate
\begin{equation}
W_{p}=\rho_F\,\Delta F_{p}+\rho_R\,\Delta R_{p}+\rho_P\,\Delta P_{p}+\upsilon_{p},
\label{eq:app-twin-duel}
\end{equation}
where \(W_p\) indicates that the focal member of pair \(p\) received the click, \(\Delta F_p\) is its compatibility advantage over its twin, \(\Delta R_p\) its display-position advantage, and \(\Delta P_p\) its price advantage, each standardized across pairs, with standard errors clustered by session. Table~\ref{tab:app-twin-duel} reports the three nested versions. The exercise asks what breaks a tie once appearance has been held nearly fixed, and it is descriptive: pairs that reach the near-twin threshold are a selected slice of the catalog, and the analysis conditions on exactly one member being clicked.

\subsection{Counterfactual click reallocation}
\label{app:counterfactual}

The estimates in the main text are per-standard-deviation associations. This section converts them into the quantity a strategy discussion needs, the number of clicks that would move under two stylized policies, and sets out the arithmetic that does the converting. Both are computed by \path{analysis/simulate_strategy.py} from the same data used in our analysis. Each is a re-expression of coefficients already reported, so the gradients that appear inside them are the goal-consistency premium and the channel moderation restated in managerial units rather than independent confirmations of either.

\begin{figure}[!ht]
\centering
\includegraphics[width=\textwidth]{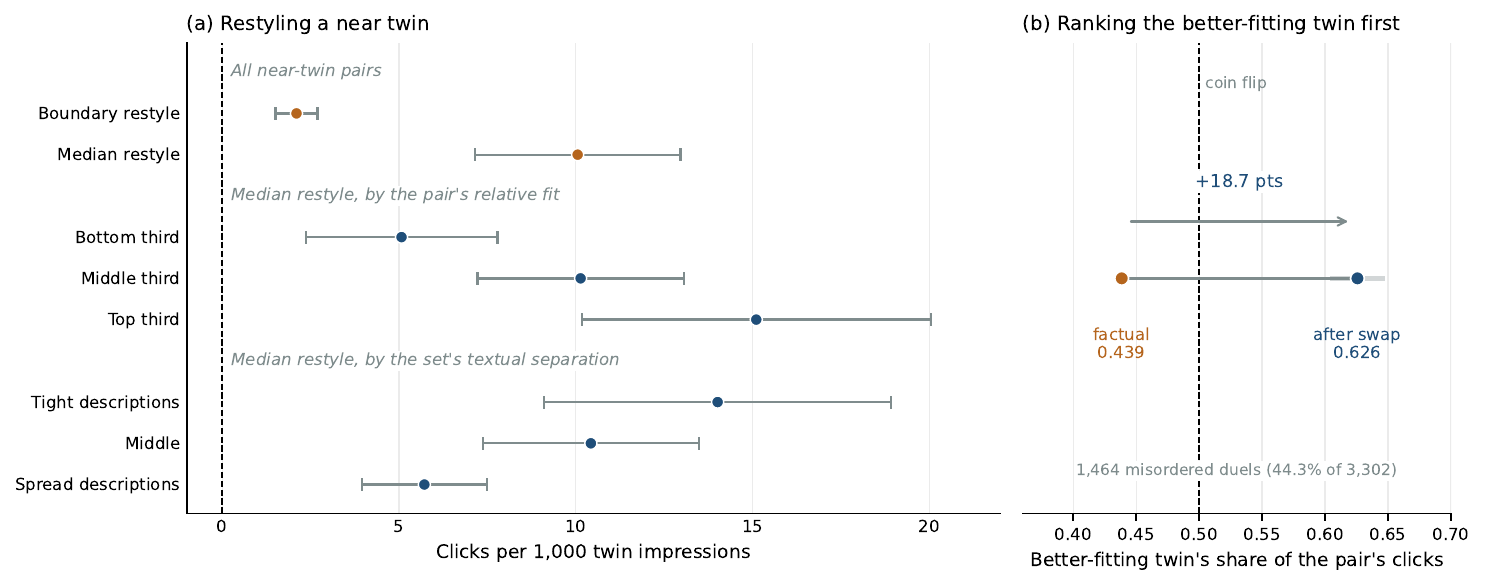}
\caption{Simulated returns to two near-twin policies.}
\label{fig:counterfactual}
\begin{minipage}{\textwidth}
\scriptsize
\emph{Notes.} Panel (a) reports the seller-side policy: the change in expected
clicks per 1{,}000 near-twin impressions when one member of a near-twin pair is
restyled so that the pair's visual distance rises to the near-twin boundary
(first row) or to the median returned pair (all remaining rows). Points are
delta-method means over 200 parameter draws and whiskers are 95\% intervals.
The two middle blocks hold the restyle target fixed at the median and split the
pairs by the restyled product's within-set relative fit and by the textual
separation of the set that returned it, so the gradients report where the same
intervention is worth most. The median restyle moves visual distance by
roughly 1.7 standard deviations, which is outside the range the estimates are
identified on, so those magnitudes are an extrapolation rather than a forecast.
Panel (b) reports the platform-side policy: among returned near-twin pairs
whose incumbent ordering placed the worse-fitting member first, the
better-fitting member's share of the pair's clicks under the observed ordering
and under a tie-break that swaps the two slots. The shaded band is the 95\%
interval. Because position is assigned by the historical ranker rather than at
random, the swap contrast absorbs whatever the ranker knew and is an upper
bound on the causal effect of the slot alone.
\end{minipage}
\end{figure}

\subsubsection*{Policy A: restyling a near twin}

Let \(\mathcal{P}\) collect the returned pairs whose members are each other's nearest visual substitute and whose distance falls in the near-twin decile, and let \(i\) index the two member rows of a pair. A restyle that separates the pair raises both members' nearest-substitute distance to a target \(\tau\), which in the units of the estimation is the shift
\begin{equation}
\Delta_{is}=\frac{\max\{0,\ \tau-D_{is}\}}{\sigma_W},
\label{eq:app-cf-shift}
\end{equation}
where \(\sigma_W\) is the standard deviation used to form \(\widetilde D^W\). Because the product mean \(\bar D_i\) and the standardization constants are held at their estimated values, a movement in \(D_{is}\) passes one for one into \(\widetilde D^W_{is}\). Two targets bracket the exercise: \(\tau\) at the near-twin decile boundary, the smallest move that leaves twin status, and \(\tau\) at the sample median nearest-substitute distance. The stored quantities do not identify each member's second-nearest peer, so the counterfactual cannot cap the move there, and the boundary target is what bounds the change from below.

The shift propagates through the channel-moderation model, Equation~\ref{eq:contest-model} with \(\widetilde S_s\) replaced by standardized textual separation \(\bar T_s\), which is the last row of Table~\ref{tab:app-contest-diagnostics}. Differentiating its conditional mean with respect to \(\widetilde D^W\) gives the row-level change
\begin{equation}
\delta_{is}=\bigl(\beta_W+\beta_{FW}\widetilde F_{is}+\beta_{WS}\bar T_{s}+\beta_{FWS}\widetilde F_{is}\bar T_{s}\bigr)\,\Delta_{is},
\label{eq:app-cf-row}
\end{equation}
so what the restyle is worth depends on the member's standing on the query and on how far apart the set's descriptions already sit. The reported quantity for a group \(\mathcal{G}\) of member rows is the event-weighted mean \(\bar\delta_{\mathcal{G}}=\sum_{i\in\mathcal{G}}w_i\delta_{is}\), with weights normalized to sum to one and the result scaled to clicks per thousand impressions. Because Equation~\ref{eq:app-cf-row} is linear in the coefficients, \(\bar\delta_{\mathcal{G}}\) is a single linear contrast \(c'\beta\) whose nonzero elements are
\begin{equation}
c_{\beta_W}=\sum_i w_i\Delta_i,\quad
c_{\beta_{FW}}=\sum_i w_i\Delta_i\widetilde F_i,\quad
c_{\beta_{WS}}=\sum_i w_i\Delta_i\bar T_s,\quad
c_{\beta_{FWS}}=\sum_i w_i\Delta_i\widetilde F_i\bar T_s,
\label{eq:app-cf-contrast}
\end{equation}
and its standard error is \(\sqrt{c'Vc}\) from the same two-way clustered covariance that produced the coefficients. No simulation noise enters Panel A of Table~\ref{tab:app-counterfactual}: its intervals are delta-method intervals for a linear function of the estimates.

\subsubsection*{Where the gained clicks come from}

Equation~\ref{eq:app-cf-row} says what a restyled pair gains and is silent about what the rest of the page loses, because a linear probability model imposes no adding-up across the returned set. A share model supplies the missing constraint. For each row of a twin-containing search let \(p^0_{is}\) be an empirical baseline, the click rate in the cell defined by recorded position and returned-set-size band over the whole panel. The conditional logit's coefficients update it multiplicatively and renormalization turns the result into within-search shares,
\begin{equation}
p^1_{is}=p^0_{is}\exp\!\bigl\{(\gamma_W+\gamma_{FW}\widetilde F_{is})\Delta_{is}\bigr\},
\qquad
\sigma^k_{is}=\frac{p^k_{is}}{\sum_{j\in C_s}p^k_{js}},\quad k\in\{0,1\}.
\label{eq:app-cf-share}
\end{equation}
Renormalizing holds the set's total clicks fixed, which is the assumption the funnel decomposition of Section~\ref{subsec:robustness-results} supports and the linear model cannot express. The reported figure is \(\sum_{i\in\text{pair}}(\sigma^1_{is}-\sigma^0_{is})\) averaged over twin-containing searches; the change for the rest of the set is its negative by construction, so the exercise reports where the clicks come from rather than assuming it. Uncertainty comes from 200 draws of \((\gamma_W,\gamma_{FW})\) from their estimated sampling distribution, with the whole reallocation recomputed on each draw.

\subsubsection*{Policy B: ranking the better-fitting twin first}

Among the pairs in \(\mathcal{P}\) with exactly one clicked member and a nonzero compatibility gap, take the better-fitting member as focal and let \(W_p\) indicate that it won the pair's click. With \(\Delta R_p\) its display-position advantage in slots, the partner's rank minus its own, \(\Delta P_p\) its price advantage and \(\Delta F_p\) its compatibility advantage, all in raw units,
\begin{equation}
W_p=\rho_R\Delta R_p+\rho_P\Delta P_p+\rho_F\Delta F_p+\upsilon_p .
\label{eq:app-cf-duel}
\end{equation}
The policy exchanges the two members' slots wherever the better-fitting one is placed lower, that is on \(\mathcal{M}=\{p:\Delta R_p<0\}\). The exchange keeps both slots occupied, leaves the rest of the returned set untouched, and holds every other covariate at its historical value, so it maps \(\Delta R_p\) to \(-\Delta R_p\) and changes nothing else. The implied change in the focal member's win probability is \(\rho_R(-2\Delta R_p)\), whose mean over the affected duels has the closed form
\begin{equation}
\bar\Delta_{\mathcal{M}}=2\,\widehat\rho_R\,\overline{|\Delta R_p|}_{\mathcal{M}},
\qquad
\operatorname{se}\bigl(\bar\Delta_{\mathcal{M}}\bigr)=2\,\overline{|\Delta R_p|}_{\mathcal{M}}\,\operatorname{se}\bigl(\widehat\rho_R\bigr).
\label{eq:app-cf-swap}
\end{equation}
The swing is therefore twice the per-slot coefficient times the average slot gap the policy closes, and multiplying by the share of duels in \(\mathcal{M}\) converts it to an effect per thousand duels overall.

\subsubsection*{What the exercise does and does not deliver}

Four limits bound the reading of Table~\ref{tab:app-counterfactual}. First, both policies are reallocations conditional on the search producing a click. They describe which product absorbs a click, never how many clicks or purchases a page generates, and the funnel decomposition is what licenses holding that margin fixed.

Second, the two Panel A targets differ in how far they extrapolate. The boundary move averages 0.38 standard deviations of within-product distinctiveness, comfortably inside the range the coefficients were estimated on. The move to the median distance averages 1.73 standard deviations, a large extrapolation of a linear approximation, and belongs in the table as an illustrative upper end rather than as a forecast.

Third, the calibration model matters and we report the sensitivity rather than burying it. Panel A propagates the channel model, whose distinctiveness coefficient at mean textual separation is 0.0055, larger than the premium model's 0.0040; calibrating on the premium model instead would scale the panel down by roughly a quarter. We use the channel model because the policy questions are precisely about how the gain varies with relative fit and with textual separation, which is the variation only that model carries.

Fourth, and most consequentially, display position in Equation~\ref{eq:app-cf-duel} is not randomly assigned. The platform ordered the two twins using signals absent from the release, and any of those signals that also predict clicks are absorbed into \(\rho_R\). Panel B is accordingly an upper bound on what a tie-break rule would deliver, not an estimate of it. What does not depend on \(\rho_R\) at all is the descriptive half of the panel, and that half is where the managerial argument of Section~\ref{subsec:managerial-implications} rests: in 44.3 percent of near-twin duels the better-fitting member was displayed lower, and in those duels it won the click 43.9 percent of the time.

\IfFileExists{exhibits/generated/tableA8_counterfactual.tex}{%
  \input{exhibits/generated/tableA8_counterfactual}
}{}

\subsection{Click quality and decision latency}
\label{app:click-quality}

A click is a weak outcome if goal-consistent differentiation merely invites curiosity. Two outcomes defined on the clicked product of each retained search address this. The first is the downstream add-to-cart or purchase indicator of Appendix~\ref{app:downstream}, on downstream-eligible searches. The second is decision latency: the seconds from the search event to the first product-detail view of the clicked product in the browsing table, counted only when that view occurs before the next search in the session and within ten minutes; unmatched clicks are excluded and the match rate is reported. Because each search contributes one row, search effects are not available; product, recorded-position, and week effects take their place, regressors keep the stage-3 standardization so that coefficients are in the units of Table~\ref{tab:main-models}, and standard errors are two-way clustered by session and product. An informative-cue account predicts that the within-product distinctiveness term and its fit interaction are positive for conversion and negative for log latency: distinctive, fitting products are recognized faster and the resulting clicks convert at least as well. A curiosity account predicts negative conversion terms. Table~\ref{tab:app-click-quality} reports both outcomes with and without the coherence terms. Latency mixes page loading and deliberation, is censored by the window and by reformulation, and is observed only for clicked products, so it speaks to the relative speed of clicks that happened rather than to the choice process itself.

\IfFileExists{exhibits/generated/tableA2_click_quality.tex}{%
  \input{exhibits/generated/tableA2_click_quality.tex}
}{%
\begin{table}[htbp]
\centering
\caption{Click quality and decision latency among clicked products}
\label{tab:app-click-quality}
\small
\begin{threeparttable}
\begin{tabularx}{\textwidth}{@{}Yrrrr@{}}
\toprule
& \multicolumn{2}{c}{Add-to-cart or purchase} & \multicolumn{2}{c}{Log seconds to detail view} \\ \cmidrule(lr){2-3}\cmidrule(lr){4-5}
& (1) & (2) & (3) & (4) \\
\midrule
Relative fit ($F$) & \TBD & \TBD & \TBD & \TBD \\
Within-product distinctiveness ($D^{W}$) & \TBD & \TBD & \TBD & \TBD \\
$F \times D^{W}$ (premium analogue) & \TBD & \TBD & \TBD & \TBD \\
$F \times D^{W} \times G$ (coherence analogue) & & \TBD & & \TBD \\
\midrule
Outcome mean & \TBD & \TBD & \TBD & \TBD \\
Clicked searches & \TBD & \TBD & \TBD & \TBD \\
\bottomrule
\end{tabularx}
\begin{tablenotes}[flushleft]
\footnotesize
\item Notes: Analysis output pending. One row per retained search; product, position, and week effects; two-way clustered standard errors. Exploratory; not part of the confirmatory tests.
\end{tablenotes}
\end{threeparttable}
\end{table}
}

\IfFileExists{exhibits/generated/tableA7_funnel.tex}{%
  \input{exhibits/generated/tableA7_funnel}
}{}

\subsection{Curvature in distinctiveness}
\label{app:differentiation-curvature}

Section~\ref{sec:theory} notes that novel designs attract interest while imposing processing costs, which implies that the favorable slope of distinctiveness need not be linear. Appendix~\ref{app:flexible-form} relaxes the fit dimension; this probe relaxes the distinctiveness dimension. Within-product distinctiveness enters through a restricted cubic spline with knots at quantiles (natural cubic regression basis with four degrees of freedom; the redundant column that sums with the others to a constant is dropped because the fixed effects absorb it), each basis column interacted with standardized fit, with the between-product interaction retained. Figure~\ref{fig:app-mechanism}(b) plots the implied change in click probability relative to the product mean over the central 90 percent of within-product distinctiveness at the 25th and 75th percentiles of fit, with the linear estimate from column 4 of Table~\ref{tab:main-models} dotted for reference, and the analysis file records the local slope at the center and at the upper end of the range at high fit. A processing-cost account predicts flattening or reversal at the top of the range; a linear specification predicts none.

\section{Model derivations}
\label{app:model}

This appendix states the individuation model of Section~\ref{subsec:model} exactly and proves the propositions. Throughout, one click occurs in the returned set \(C_s\) with \(J\) products; the search subscript is suppressed.

\subsection{Setup}

Under full individuation, choice follows merits \(m_i>0\), \(\sum_{i\in C}m_i=1\). Individuation of product \(i\) fails with probability \(q_i=q_v(D_i)\,q_t(T^{NN}_i)\), with \(q_v,q_t\in(0,1)\) continuously differentiable and strictly decreasing in their arguments. A product that fails individuation is conflated with its nearest visual substitute \(n(i)\). A conflated cluster \(\{i,j\}\) competes as one option with merit
\begin{equation}
\mu_{ij}=\max\{m_i,m_j\},
\label{eq:app-cluster-merit}
\end{equation}
capturing that a duplicated answer does not add appeal beyond its better member; a click landing on the cluster falls on member \(i\) with salience share \(\sigma_i\in(0,1)\), \(\sigma_i+\sigma_j=1\), determined by display position. When a cluster forms, the remaining options' merits renormalize by \(Z_{ij}=1-m_i-m_j+\mu_{ij}\le1\).

We work with the mutually nearest case, \(n(i)=j\) and \(n(j)=i\), which is exact for the twin analyses and is the leading case for the derivative results because only there does \(i\)'s own distance enter both failure events; the pair shares one distance \(D\) and one textual distance, so \(q_i=q_j=q\). Exact selection probabilities are
\begin{align}
P_i&=(1-q)\,m_i+q\,\frac{\sigma_i\,\mu_{ij}}{Z_{ij}},
\label{eq:app-exact}
\end{align}
and symmetrically for \(j\). The first term is the individuated branch, in which merits apply unchanged because renormalization affects only the conflated state; the second is the cluster branch. For a product whose nearest substitute is not mutual, the own-failure branch takes the same form with \(\mu\) and \(\sigma\) referring to its cluster, and the spillover it receives from other products' failures does not involve its own distances; those terms are collected in \(r_{is}\) of Equation~\ref{eq:model-reduced} and are of order \(1/J\) because each is bounded by a single product's merit.

Label the pair so that \(m_i\ge m_j\); then \(\mu_{ij}=m_i\) and \(Z_{ij}=1-m_j\le1\).

\subsection{Proof of Proposition~\ref{prop:distinctiveness}}

Differentiating Equation~\ref{eq:app-exact} in \(D\), writing \(q'=q_v'(D)\,q_t<0\),
\begin{equation}
\frac{\partial P_i}{\partial D}=-q'\Bigl(m_i-\frac{\sigma_i m_i}{1-m_j}\Bigr)
=-q'\,m_i\,\frac{1-m_j-\sigma_i}{1-m_j},
\qquad
\frac{\partial P_j}{\partial D}=-q'\Bigl(m_j-\frac{\sigma_j m_i}{1-m_j}\Bigr).
\label{eq:app-h1-exact}
\end{equation}
For the stronger member the derivative is positive whenever \(\sigma_i<1-m_j\), which holds for any interior salience share once merits are small; in a large set \(m_j=O(1/J)\). For the weaker member the sign can reverse when its salience share is large relative to its merit share, which is the sense in which conflation can favor a well-placed copy. The pair average is
\begin{equation}
\tfrac12\Bigl(\frac{\partial P_i}{\partial D}+\frac{\partial P_j}{\partial D}\Bigr)
=-\tfrac{q'}{2}\Bigl(m_i+m_j-\frac{m_i}{1-m_j}\Bigr)
=-\tfrac{q'}{2}\,\frac{m_j\,(1-m_i-m_j)}{1-m_j}>0,
\label{eq:app-h1-average}
\end{equation}
strictly positive whenever both merits are interior. Separation raises the pair's joint selection probability because a conflated pair competes with the merit of one product rather than two. In the large-\(J\) reduction \(m_j\to0\) at rate \(1/J\), Equation~\ref{eq:app-h1-exact} for the representative product becomes \(\partial P/\partial D=-q'\,m\,(1-\sigma)+O(J^{-2})\), which is the statement in the text with the salience-recapture factor \((1-\sigma)\) absorbed into the slope. \hfill\(\square\)

\subsection{Proof of Proposition~\ref{prop:scales}}

From the large-\(J\) reduction \(P_i=(1-q_i)m_i+r_i\) with \(\partial r_i/\partial D_i=0\) for non-mutual products,
\begin{equation}
\frac{\partial^2 P_i}{\partial D_i\,\partial m_i}=-q_v'(D_i)\,q_t>0 ,
\label{eq:app-crosspartial}
\end{equation}
which is constant in \(m_i\): the probability-scale interaction with any merit component is the derivative of the level effect. On the log scale, neglecting \(r_i\),
\begin{equation}
\log P_i=\log m_i+\log\bigl(1-q_v(D_i)\,q_t\bigr),
\label{eq:app-logscale}
\end{equation}
which is additively separable in \(m_i\) and \(D_i\); hence \(\partial^2\log P_i/\partial D_i\,\partial m_i=0\). For the stronger member of a mutual pair the exact expression is also multiplicative, \(P_i=m_i\bigl[(1-q)+q\sigma_i/(1-m_j)\bigr]\), so separability is exact there; for weaker members the cluster branch adds a term in \(m_j\) and the log-scale interaction is of order \(q\,\sigma\,m_j/m_i\), second order in a large set. Since the conditional logit estimates derivatives of log-odds of within-set probabilities, a null logit interaction alongside a positive linear-probability interaction is the model's joint prediction. \hfill\(\square\)

\subsection{Proof of Proposition~\ref{prop:channels}}

The magnitude of the visual-distance effect in Equation~\ref{eq:app-crosspartial} is \(\lvert q_v'(D)\rvert\,q_t(T^{NN})\,m\), increasing in \(q_t\) and hence decreasing in the textual separation of the product from its neighbor: the visual channel matters most where the textual channel fails. The symmetric statement follows by exchanging the channels. \hfill\(\square\)

\subsection{Proof of Proposition~\ref{prop:twins}}

Conditional on the click landing on the pair, from Equation~\ref{eq:app-exact},
\begin{equation}
\Pr(i \text{ wins}\mid \text{pair clicked})
=\frac{(1-q)\,m_i+q\,\sigma_i\,\mu_{ij}/Z_{ij}}{(1-q)\,(m_i+m_j)+q\,\mu_{ij}/Z_{ij}} .
\label{eq:app-duel}
\end{equation}
As \(D\to0\), \(q\to q_v(0)q_t\); if individuation essentially always fails for near-identical listings, \(q\to1\) and the expression tends to \(\sigma_i\), independent of merits. For interior \(q\) the merit advantage enters with weight proportional to \(1-q\), so between near-identical listings salience decides and merit breaks a fraction of the remaining tie. Equation~\ref{eq:app-duel} is the estimating object of the near-twin duel in Appendix~\ref{app:twin-duel}. \hfill\(\square\)

\subsection{Proof of Corollary~\ref{cor:setsize}}

With merits summing to one over \(J\) products, the representative merit is \(O(1/J)\). The slope \(-q_v'q_t\,m\) and the interquartile probability contrasts built from it are proportional to \(m\), so both shrink with \(J\); the cross-partial in Equation~\ref{eq:app-crosspartial} is scale-free in \(m\) but the interaction expressed in probability points against percentile movements of merit components again scales with the merit spread, which is \(O(1/J)\). \hfill\(\square\)

\subsection{Mapping to the estimating equations}

Log-linearizing \(q_v\) locally, \(q_v(D)\approx\bar q_v-\lambda_v D\) with \(\lambda_v>0\), the reduced form of Equation~\ref{eq:model-reduced} is \(P_{is}\approx m_{is}+\lambda_v q_t\,m_{is}\,D_{is}-\bar q_v q_t\,m_{is}+r_{is}\). Search effects absorb the set-level scale of merits, product effects absorb stable merit components and each product's average distance, and position indicators absorb the salience share's main effect, so the within-product coefficient on \(D\) estimates \(\lambda_v\,\overline{q_t\,m}\) and the fit interaction its variation with the merit component measured by fit. The model thus interprets \(\widehat\beta_W\) as the product of a perceptual slope and the average merit at stake, which is why Corollary~\ref{cor:setsize} expects both \(\widehat\beta_W\) and the probability-scale interaction to fall with returned-set size, and why neither is informative about preference for distinctive design. In the heterogeneity analysis of Appendix~\ref{app:heterogeneity}, the channel term \(q_t(T^{NN}_{is})\) is proxied by the tightness of the set's descriptions, the set-mean contextual textual distance, since the pairwise textual distance to the nearest visual substitute and the set's overall textual tightness move together.

 \clearpage
\section{Variable dictionary}
\label{app:dictionary}

Table~\ref{tab:app-variable-dictionary} lists the option-level variables. The frozen analytic file has one row per retained returned product; search-level quantities repeat across products within an event and product-level quantities across appearances.

\small
\begin{longtable}{@{}>{\raggedright\arraybackslash}p{0.23\textwidth}>{\raggedright\arraybackslash}p{0.15\textwidth}>{\raggedright\arraybackslash}p{0.54\textwidth}@{}}
\caption{Analytic variable dictionary}\label{tab:app-variable-dictionary}\\
\toprule
Variable & Level & Definition \\
\midrule
\endfirsthead
\multicolumn{3}{l}{\textit{Table \thetable\ continued}}\\
\toprule
Variable & Level & Definition \\
\midrule
\endhead
\bottomrule
\endfoot

\path{search_id} & Search & Deterministic identifier from source version and physical search-row number \\
\path{session_id} & Session & Anonymized shopping-session hash supplied by the dataset \\
\path{sku} & Product & Anonymized product identifier supplied by the dataset \\
\path{clicked} & Product-search & Indicator that the product belongs to the clicked array \\
\path{clicked_fraction} & Product-search & Clicked indicator divided by the number of distinct clicked products \\
\path{downstream_action} & Product-search & Indicator of an attributed add-to-cart or purchase event within the window \\
\path{detail_latency_s} & Search & Seconds from the search to the first detail view of the clicked product within the latency window; missing when unmatched (Appendix~\ref{app:click-quality}) \\
\path{set_size} & Search & Number of distinct returned products \\
\path{position} & Product-search & One-indexed location in the verified returned array \\
\path{semantic_fit} & Product-search & Aligned compatibility $\phi_{is}$: cosine between the retrieval-aligned query representation (Equation~\ref{eq:query-alignment}) and the description representation; the globally standardized level is a measurement contrast \\
\path{semantic_fit_raw} & Product-search & Unaligned query-description cosine; fails validation, kept for Appendix~\ref{app:fit-validation} \\
\path{fit_set_centered} & Product-search & Aligned compatibility minus the returned-set mean (relational variant) \\
\path{fit_set_rank} & Product-search & Relative fit $F_{is}$: within-set rank of aligned compatibility scaled to $[0,1]$ (primary fit measure) \\
\path{visual_diff_nn} & Product-search & Nearest-substitute distinctiveness $D_{is}$: one minus image similarity with the most similar returned competitor (primary distinctiveness measure) \\
\path{visual_diff_nn_within} & Product-search & Current nearest-substitute distinctiveness minus the product mean ($D^W_{is}$) \\
\path{visual_diff_nn_between} & Product & Product mean nearest-substitute distinctiveness minus the grand mean ($D^B_i$) \\
\path{nn_sku} & Product-search & Identifier of the nearest visual substitute in this returned set \\
\path{nn_position} & Product-search & Recorded position of the nearest visual substitute \\
\path{nn_position_gap} & Product-search & Absolute display separation from the nearest substitute (Appendix~\ref{app:adjacency}) \\
\path{nn_mutual} & Product-search & Indicator that the focal product and its nearest substitute are each other's nearest substitute \\
\path{nn_fit_gap} & Product-search & Focal minus nearest-substitute aligned compatibility \\
\path{nn_price_gap} & Product-search & Focal minus nearest-substitute price decile \\
\path{near_twin} & Product-search & Indicator that nearest-substitute distance falls in the lowest decile of the estimation sample \\
\path{fit_spread} & Search & Standard deviation of aligned compatibility within the returned set ($S_s$) \\
\path{fit_top_gap} & Search & Best minus second-best aligned compatibility within the returned set \\
\path{set_text_diff} & Search & Mean contextual textual differentiation within the returned set \\
\path{product_appearances} & Product & Eligible appearances of the product in the choice panel \\
\path{visual_diff} & Product-search & One minus average image cosine similarity with all other returned products (measurement contrast) \\
\path{visual_diff_trim} & Product-search & Average distance after removing the single most distant peer (measurement contrast) \\
\path{visual_diff_local} & Product-search & Average distance from peers within $k$ recorded positions \\
\path{visual_diff_highfit_peers} & Product-search & Average image distance from peers in the upper fit half of the returned set (Appendix~\ref{app:peer-split}) \\
\path{visual_diff_lowfit_peers} & Product-search & Average image distance from peers in the lower fit half of the returned set \\
\path{visual_diff_mean} & Product & Mean average-peer distance across eligible appearances \\
\path{visual_diff_within} & Product-search & Current average-peer distance minus the product mean \\
\path{visual_diff_between} & Product & Product mean average-peer distance minus the grand mean \\
\path{text_diff} & Product-search & One minus average description cosine similarity with all other returned products \\
\path{text_diff_within} & Product-search & Textual differentiation minus its product mean \\
\path{same_category_peers} & Product-search & Number of returned peers sharing the focal top-level category \\
\path{visual_diff_same_category} & Product-search & Average image differentiation from same-category peers when at least two exist \\
\path{category_coherence} & Product-search & Share of returned peers sharing the focal top-level category (moderation variant; nearly degenerate in this catalog) \\
\path{category_coherence_l2} & Product-search & Share of returned peers sharing the focal second-level category (primary exploratory moderator; defined when every returned product carries a usable second-level code) \\
\path{dominant_share} & Search & Share of the returned set in its largest top-level category \\
\path{homogeneous_set} & Search & Indicator that every returned product shares one top-level category \\
\path{price_rel} & Product-search & Price decile minus the returned-set mean decile \\
\path{seen_before} & Product-search & Indicator that the product was returned earlier in the session \\
\path{clicked_before} & Product-search & Indicator that the product was clicked earlier in the session \\
\path{event_weight} & Product-search & Inverse of returned-set size \\
\path{search_order} & Search & Chronological position of the search within its session \\
\path{shifted_week} & Search & Week index from the privacy-shifted timestamp \\
\path{warm_product} & Product-search & Indicator that the product appeared in the training period \\
\end{longtable}

%% file: exhibits/generated/tableA1_attrition.tex
\begin{table}[htbp]
\centering
\caption{Sequential sample construction and comparison of retained with excluded searches}
\label{tab:app-attrition}
\small
\begin{threeparttable}
\begin{tabularx}{\textwidth}{@{}Yrrr@{}}
\toprule
Step & Searches & Removed & Share of raw \\
\midrule
Raw search records & 819,516 &  & 1.000 \\
Parseable, nonempty returned arrays & 602,754 & 216,762 & 0.735 \\
Distinct returned product identifiers & 602,752 & 2 & 0.735 \\
Valid query representation & 602,752 & 0 & 0.735 \\
Exactly one distinct clicked product & 132,697 & 470,055 & 0.162 \\
Clicked product appears in returned array & 123,731 & 8,966 & 0.151 \\
At least five returned products & 93,549 & 30,182 & 0.114 \\
Complete catalog and representations & 56,931 & 36,618 & 0.069 \\
Usable top-level category & 56,931 & 0 & 0.069 \\
Final primary sample & 56,931 & 0 & 0.069 \\
Final sample: product-search rows / products / sessions & 1,049,523 & 24,383 & 53,334 \\
\midrule
\multicolumn{4}{@{}l}{\textit{Retained versus excluded searches with at least one click}}\\
& Retained & Excluded & Difference \\
Returned-set size & 18.435 & 15.464 & 2.971 \\
Number of distinct clicks & 1.000 & 1.631 & -0.631 \\
Search order within session & 1.731 & 1.893 & -0.163 \\
Weeks since first observed week & 5.138 & 7.002 & -1.864 \\
Category-homogeneous share & 0.744 & 0.769 & -0.024 \\
\bottomrule
\end{tabularx}
\begin{tablenotes}[flushleft]
\footnotesize
\item Notes: Counts at each step are cumulative and follow the fixed event-level exclusion sequence. The lower panel compares retained single-click searches with searches that had at least one click but failed a later eligibility rule; the homogeneous share is available only for searches whose full returned set could be mapped to the catalog.
\end{tablenotes}
\end{threeparttable}
\end{table}

%% file: exhibits/generated/tableA4_coherence.tex
\begin{table}[p]
\centering
\caption{Exploratory category-coherence moderation}
\label{tab:app-coherence}
\small
\begin{threeparttable}
\begin{tabularx}{\textwidth}{@{}Yrr@{}}
\toprule
& Estimate & SE \\
\midrule
Relative fit & $0.006^{***}$ & 0.001 \\
Within-product distinctiveness & $0.004^{***}$ & 0.001 \\
Category coherence & $-0.007^{***}$ & 0.001 \\
Fit $\times$ within-product distinctiveness & $0.002^{***}$ & 0.001 \\
Fit $\times$ coherence & $0.000$ & 0.001 \\
Within distinctiveness $\times$ coherence & $0.001^{*}$ & 0.000 \\
Fit $\times$ within distinctiveness $\times$ coherence & $0.000$ & 0.000 \\
Fit $\times$ between distinctiveness & $0.000$ & 0.001 \\
Between distinctiveness $\times$ coherence & $0.000$ & 0.001 \\
Fit $\times$ between distinctiveness $\times$ coherence & $0.001$ & 0.000 \\
\midrule
Observations / search events & \multicolumn{2}{r}{1,046,032 / 56,848} \\
\bottomrule
\end{tabularx}
\begin{tablenotes}[flushleft]
\footnotesize
\item Notes: Estimated on the subsample with defined second-level category coherence, with search, product, and recorded-position effects, equal-event weights, and two-way session and product clustering. Reported two-sided and unadjusted; the moderation is exploratory and its direction was not predicted. $^{*}p<0.10$, $^{**}p<0.05$, $^{***}p<0.01$.
\end{tablenotes}
\end{threeparttable}
\end{table}

%% file: exhibits/generated/tableA5_premium_diagnostics.tex
\begin{table}[p]
\centering
\caption{Is the goal-consistency premium a returned-set-size artifact?}
\label{tab:app-premium-diagnostics}
\small
\begin{threeparttable}
\begin{tabularx}{\textwidth}{@{}Yrrr@{}}
\toprule
& Estimate & SE & Searches \\
\midrule
\multicolumn{4}{@{}l}{\textit{Panel A: the premium inside narrow returned-set-size bands}}\\
Set size 5--7 & $-0.005$ & (0.005) & 8,003 \\
Set size 8--10 & $0.006$ & (0.004) & 5,521 \\
Set size 11--14 & $0.002$ & (0.003) & 6,551 \\
Set size 15--20 & $0.004^{**}$ & (0.002) & 5,265 \\
Set size 21--30 & $0.001^{**}$ & (0.000) & 31,591 \\
Set size 31+ & \multicolumn{3}{l}{insufficient} \\
\midrule
\multicolumn{4}{@{}l}{\textit{Panel B: fit and distinctiveness interacted with set-size bins}}\\
Premium, within size bins & $0.002^{***}$ & (0.000) & 1,047,903 \\
Distinctiveness, within size bins & $0.006^{***}$ & (0.002) & \\
\midrule
\multicolumn{4}{@{}l}{\textit{Panel C: catalog-wide representation placebo}}\\
Distinctiveness: observed percentile & 1.000 & (placebo mean 0.000) & 5 draws \\
Premium: observed percentile & 0.200 & (placebo mean 0.003) & 5 draws \\
\bottomrule
\end{tabularx}
\begin{tablenotes}[flushleft]
\footnotesize
\item Notes: Panel A estimates the premium model inside narrow bands of returned-set size, keeping the stage-3 standardization so the bands are comparable. Panel B returns to the full sample and interacts both relative fit and within-product distinctiveness with set-size bin indicators, so the premium is identified only within bins. Panel C permutes images across the whole catalog rather than within top-level category, which breaks the returned-set distance structure that a within-category permutation preserves when the set is category homogeneous. All models carry search, product, and recorded-position effects, equal-event weights, and two-way session and product clustering. $^{*}p<0.10$, $^{**}p<0.05$, $^{***}p<0.01$.
\end{tablenotes}
\end{threeparttable}
\end{table}

%% file: exhibits/generated/tableA3_twin_duel.tex
\begin{table}[!ht]
\centering
\caption{The near-twin duel: which member of a near-identical returned pair is chosen}
\label{tab:app-twin-duel}
\small
\begin{threeparttable}
\begin{tabularx}{\textwidth}{@{}Yccc@{}}
\toprule
& (1) & (2) & (3) \\
\midrule
Compatibility advantage over the twin & $0.029^{***}$ & $0.023^{***}$ & $0.024^{***}$ \\
 & (0.008) & (0.007) & (0.007) \\
Display-position advantage over the twin &  & $0.138^{***}$ & $0.138^{***}$ \\
 &  & (0.007) & (0.007) \\
Price advantage over the twin (cheaper) &  &  & $0.008$ \\
 &  &  & (0.008) \\
\midrule
Near-twin duels & 4,321 & 4,321 & 4,321 \\
Share won by the better-fitting twin & \multicolumn{3}{c}{0.558} \\
Share won by the higher-placed twin & \multicolumn{3}{c}{0.639} \\
Mean display-slot separation & \multicolumn{3}{c}{4.935} \\
\bottomrule
\end{tabularx}
\begin{tablenotes}[flushleft]
\footnotesize
\item Notes: The unit is a returned pair whose members are each other's nearest visual substitute, whose distance falls in the near-twin range, and exactly one of whose members was clicked. The dependent variable indicates that the focal member won the pair. Advantages are differences between the twins, standardized across pairs; a positive position advantage means the focal product was displayed higher. The query, the session, and the rest of the returned set are identical within a pair and therefore differenced out. Standard errors in parentheses are clustered by session. Exploratory; not part of the confirmatory tests. $^{*}p<0.10$, $^{**}p<0.05$, $^{***}p<0.01$.
\end{tablenotes}
\end{threeparttable}
\end{table}

%% file: exhibits/generated/tableA8_counterfactual.tex
\begin{table}[p]
\centering
\caption{Counterfactual click reallocation implied by the estimates}
\label{tab:app-counterfactual}
\small
\begin{threeparttable}
\begin{tabularx}{\textwidth}{@{}Yrr@{}}
\toprule
Policy and margin & Estimate & SE \\
\midrule
\multicolumn{3}{@{}l}{\textit{Panel A: restyling a near twin (clicks per 1,000 twin impressions)}} \\
Boundary restyle (exit the near-twin decile), all members & $2.102^{***}$ & (0.304) \\
Median restyle (to the median nearest-substitute distance), all members & $10.050^{***}$ & (1.482) \\
\quad bottom fit tercile & $5.071^{***}$ & (1.382) \\
\quad middle fit tercile & $10.134^{***}$ & (1.487) \\
\quad top fit tercile & $15.100^{***}$ & (2.515) \\
\quad text-tight sets (bottom tercile of textual separation) & $14.007^{***}$ & (2.500) \\
\quad text-spread sets (top tercile of textual separation) & $5.717^{***}$ & (0.902) \\
\midrule
\multicolumn{3}{@{}l}{\textit{Share model (median restyle, per 1,000 twin-containing searches)}} \\
Pair gain, from the rest of the set & 7.735 & [4.509, 11.051] \\
\quad of which the better-fitting member & 3.335 & \\
\midrule
\multicolumn{3}{@{}l}{\textit{Panel B: ranking the better-fitting twin first (near-twin duels)}} \\
Near-twin duels & 3,302 & \\
Share with the better-fitting twin ranked lower & 0.443 & \\
Better-fitting twin's win rate in those duels, factual & 0.439 & \\
\quad counterfactual, after the swap & 0.626 & \\
Clicks moved to the better-fitting twin per 1,000 misordered duels & 187.224 & (11.023) \\
\quad per 1,000 duels overall & 83.009 & \\
\bottomrule
\end{tabularx}
\begin{tablenotes}[flushleft]
\footnotesize
\item Notes: Model-implied, stylized counterfactuals under the historical retrieval and display policy; conditional associations, not causal effects, and conditional on the search producing a click, which the funnel analysis shows the within-set margin leaves unchanged. Panel A raises both members' nearest-substitute distance to the stated target and propagates the change through the channel-moderation linear probability model, so the implied gain varies with the member's relative fit and the set's textual separation; the stored quantities do not identify each member's second-nearest peer, so the boundary target bounds the move from below. The share model applies the conditional logit's coefficients multiplicatively to an empirical position-by-set-size baseline, renormalized within each search; the interval is a 200-draw parametric resample of the logit coefficients. Panel B estimates a linear probability model among near-twin duels from the better-fitting member's perspective and swaps the two display positions where that member sits lower, which leaves the pair's slots and the rest of the set unchanged. Fixed effects and all other covariates are held at their historical values. Two limits govern the reading and are developed in the text of this section: the move to the median distance averages 1.73 standard deviations of within-product distinctiveness and therefore extrapolates well beyond the estimation range, and display position in Panel B is chosen by the platform with signals absent from the release, so the swap figure is an upper bound on what a tie-break would deliver rather than an estimate of it. These are diagnostics for a prospective test, not deployment parameters. $^{*}p<0.10$, $^{**}p<0.05$, $^{***}p<0.01$.
\end{tablenotes}
\end{threeparttable}
\end{table}

%% file: exhibits/generated/tableA2_click_quality.tex
\begin{table}[p]
\centering
\caption{Click quality and decision latency among clicked products}
\label{tab:app-click-quality}
\scriptsize
\begin{threeparttable}
\begin{tabularx}{\textwidth}{@{}Yrrrr@{}}
\toprule
& \multicolumn{2}{c}{Add-to-cart or purchase} & \multicolumn{2}{c}{Log seconds to detail view} \\ \cmidrule(lr){2-3}\cmidrule(lr){4-5} & (1) & (2) & (3) & (4) \\
\midrule
Relative fit ($F$) & $0.006^{**}$ & $0.007^{***}$ & $0.010$ & $0.020$ \\
 & (0.003) & (0.003) & (0.032) & (0.034) \\
Within-product distinctiveness ($D^{W}$) & $-0.008^{***}$ & $-0.006^{***}$ & $0.016$ & $0.041$ \\
 & (0.002) & (0.002) & (0.028) & (0.030) \\
Category coherence ($G$) &  & $0.011^{***}$ &  & $0.003$ \\
 &  & (0.003) &  & (0.033) \\
$F \times D^{W}$ (premium analogue) & $0.002$ & $0.002$ & $-0.040$ & $-0.026$ \\
 & (0.001) & (0.002) & (0.030) & (0.029) \\
$F \times D^{B}$ & $0.002$ & $0.003$ & $-0.007$ & $0.000$ \\
 & (0.002) & (0.003) & (0.038) & (0.042) \\
$F \times G$ &  & $-0.001$ &  & $-0.021$ \\
 &  & (0.002) &  & (0.030) \\
$D^{W} \times G$ &  & $0.001$ &  & $0.030^{*}$ \\
 &  & (0.001) &  & (0.016) \\
$F \times D^{W} \times G$ (coherence analogue) &  & $0.000$ &  & $0.022$ \\
 &  & (0.001) &  & (0.022) \\
\midrule
Outcome mean & 0.116 & 0.116 & 4.566 & 4.566 \\
Clicked searches & 51,656 & 51,574 & 3,025 & 3,022 \\
Product, position, and week effects & \multicolumn{4}{c}{Yes} \\
\bottomrule
\end{tabularx}
\begin{tablenotes}[flushleft]
\footnotesize
\item Notes: Sample: the clicked product of each retained single-click search. Columns 1 and 2 use a linear probability model for an add-to-cart or purchase of the clicked product within the attribution window (Appendix D.3) on downstream-eligible searches; columns 3 and 4 use the log of seconds from the search to the first detail view of the clicked product within the latency window. Regressors are the stage-3 standardized variables, so coefficients are in the units of Table 2. Search effects are not available with one row per search; week effects replace them. Standard errors (in parentheses) are two-way clustered by session and product. Detail views were matched for 7.7 percent of clicked searches (median latency 90.8 seconds). $^{*}$, $^{**}$, $^{***}$: $p<0.10$, $0.05$, $0.01$. These estimates are exploratory and not part of the confirmatory tests.
\end{tablenotes}
\end{threeparttable}
\end{table}

%% file: exhibits/generated/tableA7_funnel.tex
\begin{table}[p]
\centering
\caption{The purchase funnel behind the clicks}
\label{tab:app-funnel}
\small
\begin{threeparttable}
\begin{tabularx}{\textwidth}{@{}Ycccc@{}}
\toprule
& Add & Purchase (window) & Purchase (session) & Abandoned \\
\midrule
Within-product distinctiveness & $-0.008^{***}$ & $-0.002^{**}$ & $-0.002^{***}$ & $-0.005^{***}$ \\
 & (0.002) & (0.001) & (0.001) & (0.001) \\
Relative fit & $0.006^{**}$ & $0.002$ & $0.002$ & $0.004^{*}$ \\
 & (0.003) & (0.001) & (0.001) & (0.002) \\
Fit $\times$ within-product distinctiveness & $0.002$ & $0.001^{**}$ & $0.001$ & $0.001$ \\
 & (0.001) & (0.001) & (0.001) & (0.001) \\
\midrule
Outcome mean & 0.116 & 0.033 & 0.039 & 0.078 \\
Clicked searches & 51,656 & 51,656 & 51,656 & 51,656 \\
\bottomrule
\end{tabularx}
\begin{tablenotes}[flushleft]
\footnotesize
\item Notes: One observation per retained single-click search, restricted to searches whose clicked product had no matching action earlier in the session. Attribution follows the pipeline rules: an event counts when it occurs after the search and before the earlier of the next retained search or the 30-minute window; the session-purchase column drops the window. Abandonment is an attributed add with no purchase for that product later in the session. Models carry product, recorded-position, and week effects; standard errors in parentheses are two-way clustered by session and product. $^{*}p<0.10$, $^{**}p<0.05$, $^{***}p<0.01$.
\end{tablenotes}
\end{threeparttable}
\end{table}